\documentclass[preprint]{revtex4}
\usepackage{amsmath}
\usepackage{amsfonts}
\usepackage{graphicx}

\begin{document}

\title{Review: Dynamics of Crowded Macromolecules/ \\ Interacting Brownian Particles}
\author{George D. J. Phillies}
\email{phillies@4liberty.net, 508-754-1859}
\affiliation{Department of Physics,	Worcester Polytechnic Institute, Worcester, MA 01609}
	
\begin{abstract}

I review theoretical treatments of diffusion in crowded (i.~e., non-dilute) solutions of globular macromolecules.  The focus is on the classical statistico-mechanical literature, much of which dates to before 1990. Classes of theoretical models include continuum treatments, correlation function descriptions, generalized Langevin equation descriptions, Smoluchowski and Mori-Zwanzig descriptions, and a brief but encouraging comparison with experimental results. The primary emphasis is on measurements made with quasi-elastic light scattering spectroscopy; I also discuss outcomes from fluorescence photobleaching recovery, fluorescence correlation spectroscopy,  pulsed-gradient spin-echo nuclear magnetic resonance, and raster image correlation spectroscopy. I close with a list of theoretical papers on the general topic.

\emph{This manuscript began as a chapter for someone else's book. For reasons not relevant here, the following will not see print there. I am therefore making a distribution to interested parties. Given the size, this is not a journal article; I am open to publication offers. In the long run, this article may become a chapter in my volume \emph{Theory of Polymer Solution Dynamics}, now in preliminary stages. Reports of typographic errors, points where the discussion is obscure, additions to Appendix C, and requests for additions are collegially welcomed.}

\end{abstract}

\maketitle

\tableofcontents

\section{Introduction}

\subsection{Plan of the Work}

Recently, there has been increased interest in crowding, the effect of intermacromolecular interactions on the dynamics of macromolecules in non-dilute solutions. There does not appear to be a firm contact between modern studies of diffusive motion in non-dilute macromolecule solutions and the classical theoretical and experimental literature on this topic. There have, after all, been more than four decades of intensive theoretical\cite{altenberger1973,phillies1974a,phillies1974b} and experimental\cite{phillies1973a,phillies1976a} studies on diffusion by non-dilute macromolecules, and by dilute probes\cite{gray1974a,turner1976a,lin1982a,ullmann1985a} in nondilute solutions of proteins or long-chain random-coil polymers.

This article focuses on one aspect of the historic literature, namely theoretical studies of diffusion by interacting macromolecules, especially as studied by quasielastic light scattering spectroscopy, fluorescence correlation spectroscopy, pulsed-gradient spin-echo NMR, raster image correlation spectroscopy, and related techniques. In particular, I treat the mutual, self, and probe diffusion coefficients $D_{m}$, $D_{s}$, and $D_{p}$ of hard spheres. A few results on bidisperse and charged systems are noted. For interacting particles, these diffusion coefficients depend on the concentration of the diffusing solutes.  The concentration dependences reflect a complicated interplay of many effects, including inter-particle direct and hydrodynamic interactions, reference frame corrections, and correlations between Brownian and driven motions. Some comparisons are made with experiment.

Calculations of diffusion coefficients may be categorized by their general approach. The discussion in this article is partitioned by those categories. Classical treatments are macroscopic, use the local concentration $c(\textbf{r},t)$ as a primary variable, and treat diffusive fluxes as being driven by so-called ``thermodynamic forces'' and hindered by ``dissipative coefficients''. The several microscopic approaches treat diffusing macromolecules as individual particles whose motions are modified by their intermacromolecular interactions. Included among the microscopic approaches are correlation function descriptions, generalized Langevin equation descriptions, and calculations based on the Mori-Zwanzig and Smoluchowski equations.

In the remainder of this section, we describe the types of diffusion coefficient. We then consider light scattering spectroscopy and methods for interpreting light scattering spectra. Section II treats continuum models for diffusion, including two- and three-component solutions and reference frame corrections.  Section III treats correlation function descriptions, which  are the starting point for microscopic calculations. Section IV uses Langevin-type equations to evaluate interesting correlation functions, including careful attention to the subtle correlations between the Brownian and direct components of the force on each particle. Section V briefly treats Smoluchowski and Mori-Zwanzig type calculations. Section VI is a discussion, including a short comparison with  experimental tests.

\subsection{Diffusion Coefficients}

The description of diffusion in terms of diffusion coefficients arises from classical experiments that observe diffusion over times that are very long relative to all microscopic molecular processes in solution. With light scattering spectroscopy, one can observe diffusive processes over short times and small distances, in which case information about diffusion more detailed than that given by the diffusion coefficient may be obtained.

Operationally, one can identify at least three different translational diffusion coefficients. The \emph{mutual} (or \emph{inter-}) diffusion coefficient $D_{m}$ characterizes the relaxation of a concentration gradient. The \emph{self} or \emph{tracer} diffusion coefficient $D_{s}$ describes the motion of a single macromolecule through a solution containing other macromolecules of the same species. The \emph{probe} diffusion coefficient $D_{p}$ determines the diffusion of an identifiable, dilute species through a complex fluid. There is also a \emph{rotational diffusion coefficient} that characterizes whole-body reorientation; this diffusion coefficient is accessible via depolarized light scattering\cite{phillies2011a}. Diffusion in solutions containing more than one solute component requires cross-diffusion coefficients for a complete characterization.

Quasi-elastic light scattering spectroscopy (QELSS) of monodisperse solutions measures $D_{m}$. $D_{m}$ can also be measured with a classical diffusion apparatus in which the disappearance of a macroscopic, artificially induced concentration gradient is observed.  The needed macroscopic, artificial concentration gradient can be produced in an analytical ultracentrifuge, permitting the mutual diffusion coefficient to be measured during a sedimentation-diffusion experiment. A true self-diffusion coefficient cannot be measured with QELSS, because laser light scattering is a coherent process.

Light scattering may be made incoherent via studying inelastic re-emission by fluorescent groups, either floating freely in solution or covalently bonded to larger molecules of interest. Fluorescence Photobleaching Recovery (FPR), Fluorescence Correlation Spectroscopy (FCS), and Raster Image Correlation Spectroscopy (RICS) take take advantage of this incoherent scattering to measure $D_{s}$ of fluorophores and fluorescently-tagged particles. Alternatively, one may measure $D_{s}$ by resorting to Pulsed-Gradient Spin-Echo Nuclear Magnetic Resonance (PGSE NMR).

Some techniques (e. g., fluorescence correlation spectroscopy) determine the diffusion coefficient of a labelled macromolecular species. If the system under study contains both dilute labelled macromolecules and also unlabeled macromolecules of the same species (perhaps at an elevated concentration $c$), and if the label does not perturb macromolecular motion, the apparent $D_{m}$ approaches closely $D_{s}$ of the macromolecules at the concentration $c$\cite{phillies1975a}.

There is no physical requirement that the labelled (``probe'') and unlabelled (``matrix'') macromolecules must except for the label be the same. If the probe and matrix species are different, one says that one is studying probe diffusion. If any of several stratagems for separating scattering due to probe particles from scattering due to the matrix solution is effective, QELSS may be used to measure the diffusion of probe particles through a background matrix solution.  For example, if the matrix molecules match the index of refraction $n$ of the solvent, so that the matrix molecules' $\frac{\partial n}{\partial c}$ vanishes, and if the probe is dilute, QELSS is readily used to measure $D_{p}$, the self-diffusion coefficient of the probes. In other cases, subtraction -- at the level of the field correlation functions -- of spectra of the matrix solution from spectra of matrix:probe solutions has permitted isolation and interpretation of the spectra of diffusing probes\cite{streletzky1995a}.

\subsection{Quasi-Elastic Light Scattering Spectroscopy}

Quasi-elastic light scattering spectroscopy, including both theoretical issues and experimental considerations, has been the subject of a series of monographs, including volumes from Berne and Pecora\cite{berne1976a}  Chu\cite{chu1974a,chu1991a}, Crosiganni, et al.\cite{crosignani1975a}, Cummins and Pike\cite{cummins1974a,cummins1977a},  Pecora\cite{pecora1985a}, and Schmitz\cite{schmitz1990a}. Readers are referred to these volumes for extended treatments of how the technique works. I present here only a very short summary.

Experimentally, in a QELSS system the liquid of interest is illuminated with a laser beam.  A series of lenses and/or irises is then used to collect the light scattered by the liquid through a narrow range of angles. The intensity $I(t)$ of the scattered light fluctuates. The intensity fluctuations are monitored using a photodetector and photon counting electronics. The actual signal being analysed is the count $n_{i}$ of photons  observed in each of a series of time intervals $(t, t+ \delta t)$.

The time-dependent intensity $I(t)$ is used to determine the intensity-intensity time correlation function
\begin{equation}
      C(\tau) = \langle I(t) I(t+\tau) \rangle.
      \label{eq:IICF}
\end{equation}
Here $\langle \cdots \rangle$ denotes an averaging process.

The information about the liquid appears in the time dependence of $C(\tau)$. Intensity measurements are made by photon counting; a real digital correlator actually determines
\begin{equation}
      C(\tau) = \sum_{i=1}^{\rm all} n_{i} n_{i+\tau}
      \label{eq:nnCF}
\end{equation}
where $n_{i}$ and $n_{i+\tau}$ are the number of photons that were counted in time intervals here labelled $i$ and $i+\tau$, and where the sum is over a large number of pairs of times $i$ and $i+\tau$ separated by the delay time $\tau$.

Scattering from a solution of Brownian particles is said to arise as scattering from a series of scattering centers.  For simple spherical particles, one has a scattering center at the center of each particle.  For polymers, which are not treated in this paper, the single scattering center is replaced with a line of scattering centers located along each polymer chain. The light that was scattered in the right direction is approximated as proceeding, without being  scattered again, to the detector. This approximation is the \emph{first-order Born approximation} for scattering. There is an extensive theoretical treatment, rarely invoked for QELSS, for scattering beyond this simple but usually adequate approximation; for a systematic treatment, see Kerker\cite{kerker1969a}.

As is shown in the standard sources, the fluctuating intensity $I(t)$ and its time correlation function $C(\tau)$ are determined by the locations of the scattering particles via the dynamic structure factor
\begin{equation}
S(q,\tau)=\left\langle \sum_{i,j,k,l=1}^{N} \sigma_{i} \sigma_{j} \sigma_{k} \sigma_{l} \exp(i\textbf{q} \cdot [\textbf{r}_{i}(t)-
\textbf{r}_{k}(t)+\textbf{r}_{l}(t+\tau)-\textbf{r}_{j}(t+\tau)])\right\rangle.
\label{eq1}
\end{equation}
In this equation, each variable in the quadruple sum proceeds over all $N$ particles, $\textbf{q}$ is the scattering vector, $\sigma_{i}^{2}$ is a scattering cross-section including all constants needed to convert from particle positions to intensities, and $\textbf{r}_{i}(t)$ and $\textbf{r}_{j}(t+\tau)$ are locations of particles $i$ and $j$ at times $t$ and $t+\tau$, respectively.

The actual correlator output is a bit more complicated than is suggested by this equation. In a simple linear correlator, in which the $n_{i}$ are all collected over equal time intervals, the actual output $C(\tau)$ of a digital autocorrelator is a wedge-weighted average of $S(q,\tau)$ over a range of delay times, the center of the wedge being the nominal correlation channel location and the width at half-height of the wedge being the time spacing $\Delta\tau$ between correlator channels. As an exception, $C(0)$ is not a good approximation to $S(q,0)$. In a modern \emph{multi-tau} correlator, with increasing $\tau$ the $n_{i}$ are aggregated into counts covering longer and longer time intervals. In a multitau correlator the averaging over delay times is more complicated. The delay time $\tau$ to be assigned to a channel requires careful analysis\cite{phillies1996z}. With respect to the importance of careful analysis, for which see ref.\ \onlinecite{phillies1996z}, suffice it to say that the so-called 'half-channel correction' is totally wrong for linear correlators, and at best a crude approximation for multitau correlators. Identifying $C(\tau)$ with $S(q,\tau)$ is possible, so long as the two values of $\tau$ are properly adjusted. This averaging matter is a purely experimental issue; in the following we assume that it has been handled correctly.

Closely related to the dynamic structure factor is the field correlation function or intermediate structure factor $g^{(1)}(q,\tau)$.
\begin{equation}
g^{(1)}(q,\tau)  = \left\langle \sum_{i,j=1}^{N} \sigma_{i} \sigma_{j} \exp(i\textbf{q} \cdot [\textbf{r}_{i}(t)-\textbf{r}_{j}(t+\tau)])\right\rangle.
\label{eq2A}
\end{equation}
Under normal experimental conditions (this restriction does not include some modern microscopic techniques) all information in $S(q,\tau)$ is actually contained in $g^{(1)}(q,\tau)$, so it is with the evaluation of the much simpler $g^{(1)}(q,\tau)$ in various guises that we are concerned in this paper. This article confines itself to scattering from large-volume systems, for which $g^{(1)}(q,\tau)$ can be extracted from $S(q,\tau)$ via
\begin{equation}
   S(q,\tau) = A(g^{(1)}(q,\tau))^{2} + B.
   \label{eq2}
\end{equation}
$A$ is an instrumental constant. $B$ is the baseline, the numerical value of $S(q,\infty)$ to which the correlation function decays.

The transition in eq.\ \ref{eq2} from $S(q,\tau)$ to $g^{(1)}(q,\tau)$ is an extremely good approximation. Crosignani, et al.\cite{crosignani1975a} provides the justification. Terms of the quadruple sum of eq.\ \ref{eq1} are only non-zero if the particles are close enough for their positions or displacements to be correlated.  Terms that put one or three particles in a scattering volume average to zero. So long as the regions over which particle positions and displacements are correlated -- the correlation volume -- are much smaller than the region from which scattered light is being collected, the sum has many more terms that put two particles in each of two correlation volumes than it has terms that put four particles in a single correlation volume. The error in the approximation in eq.\ \ref{eq2} is in the mistreatment of terms putting four particles in the same correlation volume; this error is small.

The $\tau \rightarrow 0$ limit of the pair correlation function is the static structure factor
\begin{equation}
g^{(1)}(q)\equiv g^{(1)}(q,0)=\left\langle \sum_{i=1}^{N}\sigma_{i}^{2} + \sum_{i\neq j=1}^{N} \sigma_{i} \sigma_{j} \exp(i\textbf{q} \cdot [\textbf{r}_{i}(t) - \textbf{r}_{j}(t)])\right\rangle,
\label{eq3}
\end{equation}
which describes equal-time correlations between the positions of particles in the solution. In eq. \ref{eq3}, the first and second summations are, respectively, the \emph{self} and \emph{distinct} terms of $g^{(1)}(q)$.

\subsection{Spectral Analysis}

How does one extract experimental parameters from $S(q,t)$? The first and most important issue is that light scattering spectra are relatively featureless, and therefore the number of parameters that can be extracted from a spectrum is extremely limited\cite{phillies1988z}. From one spectrum, a half-dozen parameters is often optimistic.  A spectrum having well-separated relaxations spread over four orders of magnitude in time and a very high signal-to-noise ratio may yield as many as eight parameters, though the scatter in repeated measurements will be substantial.

Four noteworthy approaches to analyzing $S(q,\tau)$ are cumulant analysis, lineshape fitting, moment analysis, and inverse Laplace transformation. We consider these seriatim. Cumulant analysis begins with the observation that the field correlation function can be written  formally as a sum of exponentials
\begin{equation}
      g^{(1)}(q,\tau) = \int_{0}^{\infty} d \Gamma A(\Gamma) \exp(-\Gamma \tau)
   \label{eq:laplace}
\end{equation}
Eq.\ \ref{eq:laplace} is the statement that $g^{(1)}(q,\tau)$ can be represented as a Laplace transform; it has no physical content.  The relaxation distribution $A(\Gamma)$ is normalized so that
\begin{equation}
      \int_{0}^{\infty} d \Gamma A(\Gamma) = 1.
   \label{eq:laplace1}
\end{equation}

The first moment of $A(\Gamma)$ is
\begin{equation}
       \bar{\Gamma} =\int_{0}^{\infty} d \Gamma A(\Gamma) \Gamma .
   \label{eq:gammabar}
\end{equation}
The central moments of $A(\Gamma)$ are
\begin{equation}
     \mu_{n} = \int_{0}^{\infty} d \Gamma A(\Gamma) (\Gamma - \bar{\Gamma})^{n}.
   \label{eq:gammabar1}
\end{equation}

In cumulant analysis, as introduced by Koppel\cite{koppel1972} and greatly improved by Frisken\cite{frisken2001a}, the spectrum is represented by
\begin{equation}
     g^{(1)}(q,\tau) = \exp\left(\sum_{n=0}^\infty\frac{K_{n}(-\tau)^{n}}{n!}\right)
\label{eq4}
\end{equation}
 The $K_{n}$ are the cumulants, which are not the same as the central moments except for small $n$, $K_{0}$ being the spectral amplitude and $K_{1}=\bar{\Gamma}$ being the average decay rate.

A traditional advantage of the cumulant expansion is that it can be applied by making a weighted linear least-squares fit, namely
\begin{equation}
\ln(g^{(1)}(q,\tau))\equiv \ln((S(q,\tau)-B))^{1/2}=\sum_{n=0}^{N} \frac{K_{n}(-\tau)^{n}}{n!}
\label{eq6}
\end{equation}
The $\ln$ and square root change the statistical weights to be assigned by the spectral fitting program to different data points. $N$ is a truncation parameter, the \emph{order} of the fit; it fixes the highest-order cumulant to be included in the fitting process.

Frisken\cite{frisken2001a} provides an alternative expansion
\begin{equation}
S(q,\tau)=\left[\exp(- K_{1} t) \left(1 + \frac{\mu_{2} t^{2}}{2!} +\frac{\mu_{3} t^{3}}{3!}+\ldots\right) \right]^{2}+B
\label{eq4F}
\end{equation}
in which the $\mu_{n}$ are the central moments.  In this expansion, $K_{1}$ and the $\mu_{n})$ are to be extracted from $S(q,\tau)$ via non-linear least sqare fitting methods, e.g., the simplex algorithm. (There are two very different classes of simplex algorithm. The useful one here is the Nelder-Mead\cite{nelder1965a} functional minimization approach.  The other simplex approach is Dantzig's linear programming method\cite{dantzig1998a}.) The first three $K_{n}$ are the same as the first three $\mu_{n}$, but the higher-order $K_{n}$ and $\mu_{n}$ are different.

Before the series are truncated, equations \ref{eq4}-\ref{eq4F} are all  ways to write $g^{(1)}(q,\tau)$ as a convergent exact power series, using expansion coefficients that are readily described in terms of the Laplace transform $A(\Gamma)$ of $g^{(1)}(q,\tau)$. The infinite series are therefore appropriate to describe an arbitrary $A(\Gamma)$. Some truncated series may have numerical stability issues if $\tau$ covers a wide range of times; the Frisken expansion of eq.\ \ref{eq4F} avoids these. Under practical conditions, one fits the complete $g^{(1)}(q,\tau)$ or $S(q,\tau)$ to eq. \ref{eq4}, eq.\ \ref{eq6}, or eq.\ \ref{eq4F} while varying $N$ upwards from 1, and uses the smallest $N$ such that further increases in $N$ do not substantially improve the quality of the fit.

The mutual diffusion coefficient obtained by light scattering is defined to be
\begin{equation}
D_{m}=\frac{K_{1}}{q^{2}}.
\label{eq5}
\end{equation}

Cumulants can be written as the logarithmic derivatives of $g^{(1)}(q,\tau)$ in the limit of small time delays, so that formally
\begin{equation}
K_{n}=\lim_{\tau\rightarrow 0}\left(-\frac{\partial}{\partial \tau}\right)^{n}\ln[g^{(1)}(q,\tau)].
\label{eq7}
\end{equation}
The presence of the limit in front of the derivative sometimes leads to the false assertion that cumulants only capture fast relaxations. Literal application of this formula in theoretical calculations of the $K_{n}$ is complicated by time scale issues. The physically observable $g^{(1)}(q,\tau)$ is only obtained for $\tau$ greater than the correlator channel width $\Delta t$, so the small-time limit in eq.\ \ref{eq7} does not include phenomena that relax completely in times $\ll \Delta t$.

An alternative to cumulants analysis is a forced fit of an assumed form to $S(q,\tau)$ or $g^{(1)}(q,\tau)$. In the earliest days of the field $S(q,\tau)$ was often force-fit to a single exponential.  In complex fluids and glassy liquids, relaxations often take the form of a stretched exponential $\exp(- \alpha \tau^{\beta})$ in time, or a sum of several stretched exponentials in time\cite{kob2007a,phillies2011a}. For interacting systems, the long-time part of the spectrum is sometimes found to follow a power-law decay
\begin{equation}
S(q,\tau)=a\tau^{\nu},
\label{eq8}
\end{equation}
$a$ and $\nu$ being fitting parameters. Long-time power-law decays plausibly arise from mode-coupling behavior. Power-law tails have been observed experimentally for light scattering spectra of strongly-interacting charged polystyrene spheres at very low salt concentrations.\cite{phillies1983z}.

For some purposes it is useful to determine the moments $M_n$ of the field correlation function, where for $n \geq 1$
\begin{equation}
M_n=\frac{\int_{0}^{\infty}d\tau \tau^{n-1}g^{(1)}(q,\tau)}{g^{(1)}(q,0)}.
\label{eq9}
\end{equation}
The moments have the physical advantage that they are defined as integrals, not polynomial fits or derivatives, so that (if a functional fit can be used to extrapolate $g^{(1)}(q,\tau)$ to $\tau \rightarrow \infty$) moments and cumulants are susceptible to different sorts of noise.

Moments give average diffusion coefficients, with the slowest decays weighted most heavily in the average. If the spectrum is written $g^{(1)}(q,\tau) = \int d\Gamma A(\Gamma)\exp(-\Gamma\tau)$, with $A(\Gamma)$ having been normalized so that $\int d\Gamma A(\Gamma)=1$, then $M_{1}= \langle 1/\Gamma\rangle$, and similarly for the higher-order moments.

In principal, spectra of interacting systems can also be fit to sums of exponentials via inverse Laplace transform methods. However, inverse Laplace transforms are ill-posed, so the outcomes are highly sensitive to noise. Some software reports the smoothest (in some sense) function that is consistent with the observed spectrum. The relationship between the smoothest function and the actual Laplace Transform can be obscure. Furthermore, most theories do not naturally lead to multi-exponential forms for spectra of monodisperse interacting systems.

We now advance to treatments of diffusion, beginning with the continuum treatments.

\section{Continuum Treatment of Light Scattering}

\subsection{Introduction to Continuum Treatments}

Continuum descriptions of concentration may be traced back to Fick's original memoir on diffusion. Continuum descriptions have been used extensively to analyse diffusion problems, and are sketched here. Continuum models generally omit details of the interactions between individual particles.

Many continuum models are based on non-equilibrium thermodynamics. Non-equilibrium thermodynamics asserts that diffusion currents may be written as products of forces and dissipation coefficients. This level of refinement was apparently adequate for early 20$^{th}$ century treatments of sedimentation and electrophoresis, in which particles moved under the influence of an external field that was independent of molecular coordinates, while dissipation coefficients were described in terms of averages over molecular coordinates. Some non-equilibrium thermodynamic models postulate non-mechanical ``thermodynamic driving forces'', whose magnitudes are not computed from classical statistical mechanics as applied to microscopic molecular systems.

Non-equilibrium thermodynamics is not consistent with the microscopic treatments given in later sections. If both the force and the dissipation are determined by molecular coordinates, then the current should be given by an average over the product of an instantaneous force and an instantaneous mobility, not by the product of an averaged force and an averaged mobility coefficient. Nonetheless, continuum models are invoked frequently. Having provided a necessary caveat for the following discussion, we turn to continuum models.

\subsection{Formulation of Continuum Treatments; Einstein Model}

In continuum treatments of diffusion, a solution is treated as having a continuum concentration $c(\textbf{r},t)$. In the continuum treatments, $\textbf{r}$ refers to a location in space, while $c(\textbf{r},t)$ refers to the density of scattering particles near $\textbf{r}$ at time $t$. This description has been extensively used to analyze diffusion problems. Note that in this Section the coordinate $\textbf{r}$ has entirely changed its meaning.  In the previous section, the $N$ scattering particles were assigned time-dependent coordinates $\{\textbf{r}_{1}, . . .,\textbf{r}_N\}$ that specified their locations as functions of time. In this section, each $\textbf{r}$ is a fixed location in space. In older works on diffusion and experimental methods for its study, authors will sometimes jump back and forth between these two meanings of $\textbf{r}$.

A simple continuum treatment of diffusion uses the continuity equation
\begin{equation}
\frac{\partial c(\textbf{r},t)}{\partial t}=-\nabla\cdot\textbf{J}(\textbf{r},t)
\label{eq71}
\end{equation}
and Fick's diffusion equation
\begin{equation}
\textbf{J}(\textbf{r},t)=-D_{F}\nabla c(\textbf{r},t)
\label{eq72}
\end{equation}
to obtain
\begin{equation}
\frac{\partial c(\textbf{r},t)}{\partial t}=D_{F}\nabla^{2}c(\textbf{r},t).
\label{eq73}
\end{equation}
Here $\textbf{J}(\textbf{r},t)$ is the time- and position-dependent concentration current and $D_{F}$ is the Fick's Law diffusion coefficient. In eq.\ \ref{eq73}, terms in $\nabla D_{F}=\frac{\partial D_{F}}{\partial c}\nabla c$ have been omitted, which is appropriate if the concentration dependence of $D_{F}$ is small enough.

The simplest continuum model for diffusion is due to Einstein.  Despite its simplicity, the model is adequate to predict a diffusion coefficient for dilute particles in terms of mechanical properties of the solute and standard statistico-mechanical considerations. Einstein considered a solution of highly dilute particles, each having buoyant mass $m$, floating in solution.  There is a gravitational field, strength $g$, so the potential energy of each particle is
\begin{equation}
      U = m g z,
      \label{eq:umgz}
\end{equation}
with $z$ being the vertical coordinate. From standard statistical mechanics, the equilibrium concentration of particles in solution is
\begin{equation}
      c(z,t)  = c_{0} \exp( - \beta m g z),
      \label{eq:umgz2}
\end{equation}
in which $c_{0}$  is the concentration of particles at $z=0$, which is presumed to be inside the solution. Here  $\beta = (k_{B} T)^{-1}$, where $k_{B}$ is Boltzmann's constant and $T$ is the absolute temperature.

However, from Fick's Law there will be a diffusion current
\begin{equation}
      J_{zD}(z,t)  =  D_{F} \beta m g c_{0} \exp( - \beta m g z),
      \label{eq:umgz3}
\end{equation}
the current being upward (for $m_{g} > 0$, as is not always the case) because the concentration is larger as one goes farther below the surface of the liquid.  As a result of the gravitational force $m g$, the solute particles fall at their terminal velocity $m g/f$, leading to an induced sedimentation current
\begin{equation}
      J_{zS}(z,t)  =  \frac{ m g}{f}  c_{0} \exp( - \beta m g z),
      \label{eq:umgz4}
\end{equation}
In this equation $f$ is the drag coefficient for motion at constant velocity. Even though each particle falls at the same terminal velocity, the sedimentation current depends on $z$ because the concentration of particles depends on $z$.

We are at equilibrium, so the concentration does not depend on time, meaning the diffusion and sedimentation currents must add to zero.  The only way the two currents can add to zero is if
\begin{equation}
      D_{F}  =  \frac{ k_{B} T}{f}.
      \label{eq:umgz5}
\end{equation}
Equation \ref{eq:umgz5} is the Einstein diffusion equation.  When combined with Stokes' Law \begin{equation}
       f = 6 \pi \eta a
      \label{eq:umgz6}
\end{equation}
for the drag coefficient of a sphere, one obtains the Stokes-Einstein equation
\begin{equation}
      D_{F}  =  \frac{k_{B} T}{6 \pi \eta a}
      \label{eq:umgz7}
\end{equation}
for the diffusion coefficient of a sphere. Here $\eta$ is the solution viscosity and $a$ is the sphere radius. Observe that the derivation is a 'just-so' story.  The value of $D_{F}$ was not calculated directly, for example from Newtonian mechanics.  Instead, it was shown that $D_{F}$ had to have a certain value, or the laws of classical mechanics and statistical mechanics would be violated.

How is the scattering intensity related to the concentration of particles in solution? The key step is to recognize that the scattered light is determined by the q$^{\rm th}$ spatial Fourier component $a_{q}(\tau)$ of the solution concentration, namely

\begin{equation}
   a_{q}(\tau)  =   \sum_{i=1}^{N} \sigma_{i}  \exp(i\textbf{q} \cdot \textbf{r}_{i}(\tau))
\label{eqfourier}
\end{equation}
so if there is only one solute component the intermediate structure factor becomes
\begin{equation}
g^{(1)}(q,\tau) \sim \sigma^{2} \langle a_{q}(0) a_{-q}(\tau) \rangle.
\label{eq74}
\end{equation}
Here $\sigma^{2}$ is proportional to the particle scattering cross-section. $\sigma^{2}$ includes all the constants, some substance-dependent, that show how the scattered light intensity depends on the size of the concentration fluctuations.

The Fourier components satisfy
\begin{equation}
c(\textbf{r},t)=\sum_{q} a_{q}(t) \cos(\textbf{q}\cdot\textbf{r}+\phi_{q}(t)),
\label{eq75}
\end{equation}
$\phi_q$ being a time-dependent phase unique to each Fourier component. The phase disappears if: The $a_{q}(t)$ are made complex, the cosine is replaced with a complex exponential, and the real part is taken.

Substitution of eq.\ \ref{eq75} into eq.\ \ref{eq73} gives the temporal evolution of the $a_{q}$ as
\begin{equation}
a_{q}(t) = a_{q}(0) \exp(-D_{F}q^{2}t).
\label{eq76}
\end{equation}

The scattered field mirrors the behavior of the concentration fluctuations via
\begin{equation}
g^{(1)}(q,\tau) \sim  \sigma^{2} \langle [a_{q}(0)]^{2} \rangle \exp(-D_{F}q^{2}\tau).
\label{eq77}
\end{equation}
Comparison with the models below identifies $D_{F}$ as the mutual diffusion coefficient $D_{m}$.

\subsection{Three-Component Systems}

By extending this treatment to a three-component solute:solute:solvent mixture, self- and probe diffusion can be discussed\cite{phillies1974a,phillies1974b}. Denoting the solute components as $A$ and $B$ with concentrations $c_{A}$ and $c_{B}$, respectively, the corresponding scattering cross-sections as $\sigma_{A}$ and $\sigma_{B}$, and the amplitudes of the corresponding $q^{th}$ spatial Fourier components of the two concentrations as $a_{qA}(t)$ and $a_{qB}(t)$, respectively, the electric field of the scattered light can be written
\begin{equation}
E_{s}(q,t) \sim \sigma_{A} a_{qA}(t)+\sigma_{B} a_{qB}(t).
\label{eq78}
\end{equation}
In general, the field correlation function is
\begin{equation}
    g^{(1)}(q,\tau) =\langle E_{s}(q,t)E_{s}(q,t+\tau) \rangle.
    \label{eq:g1e}
\end{equation}
Within the continuum models the temporal evolution of the $a_{qi}(t)$ is determined by diffusion equations. In a macroscopic description of a non-dilute solution, a diffusion current of either species is driven by concentration gradients of both species, so that
\begin{equation}
   \textbf{J}_A(\textbf{r},t) = -D_A \nabla c_{A}(\textbf{r},t) -D_{AB} \nabla  c_{B}(\textbf{r},t)
   \label{eq79}
\end{equation}
and
\begin{equation}
     \textbf{J}_B(\textbf{r},t) = -D_{BA} \nabla c_{A}(\textbf{r},t) - D_{B} \nabla  c_{B}(\textbf{r},t),
    \label{eq80}
\end{equation}
$\textbf{J}_{i}$ being the current of species $i$. In the above equations, the $D_{i}$ are single-species diffusion coefficients, while the $D_{ij}$ are cross-diffusion coefficients.

Applying the continuity equation, for concentration fluctuations one has
\begin{equation}
    \frac{\partial a_{qA}(t)}{\partial t} = -\Gamma_{A} a_{qA}(t) -\Gamma_{AB} a_{qB}(t)
    \label{eq81}
\end{equation}
and
\begin{equation}
    \frac{\partial a_{qB}(t)}{\partial t} = -\Gamma_{BA} a_{qA}(t) -\Gamma_{B} a_{qB}(t).
    \label{eq82}
\end{equation}
Here $\Gamma_{i}=D_{i} q^{2}$. These simultaneous equations have as solutions
\begin{equation}
a_{qA}(t) = \frac{a_{qA}(0)}{\Gamma^{+} - \Gamma^{-}} \left[(\Gamma_{A}-\Gamma^{-})\exp(-\Gamma^{+}t)+(\Gamma^{+}-\Gamma_{A})\exp(-\Gamma^{-}t)
     \right]\notag
\end{equation}
\begin{equation}
+\left(\frac{a_{qB}(0)\Gamma_{AB}}{\Gamma^{+} - \Gamma^{-}}\right) \left[\exp(-\Gamma^{+}t)-
\exp(-\Gamma^{-}t)\right]
\end{equation}
and
\begin{equation}
a_{qB}(t)=\frac{a_{qB}(0)}{\Gamma^{+}-\Gamma^{-}}\left[(\Gamma_{B}-\Gamma^{-})
\exp(-\Gamma^{+}t)+(\Gamma^{+}-\Gamma_{B})\exp(-\Gamma^{-}t)\right]\notag
\end{equation}
\begin{equation}
+\left(\frac{a_{qA}(0)\Gamma_{BA}}{\Gamma^{+}-\Gamma^{-}}\right)\left[\exp(-\Gamma^{+}t)-
\exp(-\Gamma^{-}t)\right].
\label{eq83}
\end{equation}
The predicted spectrum is
\begin{equation}
g^{(1)}(q,t)=I_{+}\exp(-\Gamma^{+}t)+I_{-}\exp(-\Gamma^{-}t).
\label{eq84}
\end{equation}
The spectrum of a two-macrocomponent mixture is thus a sum of two exponentials, corresponding to two independent relaxational modes of the system. The decay constants of the two modes are
\begin{equation}
\Gamma^{\pm}=\frac{1}{2}(\Gamma_{A}+\Gamma_{B})\pm
\left[\left(\frac{\Gamma_{A}-\Gamma_{B}}{2}\right)^{2}+
\Gamma_{AB}\Gamma_{BA}\right]^{1/2}.
\label{eq85}
\end{equation}
The $\Gamma^{\pm}$ in general depend on the diffusion coefficients of both solute species.

The relaxation modes do not correspond one-to-one to solution components. Each mode intensity depends on the scattering powers and interactions of both species. Defining $\alpha=\langle a_{qA}(0)^{2}\rangle$, $\beta=\langle a_{qB}(0)^{2}\rangle$, $\gamma = \langle a_{qA}(0)a_{qB}(0)\rangle$, and $A=\sigma_{A}^{2}\alpha+\sigma_{B}^{2}\beta+2\sigma_{A}\sigma_{B}\gamma$, the intensities are
\begin{equation}
I_+=\frac{1}{A(\Gamma^{+}-\Gamma^{-})}\left[(\Gamma_{A}-\Gamma^{-})(\sigma_{A}^{2}\alpha
+\sigma_{A}\sigma_{B}\gamma)+(\Gamma^{+}-\Gamma_{A})(\sigma_{B}^{2}\beta+\sigma_{A}\sigma_{B}
\gamma)\right.\notag
\end{equation}
\begin{equation}
\left.+\Gamma_{AB}(\sigma_{A}^{2}\gamma+\sigma_{A}\sigma_{B}\beta)+\Gamma_{BA}(\sigma_{B}^{2}
  \gamma+\sigma_{A}\sigma_{B}\alpha)\right]
\label{eq86}
\end{equation}
and
\begin{equation}
I_-=\frac{1}{A(\Gamma^{+}-\Gamma^{-})}\left[(\Gamma^{+}-\Gamma_{A})(\sigma_{A}^{2}\alpha
 +\sigma_{A}\sigma_{B}\gamma)+(\Gamma_{A}-\Gamma^{-})(\sigma_{B}^{2}+\sigma_{A}\sigma_{B}
   \gamma)\right.\notag
\end{equation}
\begin{equation}
\left.-\Gamma_{AB}(\sigma_{A}^{2}\gamma+\sigma_{A}\sigma_{B}\beta)-
 \Gamma_{BA}(\sigma_{B}^{2}\gamma+\sigma_{A}\sigma_{B}\alpha)\right].
\label{eq87}
\end{equation}
The intensity of each mode depends on the diffusion coefficients and scattering powers of both species. If only one species scatters light significantly but both species are non-dilute, the spectrum is a double exponential. If both solutes are dilute, $\gamma$ and the $\Gamma_{ij}$ vanish. In this limit eqs.\ \ref{eq84}-\ref{eq87} reduce correctly to the low-concentration normalized form
\begin{equation}
g^{(1)}(q,\tau)=\frac{\sigma_{A}^{2}\alpha \exp(-\Gamma_{A}\tau)+\sigma_{B}^{2}\beta\exp(-\Gamma_{B}\tau)}{\sigma_{A}^{2}\alpha+
\sigma_{B}^{2}\beta}
\label{eq88}
\end{equation}
in which each exponential corresponds to the diffusion of a particular chemical species.

\subsection{Mutual, Self, and Probe Diffusion Coefficients}

The above results actually describe the mutual, self, and probe diffusion coefficients. To measure $D_{m}$, one examines a system containing a single macrocomponent $A$, in which case $\epsilon_{B}$, $\beta$, $\gamma$, and the $\Gamma_{ij}$ are zero, so $\Gamma^{+}=\Gamma_{A}$, $I_{+}=1$, and $I_{-}=0$. The field correlation function reduces to a single exponential $\exp(-\Gamma_{A}t)$. The continuum theory thus predicts that a mutual diffusion experiment on a binary system measures $\Gamma_{A}$. The exponential decay constant $\Gamma_{A}/q^{2}$ may be identified with $D_{F}$ or with $D_{m}$ of the microscopic theories.

In a probe diffusion experiment, one of the components $A$ is dilute, while the other component $B$ scatters no light. Identifying $A$ as the probe, the model requires $\sigma_{B}=0$, $\gamma=0$, and $\Gamma_{AB}=0$. The final two equalities arise because $A$ is dilute, so almost all $B$ particles are distant from any $A$ particle and do not have their motions or positions perturbed by the presence of $A$. With these values for model parameters, the continuum treatment predicts $\Gamma^{+}=\Gamma_{A}$, $\Gamma^{-}=\Gamma_{B}$, $I^{+}=1$, and $I^{-}=0$, so
\begin{equation}
g^{(1)}(q,\tau)=\exp(-\Gamma_{A}\tau)
\label{eq89}
\end{equation}
Under probe conditions, according to the continuum model QELSS obtains $D_{A}$ of the probe, $D_{A}$ now being the diffusion coefficient of the dilute probe particles through a pseudobinary solution. No matter the concentration of the matrix, neither $\Gamma_{B}$ nor $\Gamma_{AB}$ enters the spectrum. In favorable cases, the requirement ''...scatters no light...'' can be relaxed, as discussed below.

In a probe experiment, fluctuations in the concentration of a concentrated matrix $B$ may create currents of species $A$. Why don't these currents contribute to the spectrum? Fluctuations $a_{qB}(0)$ do act on $A$ particles, contributing a non-zero term to $\frac{\partial a_{qA}}{\partial t}$. However, in the tracer limit, the currents that $B$ induces in $A$ are uncorrelated with the initial concentration fluctuations in $A$, so they are equally likely to enhance or diminish the fluctuations $a_{qA}(t)$ of species $A$.  The concentration fluctuations driven by the $B$-$A$ cross-coupling are not correlated with $a_{qA}(0)$ and do not within the continuum picture affect the spectrum $\langle I(0)I(\tau)\rangle$.

In addition to the mutual, self, and probe diffusion coefficients, the classical literature includes references to a ``tracer'' diffusion coefficient. A tracer diffusion experiment uses an unlabeled species $A$ and its labelled twin $A^{*}$. The label allows one to identify the $A^{*}$ molecules, but $A$ and $A^{*}$ are elsewise identical. The requirement that the two species be elsewise identical is more demanding than it sounds, especially for large molecules.  For long-chain hydrocarbon polymers, even perdeuteration may lead to issues.

In a classical tracer experiment, one creates a non-equilibrium system containing macroscopic, countervailing gradients in the concentrations of $A$ and $A^{*}$, the gradients being so arranged that the total concentration $c_{A}+c_{A^*}$ is everywhere the same. This arrangement of concentrations arises in fluorescence photobleaching recovery, in which $A$ and $A^{*}$ correspond to the bleached and unbleached molecules under consideration, with the total concentration $c_{A}+c_{A^*}$ being the same as the pre-bleaching concentration of unbleached and not-yet-bleached molecules.  The flux of $A^{*}$ down its concentration gradient is measured, the tracer diffusion coefficient being
\begin{equation}
\textbf{J}_{A^*}=-D_{T}\nabla c_{A^*}(\textbf{r},t)
\label{eq90}
\end{equation}
As seen in the next section, the use of countervailing concentration gradients cancels any gradients in the non-ideal parts of the chemical potentials of $A^{*}$ and $A$. A discussion of the physical interpretation of the tracer diffusion coefficient resumes after a treatment of reference frames and irreversible thermodynamics.

\subsection{Reference Frames,  Irreversible Thermodynamics}

This subsection considers reference frames and their implications, following closely the discussion of Kirkwood et al.\cite{kirkwood1960}. We discuss \emph{practical diffusion coefficients} $D_{ij}$ and also \emph{fundamental diffusion coefficients} $\Omega_{ij}$.  We also consider implications of the Onsager reciprocal relations, as they arise in irreversible thermodynamics.

The practical diffusion coefficients relate the diffusion currents to the concentration gradients via
\begin{equation}
\textbf{J}_{i}=\sum_{j=1}^{q} D_{ij}\nabla c_{j}(\textbf{r},t).
\label{eq91}
\end{equation}

The fundamental diffusion coefficients relate the diffusion currents to the chemical potential gradients via
\begin{equation}
\textbf{J}_{i}=\sum_{j=0}^{q} \Omega_{ij} \nabla \mu_{j}(\textbf{r},t).
\label{eq92}
\end{equation}

The Gibbs-Duhem equation\cite{gibbs1876} constrains the chemical potentials $\mu_{j}$ by
\begin{equation}
\sum_{j=0}^{q} c_{j}\frac{\partial \mu_{j}}{\partial x}=0.
\label{eq93}
\end{equation}

Here sums pass over thermodynamic components $\{0,1, . . . ,q\}$, $0$ denoting the solvent. Gibbs\cite{gibbs1876} emphasizes that choosing a particular component as the solvent is an arbitrary act. Any one component may be identified as the solvent; the other components are then solutes. Thermodynamic chemical components need not correspond to chemical compounds in a simple way, a result particularly useful in treating ionic solutions. In the following discussion it is convenient to use mass units, so that $c$ is in gram$\cdot$cm$^{-3}$, $\bar{v}_{i}$ is the partial volume per gram of solute $i$, etc.

Reference frames appear implicitly in the continuum analysis. Eq.\ \ref{eq91} relates diffusive currents $\textbf{J}_{i}(\textbf{r},t)$ to local concentration gradients $\nabla c_{j}(\textbf{r},t)$. Elementary physical considerations show that velocities (and therefore currents) only have meaning when the local zero of velocity\textemdash{}the reference frame\textemdash{}is specified. As a notational matter, reference frames are indicated by an exterior subscript $R$, so $(\textbf{J}_{i})_{R}$ denotes the current of $i$ as observed in the $R$ reference frame. To determine a current in a frame $A$ from the current in a frame $B$, one has
\begin{equation}
(\textbf{J}_{i})_{A}=(\textbf{J}_{i})_{B}-c_{i}\textbf{v}_{AB},
\label{eq94}
\end{equation}
$\textbf{v}_{AB}$ being the velocity of frame $A$ as measured in frame $B$.

At least four frames are useful in a continuum description of diffusion. In the mass-fixed frame $M$, the center of mass of the system does not move, so
\begin{equation}
\sum_{i=0}^{q}(\textbf{J}_{i})_{M} = 0.
\label{eq95}
\end{equation}
Onsager\cite{onsager1931} proposed that the fundamental diffusion coefficients are subject to symmetry constraints\textemdash{}the ``Onsager reciprocal relations''\textemdash{}applicable in the mass-fixed frame:
\begin{equation}
(\Omega_{ij})_{M}=(\Omega_{ji})_{M}.
\label{eq96}
\end{equation}

In the solvent-fixed frame $0$, the solvent is stationary:
\begin{equation}
(\textbf{J}_{0})_{0}=0.
\label{eq97}
\end{equation}
The solvent current in question is the bulk flow of solvent. Interparticle hydrodynamic interactions are hidden in the diffusion tensors. Since the fundamental hydrodynamic interaction tensors $\textbf{b}$ and $\textbf{T}$ as discussed in Section IV are derived by requiring that solvent flow vanishes as $r\rightarrow\infty$, it is often assumed that hydrodynamic calculations are made in the solvent-fixed frame. Microscopic calculations of later sections refer to results in the experimentally-accessible volume-fixed frame.

The relative velocity of the mass- and solvent-fixed frames is obtained by writing eq.\ \ref{eq94} for component $0$, and applying eq.\ \ref{eq97}, showing
\begin{equation}
v_{0M}=  \frac{(\textbf{J}_{0})_{M}}{c_{0}}.
\label{eq98}
\end{equation}
One can define a new set of fundamental diffusion coefficients $(\tilde\Omega_{ij})_{0}$, which satisfy the Onsager reciprocal relations in the solvent-fixed frame if the $(\Omega_{ij})_{M}$ satisfy these relations. From eq.\ \ref{eq98}
\begin{equation}
(\textbf{J}_{i})_{0}=(\textbf{J}_{i})_{M}-\frac{c_{i}}{c_{0}}(\textbf{J}_{0})_{M}.
\label{eq99}
\end{equation}

Using eq.\ \ref{eq92} to express the $(\textbf{J}_{i})_{0}$ in terms of the fundamental diffusion coefficients in the mass-filled frame, and applying the Gibbs-Duhem equation to eliminate reference to the chemical potential of the solvent, one may write
\begin{equation}
(\textbf{J}_{i})_{0}=\sum_{j=1}^{q}(\tilde\Omega_{ij})_{0}\nabla \mu_{j},
\label{eq100}
\end{equation}
in which the $(\tilde\Omega_{ij})_{0}$, defined by
\begin{equation}
(\tilde\Omega_{ij})_{0}=(\Omega_{ij})_{M}-\frac{c_{i}}{c_{0}}(\Omega_{0j})_{M}
-\frac{c_{j}}{c_{0}}(\Omega_{i0})_{M}+\frac{c_{i}c_{j}}{c_{0}^{2}}(\Omega_{00})_{M},
\label{eq101}
\end{equation}
manifestly have the same symmetry as the $(\Omega_{ij})_{M}$.

Actual experimental data refers not to the solvent-fixed frame but to the cell-fixed frame $c$. In many systems, volume of mixing effects are small. In this case, the cell-fixed frame and the volume-fixed frame $V$, defined by
\begin{equation}\sum_{j=0}^{q}(\textbf{J}_{j})_{V}\bar{v}_{j}=0,
\label{eq102}
\end{equation}
are the same. The partial volume of component $j$ is $\bar{v}_{j}$. If volume-of-mixing effects are not small, in a classical gradient diffusion experiment the interdiffusion of the two diffusing components is accompanied by bulk flow and a change in the volume of the solution.

To unite experimental data with theoretical results and with ratioanles dependent upon the Onsager reciprocal relations, the volume- and solvent-fixed frames must be linked. The relative velocity of the solvent- and volume-fixed frames is obtained from eq.\ \ref{eq94} by multiplying by $\sum_{i}\bar{v}_{i}$ and applying eqs.\ \ref{eq97} and \ref{eq102}, giving
\begin{equation}
v_{V0}=-\sum_{i=1}^{q}\bar{v}_{i}(\textbf{J}_{i})_{V},
\label{eq103}
\end{equation}
from which follows the relationship between the practical diffusion coefficients in the two frames. Using eq.\ \ref{eq91} to replace the $\textbf{J}_{i}$ with the $D_{ij}$
\begin{equation}
(D_{ij})_{V}=(D_{ij})_{0}-c_{i}\sum_{k=1}^{q}\bar{v}_{k}(D_{kj})_{0}, {} i,j\in(1,q)
\label{eq104}
\end{equation}
or
\begin{equation}
(D_{ij})_{0}=(D_{ij})_{V}+\frac{c_{i}}{c_{0}\bar{v}_{0}}\sum_{k=1}^{q}\bar{v}_{k}
(D_{kj})_{0}, \quad i,j\in(1,q)
\label{eq105}
\end{equation}
To obtain the $\Omega_{ij}$ from the $D_{ij}$, note that eq.\ \ref{eq92} can be written
\begin{equation}
(\textbf{J}_{i})_{0}=\sum_{k,l=1}^{q}(\tilde\Omega_{ik})_{0}\frac{\partial\mu_{k}}{\partial c_{l}}\nabla c_{l}.
\label{eq106}
\end{equation}
Eq.\ \ref{eq91} then implies that the practical and fundamental diffusion coefficients obey
\begin{equation}
(D_{il})_{0}=\sum_{k=1}^{q}(\tilde\Omega_{ik})_{0}\frac{\partial\mu_{k}}{\partial c_{l}}.
\label{eq107}
\end{equation}
Matrix inversion techniques applied to the $\frac{\partial\mu_{k}}{\partial c_{l}}$ can be used to calculate the $(\tilde\Omega_{ij})_{0}$ in terms of the $(D_{ij})_{0}$, permitting experimental tests of the Onsager reciprocal relations\cite{kirkwood1960}.

As we will see below, the reference frame correction goes away if one uses the correct form for the interparticle hydrodynamic interaction tensors, but the reference frame term connecting the solvent- and volume-fixed frames reappears in a new guise with very nearly the same net effect.

\subsection{Generalized Stokes-Einstein Equation}
We now turn to the so-called Generalized Stokes-Einstein equation, which connects the mutual diffusion coefficient to the self diffusion coefficient at the level of approximation of the continuum models. We begin by showing that reference frame treatments serve to connect $D_{m}$ and $D_{s}$ to the $\tilde\Omega_{ij}$ and to the concentration dependences of the chemical potentials.

Specializing eq.\ \ref{eq104} to a system with $q=1$ shows $(D_{11})_{V}$ in a binary solvent:solute system is the usual mutual diffusion coefficient:
\begin{equation}
(D_{11})_{V}=(D_{11})_{0}(1-\phi),
\label{eq108}
\end{equation}
where $\phi=c_{1}\bar{v}_{1}$ is the volume fraction of the macrocomponent in solution. The $1-\phi$ factor is the well-known reference frame correction to $D_{m}$.

The self-diffusion coefficient may be measured by labeling a few solute molecules, so that we have a labelled species $1$ and an unlabeled species $2$, establishing in the system countervailing gradients $\nabla c_{1}(\textbf{r},t)$ and $\nabla c_{2}(\textbf{r},t)$ so arranged that the total concentration   $c_{1}(\textbf{r},t) +  c_{2}(\textbf{r},t)$ is everywhere the same, and measuring the flux of the dilute labelled species $1$.

The continuum model was applied to this problem in refs.\ \onlinecite{phillies1974a} and \onlinecite{phillies1974b}, whose treatment is now followed. Under these conditions, $D_{s}$ may be defined
\begin{equation}
\frac{\partial c_{1}}{\partial t}=(D_{s})_{V}\nabla ^{2}c_{1}(\textbf{r},t).
\label{eq109}
\end{equation}
In terms of eq.\ \ref{eq91} (written for two components), eq.\ \ref{eq109} becomes
\begin{equation}
\frac{\partial c_{1}}{\partial t}=((D_{11})_{V}-(D_{12})_{V})\nabla ^{2}c_{1}(\textbf{r},t).
\label{eq110}
\end{equation}

The $(D_{ij})_{V}$ can be expressed in terms of fundamental diffusion coefficients in the solvent-fixed frame by applying eqs.\ \ref{eq104} and \ref{eq107}, showing
\begin{equation}
(D_{ij})_{V}=\sum_{l=1}^{q}\left[(\tilde\Omega_{il})_{0}-\sum_{k=1}^{q}
  c_{i}\bar{v}_{k}(\tilde\Omega_{kl})_{0}\right]\frac{\partial\mu_{l}}{\partial c_{j}}.
\label{eq111}
\end{equation}

As seen in refs.\ \onlinecite{phillies1974a} and \onlinecite{phillies1974b}, the chemical potential derivatives can be written
\begin{equation}
\frac{\partial\mu_{l}}{\partial c_{j}}=\delta_{ij}\frac{k_{B}T}{c_{j}}+I
\label{eq112}
\end{equation}
to first order in $c_{j}$, $\delta_{ij}$ being a Kronecker delta and $I$ being an interaction integral. Similarly, the $\Omega_{ij}$ may \emph{formally} be written
\begin{equation}
\Omega_{ii}=\frac{c_{i}}{f_{i}}
\label{eq113}
\end{equation}
and
\begin{equation}
\Omega_{ij}=\frac{c_{i}c_{j}}{f_{ij}},  i\neq j
\label{eq114}
\end{equation}
$f_{i}$ and $f_{ij}$ being formal dissipative factors. The concentration dependences of $\mu_{l}$ and $\Omega_{ij}$ ensure that the Onsager reciprocal relations are satisfied, and that in a $1$-component system
\begin{equation}
(D_{11})_{V}=(\tilde\Omega_{11})_{0}\left(\frac{\partial\mu_{1}}{\partial c_{1}}\right)(1-\phi)
\label{eq115}
\end{equation}
or equivalently that
\begin{equation}
(D_{m})_{V}=\frac{c_{1}\frac{\partial\mu_{1}}{\partial c_{1}}}{f_{1}}(1-\phi).
\label{eq116}
\end{equation}
On the other hand, from the above
\begin{equation}
D_{T}=\left((\tilde\Omega_{11})_{0}-(\tilde\Omega_{12})_{0}\right)\frac{k_{B}T}{c_{1}}-\phi_{1}\left(\frac{k_{B}T}{c_{1}}-\frac{k_{B}T}{c_{2}}\right)(\tilde\Omega_{12})_{0}
\label{eq117}
\end{equation}
and
\begin{equation}
\lim_{c_{1}\rightarrow 0}D_{T}=\frac{k_{B}T}{f_{1}}.
\label{eq118}
\end{equation}
If $f_{1}$ is determined by the total concentration $c_{T}=c_{1}+c_{2}$, and if $(D_{11})_{V}$ and $D_{T}$ are measured at the same total concentration $c_{T}$, equations \ref{eq118} and \ref{eq116} lead to the generalized Stokes-Einstein equation
\begin{equation}
(D_{m})_{V}=D_{T} c_{1}\frac{\partial\mu_{1}}{\partial c_{1}}\frac{(1-\phi)}{k_{B}T}\equiv D_{T}\left(\frac{\partial\Pi}{\partial c}\right)_{P,T}(1-\phi).
\label{eq119}
\end{equation}

Here $\Pi$ is the osmotic pressure at constant temperature and total pressure. The generalized Stokes-Einstein equation requires that the mutual and self diffusion coefficients share a single friction factor. That is, if as explained below one writes
\begin{equation}
D_{m}=[g^{(1)}(k,0)]^{-1}\frac{k_{B}T(1-\phi)}{f_{M}}
\label{eq120}
\end{equation}
and
\begin{equation}
D_{s}=\frac{k_{B}T}{f_{s}},
\label{eq121}
\end{equation}
then according to the continuum model the drag coefficients for $D_{m}$ and $D_{s}$ are equal, i.e., $f_{M}=f_{s}$.

The continuum model does not agree with the microscopic models given below, because the microscopic models predict that $f_{s}$ and $f_{M}$ have unequal concentration dependences. In terms of results in later sections: If one expands $f_{i}=f_o(1+\alpha_{i}\phi)$, the microscopic prediction is $\alpha_{s}\neq\alpha_{M}$ for hard sphere suspensions. If the hard spheres gain an electric charge, adding a Debye potential to their interactions, microscopic models predict that the difference between $f_{M}$ and $f_{s}$ increases. Charging the diffusing spheres reduces\cite{phillies61} $|\alpha_{s}|$ towards zero, because the spheres stay further apart, weakening the hydrodynamic forces that retard self-diffusion. Charging the spheres makes $|\alpha_{M}|$ larger, because the integral in eq.\ \ref{eq50} over the Oseen tensor increases.

\section{Correlation Function Descriptions}

Correlation functions provide the fundamental description within statistical mechanics for time-dependent processes and transport coefficients. This section treats correlation function descriptions of diffusion, including correlation functions that describe QuasiElastic Light Scattering Spectroscopy (QELSS), Fluorescence Correlation Spectra (FCS) \cite{magde1974a,schwille1999a,phillies1975a}, Raster Image Correlation Spectroscopy (RICS)\cite{digman2005a,digman2005b}, and Pulsed-Gradient Spin-Echo Nuclear Magnetic Resonance (PGSE NMR\cite{phillies2015b}) The analysis here follows closely our prior papers, including refs.\ \onlinecite{phillies2015b,phillies2005a,phillies2011b,phillies2012a}.

The reader will notice that this approach takes us from the observed relaxation functions through to the moments of two displacement distribution functions and then grinds to a halt. Calculations based on microscopic models that are powerful enough to give quantitative time and concentration dependences appear in a later section.

\subsection{Quasielastic Light Scattering Spectroscopy}

The starting point is the field correlation function, eq.\ \ref{eq2}, which gives the spectrum in the form an average over the positions of a particle $i$ at time $t$ and a particle $j$ at time $t+\tau$. The average is in principle calculated as an average over a probability distribution function $P(\textbf{r}_{i}(t), \textbf{r}_{j}(t+\tau))$, which gives the probability of finding particles $i$ and $j$ (which may be the same particle, $i=j$ being allowed) at the indicated locations at the times $t$ and $t+\tau$, respectively. However, it is more effective to divide $P(\textbf{r}_{i}(t), \textbf{r}_{j}(t+\tau))$ into its self and distinct parts, and then to analyse these two parts separately. N.~B.: We have switched back to using the $\textbf{r}_{j}$ to represent the locations of individual scatterers, not to label locations in the scattering volume.

We begin with eq.\ \ref{eq2}, namely
\begin{equation}
g^{(1)}(q,\tau)  = \left\langle \sum_{i,j=1}^{N} \sigma_{i} \sigma_{j} \exp( i\textbf{q} \cdot [\textbf{r}_{i}(t+\tau) - \textbf{r}_{j}(t)])\right\rangle.
\label{eq:101}
\end{equation}
The terms in the sum partition into self and distinct parts, $g^{(1s)}(q,t)$ and $g^{(1d)}(q,t)$, respectively, with
\begin{equation}
g^{(1)}(q,\tau) = g^{(1s)}(q,t)+ g^{(1d)}(q,t).
\label{eq:104}
\end{equation}

Here
\begin{equation}
g^{(1s)}(q,\tau) = \left\langle \sum_{i=1}^{N} \sigma_{i}^{2} \exp(i\textbf{q} \cdot [\textbf{r}_{i}(t+\tau)-\textbf{r}_{i}(t)])\right\rangle
\label{eq:102}
\end{equation}
for the self part, and
\begin{equation}
g^{(1d)}(q,\tau) = \left\langle \sum_{i,j=1, i\neq j}^{N} \sigma_{i} \sigma_{j} \exp(i\textbf{q} \cdot [\textbf{r}_{i}(t+\tau)-\textbf{r}_{j}(t)])\right\rangle.
\label{eq:103}
\end{equation}
for the distinct part. In the distinct part, the summation indices $i$ and $j$ are each taken separately from $1$ to $N$, but the $N$ terms in which $i$ and $j$ happen to be the same are omitted.

One usefully introduces new variables:
\begin{equation}
     \label{DeltaXdef}
     X_{i}(\tau) = x_{i}(t+ \tau) -  x_{i}(t)
\end{equation}
represents how far particle $i$ moves parallel to the scattering vector ${\bf q}$ during time $\tau$, while
 \begin{equation}
    \label{eq:xij0def}
    R_{ij}(t) = x_{i}(t) -  x_{j}(t)
\end{equation}
is the component parallel to the scattering vector ${\bf q}$ of the distance between particles $i$ and $j$ at the initial time $t$.

We now consider how to evaluate or expand the two parts of the field correlation function. We apply the substitution
\begin{equation}
     \label{newvariables}
{\bf \hat{q}} \cdot (\textbf{r}_{i}(t+\tau) - \textbf{r}_{j}(\tau) )=  X_{i}(\tau) + R_{ij}(t).
 \end{equation}

 The self part of the field correlation function may be written in terms of the new variables as
\begin{equation}
       g^{(1s)}(q,\tau) = \left\langle \sum_{i = 1}^{N} \exp(+ \imath q X_{i}(\tau))\right\rangle.
       \label{eq:g1qtself}
\end{equation}
The particles are all the same, so the sum on $i$ can be replaced with a count $N$ of the number of identical terms in the sum; the label $i$ is now irrelevant. A Taylor series expansion followed by interchange of the sum and the ensemble average gives
\begin{equation}
      g^{(1s)}(q,\tau) =  N \left\langle \sum_{n=0}^{\infty} \frac{(\imath q X(\tau))^{n}}{n!} \right\rangle.
            \label{eq:g1qtcomplex}
\end{equation}
To advance farther, one notes that the average is over the displacement distribution function $P(X, \tau)$, this being the function that gives the probability that a particle will move through $X$ during time $\tau$. Up to constants, the averages
\begin{equation}
          \langle (\imath q X(\tau))^{n} \rangle = \int dX (\imath q X(\tau))^{n} P(X,\tau),
         \label{eq:pxtuse}
\end{equation}
are the moments of $P(X,t)$. The odd moments vanish by symmetry, namely the likelihoods of observing displacements $+X_{i}$ and $-X_{i}$ are equal. $P(X, \tau)$ is influenced by whatever else is in the system, e.g., non-scattering components, and represents an ensemble average over positions and momenta of those components.

Further rearrangements lead to
\begin{equation}
      g^{(1s)}(q,\tau) = N \exp\left( - \frac{1}{2} q^2 \langle X(\tau)^{2} \rangle + \frac{1}{24} q^{4}( \langle X(\tau)^{4} \rangle - 3\langle X(\tau)^{2} \rangle^{2}) - {\cal O}(q^{6})\right).
      \label{eq:g1sanddisplacements}
\end{equation}
The relaxation of $g^{(1s)}(q,\tau)$ is thus determined by the even moments of $P(X,\tau)$.

A similar approach may be used to evaluate $g^{(1d)}(q,\tau)$. The first step is to introduce a fine-grained displacement distribution function ${\cal P}(X, \tau, \{ {\bf r}^{M} \} )$.  This function gives the probability that a particle 1 will have a displacement $X$ during a time interval $\tau$, given that the coordinates $\{ {\bf r}^{M} \}$ of the other particles in the system at the initial time $t$ are specified. The list of $M$ other particles includes the $N-1$ other scattering particles in the system and the $M-N+1$ other non-scattering particles in the system; $M=N-1$ and $N=1$ are allowed. In studies of simple mutual diffusion, $M=N$; all particles are scattering particles. In probe diffusion experiments, $N$ is made sufficiently small relative to the size of the container that interactions between the scattering particles are quite small. One can also study mutual diffusion of \emph{non-dilute} scattering particles in the presence of a background non-scattering matrix macromolecule. A few experimental studies of this circumstance have been made; see Phillies\cite{phillies2011a}, Section 11.6, for a review of these.

The displacement distribution function $P(X,\tau)$ is related to $\cal{P}$  by
\begin{equation}
    P(X, t) =   \int d{\bf r}^{M}   {\cal P}(X, \tau, \{ {\bf r}^{M} \}) \exp(-\beta (W_{M} -A) ).
    \label{eq:Paverage}
\end{equation}
Here the integral is over the positions at time $t$ of all scattering and non-scattering particles other than the particle of interest, while $W_{M}$ is the total potential energy including the particle of interest and the other $M$ particles, and $A$ is the nonideal part of the Helmholtz free energy.

So long as the particles are all identical, $g^{(1d)}(q,\tau)$ is a sum of $N(N-1)$ identical terms. It is convenient to make a Taylor series expansion on $X_{1}(\tau)$ while leaving $R_{12}$ in an exponential,
\begin{equation}
       g^{(1d)}(q,\tau) = N(N-1) \left\langle \sum_{n=0}^{\infty} \frac{ (\imath q X_{1}(\tau))^{n}}{n!}  \exp(\imath q R_{12})\right\rangle.
       \label{eq:g1distinctqt3}
\end{equation}

On replacing the formal average $\langle \cdots \rangle$ with the fine-grained distribution function,
\begin{displaymath}
      g^{(1d)}(q,\tau) = N(N-1) \int d X_{1}  \sum_{n=0}^{\infty} \frac{ (\imath q X_{1}(\tau))^{n}}{n!}
\end{displaymath}
\begin{equation}
  \times   \int d{\bf r}^{M}  \exp(- \imath q R_{12}){\cal P}(X_{1}, \tau, \{ {\bf r}^{M} \}) \exp(-\beta (W_{M} -A) ) .
     \label{eq:g1distinctqt4}
\end{equation}

The second line of eq.\ \ref{eq:g1distinctqt4} implicitly defines a reduced distribution function
\begin{equation}
    P_{2}(X_{1},\tau, R_{12}) = \int d{\bf r}_{1} d{\bf r}_{12P}   d{\bf r}_{3} d{\bf r}_{4}\ldots d{\bf r}_{N}  {\cal P}(X_{1}, \tau,\{ {\bf r}^{M} \}) \exp(-\beta (W_{M} -A) ),
    \label{eq:tildeR12}
\end{equation}
Here $d{\bf r}_{12P}$ represents the integral over the two components of $\textbf{R}_{12}$ that are perpendicular to the scattering vector. We actually need the spatial Fourier transform
\begin{equation}
      \tilde{P}_{2}(X_{1},\tau,q) = \int d \textbf{R}_{12}  \exp(- \imath q R_{12}) P_{2}(X_{1},\tau, R_{12}),
      \label{eq:tildeq}
\end{equation}
in which $R_{12}$ is the component of   $\textbf{R}_{12}$ that lies along the scattering vector.

The distinct part of the field correlation function is finally reduced to
\begin{equation}
       g^{(1d)}(q,\tau) = N(N-1) \int d X_{1}  \sum_{n=0}^{\infty} \frac{ (\imath q X_{1}(\tau))^{n}}{n!} \tilde{P}(X_{1},\tau,q).
 \label{eq:g1dtilde}
\end{equation}

Comparison of eqs.\ \ref{eq:pxtuse}  and \ref{eq:g1dtilde} reveals an important result.  $g^{(1s)}(q,\tau)$ and $g^{(1d)}(q,\tau)$ are obtained as averages of the same variable $\imath q X_{i}(\tau)$ over different correlation functions, namely $P(X, \tau)$ and $\tilde{P}(X_{1},\tau,q)$, so their information contents are not the same.  We will henceforth drop subscripts on $X$ when they are not significant.

Why did we start with $P(X, \tau)$ when we calculated $g^{(1s)}(q,t)$, but with ${\cal P}(X_{1}, \tau, \{ {\bf r}^{M} \})$ when we calculated $g^{(1d)}(q,\tau)$? First, $P(X,\tau)$ is obtained as an average over ${\cal P}(X_{1}, \tau, \{ {\bf r}^{M} \})$, but in that average all of the $\{R^{M} \}$ are treated the same way. When we calculate $g^{(1d)}(q,t)$, the object being averaged is $X \exp(i q R_{12})$.  A single coordinate $R_{12}$ is treated differently from all the rest, so the average must be done explicitly. Also, the count of terms in the intermediate step replacing $R_{ij}$  with $R_{12}$ was facilitated by invoking ${\cal P}$.

\subsection{Alternative Methods for Measuring Single Particle Diffusion Coefficients}

This section treats some other methods for measuring single particle diffusion coefficients, namely Pulsed-Gradient Spin-Echo NMR (PGSE NMR), Fluorescence Correlation Spectroscopy (FCS), and Raster Image Correlation Spectroscopy (RICS). The first of these measures the same $g^{(1s)}(q,t)$ as does QELSS when QELSS is used in probe diffusion mode, but on different time and distance scales.  The other two techniques also lead back to averages and moments for $P(X,\tau)$, but by slightly different paths.

The physical quantity measured in PGSE NMR is quite different from the quantity measured in FCS, but the mathematical form is the same, namely the PGSE NMR relaxation spectrum is given by
\begin{equation}
      {\bf M}(2 T)  = {\bf M}(0) \exp(- i {\bf q}\cdot ({\bf r}(t +  \tau) - {\bf r}(t)) ).
        \label{eq:spinechofinal}
\end{equation}
Here $2 T$ is the time interval required for the formation of the spin echo, namely twice the time interval $T$ between the initial $\pi/2$ magnetizing RF pulse and the $\pi$ pulse that reverses the chirality of the spin magnetization.\cite{phillies2015b}.  The wave vector has a new meaning. ${\bf q} =\gamma \delta {\bf g}$, where $\gamma$ is the gyromagnetic ratio of the spin being observed, $\delta$ is the duration of a gradient pulse, and ${\bf g}$ is the field gradient of the superposed gradient pulse. Finally, $\tau$ is the time interval between the two gradient pulses; it is the time over which diffusion is observed. The notations in use the QELSS and NMR subdisciplines are not entirely compatible. For ease of reading we have forced eq.\ \ref{eq:spinechofinal} into QELSS notation. The meaning of ${\bf q}$ in eq.\ \ref{eq:spinechofinal} is very different from the meaning of ${\bf q}$ in the previous section, but the mathematical structure for the value of the PGSE NMR relaxation function is precisely the same as the mathematical structure was for $g^{(1s)}(q,t)$ in light scattering.  The time dependence of ${\bf M}(2 \tau)$ from the correlation function approach therefore looks precisely the same the expression for the time dependence of $g^{(1s)}(q,t)$ using the same correlation function approach.

Fluorescence Correlation Spectroscopy and Raster Image Correlation Spectroscopy are very similar in their physical description, but differ at one key point.  In each method, one observes the motions of fluorophores, which may be free-floating molecules or may be fluorescent groups physically or chemically bound to the diffusing molecules of interest. During the experiment, a small volume of solution is illuminated with a laser beam. In modern instruments, the diameter $w$ of the illuminating laser beam may be as small as a few hundred nm across. The fluorescent intensity is then measured at a series of times, and the intensity-intensity time correlation function of the emitted light is measured. FCS and RICS differ in that in FCS one repeatedly observes fluorescence emitted by a single volume of solution, while in RICS one repeatedly observes fluorescence emitted from a series of neighboring volumes of solution.

FCS and RICS sound much like QELSS, except the re-emission process  is fluorescence rather than quasielastic scattering. Because all phase information is lost during the fluorescent re-emission,  in FCS and RICS  one captures information on the number of fluorophores in the illuminated volume, but loses most information on the relative positions of the fluorophores within the volume. Qualitatively, in FCS one waits for moments when the fluorescent intensity is particularly bright or dim, and then waits to see how long is typically required for the fluorescent intensity to return to its average level. Qualitatively, in  RICS, one waits for moments when the fluorescent intensity from the first volume is particularly bright or dim, and then asks how rapidly the brightness or dimness spreads to neighboring solution volumes.

The FCS and RICS spectra $G(\tau)$ are given by the same expression
\begin{equation}
     G(\tau) = \int_{V} d{\bf r}  \int_{V} d{\bf r'} {\cal I}({\bf r}) {\cal I}({\bf r'}) P({\bf r'} - {\bf r}, \tau).
     \label{eq:FCSform}
\end{equation}
In this equation, ${\cal I}({\bf r})$ and ${\cal I}({\bf r'})$ refer to the position-dependent intensities of the illuminating laser at the times $t$ and $t+\tau$, respectively. For the model calculation here, the intensities are approximated as being Gaussian cylinders. In FCS, the two cylinders have the same center. In RICS, the two cylinders are displaced by a distance $a$ that is for mathematical convenience taken to lie along the $x$ axis. $P({\bf r'} - {\bf r}, \tau)$ is the displacement distribution function, expressed in terms of the initial and final positions ${\bf r}$ and ${\bf r'}$ of a diffusing fluorophore.

Equation \ref{eq:FCSform} is a convolution integral, which can be evaluated via Fourier transform techniques. For FCS, the spatial Fourier transform of the intensity
\begin{equation}
     {\cal I}({\bf r}) = I_{o} \exp(- r^{2}/w^{2})
   \label{eq:electric}
\end{equation}
is
\begin{equation}
    {\cal I}(q)  = I_{0} \exp(-q^{2} w^{2}/4)
    \label{eq:Iqcalculated}
\end{equation}
where $r$ is the distance from the center of the cylinder and ${\bf q}$ is the spatial Fourier transform vector. The spatial Fourier transform for each dimension of the displacement distribution function is
\begin{equation}
     F(q,\tau) = \int d x \exp(i q x) P(x , \tau).
     \label{eq:Fqtdef}
\end{equation}
After a Taylor series expansion of $\exp(i q X)$, enforcing the requirement that that $P(x,\tau)$ is symmetric in $x$, and extracting a leading Gaussian term $\exp(- q^{2} \langle X^{2} \rangle/2)$ from the expansion, $F(q,\tau)$ is found to be
\begin{equation}
F(q,\tau) = \exp(- q^{2} K_{2}(\tau)/2) (1 + q^{4} K_{4}(\tau)/24 + q^{6} K_{6}(\tau)/720 + {\cal(O)}(q^{8})).
     \label{eq:Fktexpand}
\end{equation}
The first few $K_{2n}$ are
\begin{equation}
    K_{2}(\tau)) = \langle (x(\tau))^{2} \rangle,
     \label{eq:K2def}
\end{equation}

\begin{equation}
   K_{4}(\tau)) = \langle (x(\tau))^{4} \rangle - 3 \langle (x(\tau))^{2} \rangle^{2},
     \label{eq:K4def}
\end{equation}

and

\begin{equation}
    K_{6}(\tau)) =  - \langle (x(\tau))^{6} \rangle +15 \langle (x(\tau))^{2} \rangle \langle (x(\tau))^{4} \rangle - 30 \langle (x(\tau))^{2}\rangle^{3}.
     \label{eq:K6def}
\end{equation}
In the above three equations, $x(\tau)$ is the (time-dependent) displacement of the particle along the $x$ axis. The brackets $\langle \ldots \rangle$ denote an average over $P(x , \tau)$. The $K_{2n}$ can all be written entirely in terms of the even spatial moments of $P(x , \tau)$.

Substituting into the spatial Fourier transform form of eq.\ \ref{eq:FCSform},
\begin{equation}
    G(\tau) = N \int d{\bf q} ({\cal I}(q))^{2} F(q,\tau),
    \label{eq:Fourier}
\end{equation}
the FCS spectrum may be written
\begin{equation}
     G(\tau) =    \frac{2 \pi I_{0}^{2}}{w^{2}+ K_{2}(\tau)} \left(1+\frac{K_{4}(\tau)}{3(w^{2}+ K_{2}(\tau))^{2}} + \frac{K_{6}(\tau)}{15(w^{2}+ K_{2}(\tau))^{3}}+\ldots\right).
     \label{eq:FCSSpectrum}
\end{equation}
The natural variables for writing this form are the  $K_{n}(t)/w^{2n}$, for which
\begin{equation}
     G(\tau) =    \frac{2 \pi I_{0}^{2}}{w^{2}} \frac{1}{1+ K_{2}(\tau)/w^{2}} \left(1+\frac{K_{4}(\tau)/w^{4}}{3(1+ K_{2}(\tau)/w^{2})^{2}} + \frac{K_{6}(\tau)/w^{6}}{15(1+ K_{2}(\tau)/w^{2})^{3}}+\ldots\right).
     \label{eq:FCSSpectrum2}
\end{equation}

Many published discussions of FCS invoke the Gaussian diffusion approximation.  For Gaussian diffusion, $K_{2}$ is the mean-square displacement, while $K_{4}$, $K_{6}$, and higher are all zero. In this case, the spectrum reduces to the form of Magde, et al.\cite{magde1974a}
\begin{equation}
     G(\tau) =    \frac{2 \pi I_{0}^{2}}{w^{2}} \frac{1}{1+ K_{2}(\tau)/w^{2}}.
     \label{eq:originalGt}
\end{equation}

The calculation of the RICS spectrum is extremely similar to the calculation of the FCS spectrum, except that in RICS one cross-correlates the intensities observed at two times in two different volumes of solution.

The RICS spectrum may be written
\begin{equation}
     G(\tau) = \int_{V} d{\bf r}   \int_{V'} d{\bf r'} {\cal I}({\bf r}) {\cal I}({\bf r'}) P({\bf \Delta r},\tau).
     \label{eq:RICSform}
\end{equation}
Coordinates are chosen so that the origins of ${\bf r}$ and ${\bf r'}$  are at the centers of the two illuminated regions, in which case the particle displacement is
\begin{equation}
      {\bf \Delta r} = {\bf r'} + a \hat{i} - {\bf r}.
    \label{eq:displacement}
\end{equation}
with $\hat{i}$ being the unit vector along the $x$ axis.

For a cylindrical beam whose center is parallel to the $z$-axis, the intensity can be written in terms of spatial Fourier transforms as
\begin{equation}
      {\cal I}({\bf r}) = (2 \pi)^{-2} \int dq_{x} dq_{y}   I_{0} \exp(-((q_{x})^{2}+(q_{y})^{2}) w^{2}/4) \exp(i (x q_{x} + y q_{y}))
      \label{eq:intensityq1}
\end{equation}
and
\begin{equation}
      {\cal I}({\bf r'}) = (2 \pi)^{-2}  \int dq'_{x} dq'_{y}   I_{0} \exp(-((q'_{x})^{2}+(q'_{y})^{2}) w^{2}/4)  \exp(i (x' q'_{x} + y' q'_{y})).
      \label{eq:intensityq2}
\end{equation}
The two components of the spatial Fourier transform vectors are $(q_{x}, q_{y})$ and $(q'_{x}, q'_{y})$, respectively.

Diffusive motions in the $x$ and $y$ directions are independent, so the RICS spectrum may be written
\begin{equation}
    G(\tau) = G_{x}(\tau) G_{y}(\tau)
    \label{eq:RICSspectrum2}
\end{equation}
The component $G_{y}(\tau)$ is the same as the $y$-component $G(\tau)$ for FCS, namely
\begin{equation}
    G_{y}(\tau) = N \int_{-\infty}^{\infty} dq_y ({\cal I}(q_y))^{2} F(q_y,\tau)
     \label{eq:RICSspectrumy}
\end{equation}
while the $x$-component is
\begin{equation}
    G_{x}(\tau) = N \int_{-\infty}^{\infty} dq ({\cal I}(q))^{2} F(q,\tau) \cos(qa).
     \label{eq:RICSspectrumx}
\end{equation}

The integral for the $x$-component gives
\begin{displaymath}
G_{x}(\tau) = N  \left(\exp\left(-\frac{a^{2}}{2 \left(K_{2}(\tau)+w^{2}\right)}\right)(2 \left(K_{2}(\tau)+w^{2}\right))^{-1}\right.
\end{displaymath}
\begin{displaymath}
\times \left(1+\frac{K_{4}(\tau) \left(a^{4}-6 a^{2} \left(K_{2}(t)+w^{2}\right)+3
\left(K_{2}(\tau)+w^{2}\right)^{2}\right)}{24 \left(K_{2}(\tau)+w^{2}\right)^{4}} \right.
\end{displaymath}
\begin{equation}
\left. -\frac{K_{6}(\tau) \left(a^{6}-15 a^{4} \left(K_{2}(\tau)+w^{2}\right)+45 a^{2} \left(K_{2}(\tau)+w^{2}\right)^{2}-15
\left(K_{2}(\tau)+w^{2}\right)^{3}\right)}{720 \left(K_{2}(\tau)+w^{2}\right)^{6}}\right)
\label{eq:RICSGxt}
\end{equation}
This form is not quite the same as that seen in Digman, et al.\cite{digman2005a,digman2005b} because: Digman, et al., only allowed for Gaussian Diffusion; the displacement $a$ between the two laser beams is here taken to be a continuous variable; and the time displacement $t$ is here used as an independent variable relative to $a$.  The observed spectrum will be a product of eq.\ \ref{eq:RICSGxt} and the FCS relaxation forms for the $y$ and $z$ directions.

\subsection{Partial Solutions for $g^{(1)}(q,\tau)$ and  $g^{(1s)}(q,\tau)$}

This section treats partial solutions in which $g^{(1)}(q,\tau)$ is calculated up to some point in terms of molecular parameters. We start with the simplest approaches.

The first case refers to systems in which the $N$ scattering particles are highly dilute. There may also be $M-N+1$ matrix particles, which may not be dilute. If the scattering particles are adequately dilute or elsewise noninteracting, then $g^{(1d)}(q,t)$ vanishes.  The disappearance is seen in eq.\ \ref{eq:g1distinctqt3}, in which the factor $\exp(\imath q R_{12})$ refers to a distinct pair of scattering particles. If particles 1 and 2 essentially never interact with each other, then all phases of the complex exponential are for all practical purposes equally likely, in which case the complex exponential averaged over all pairs of scattering particles averages to zero, taking the integral with it.  In this case, then, $g^{(1)}(q,t)$ is very nearly determined by $g^{(1s)}(q,t)$, eq.\ \ref{eq:g1sanddisplacements}.

At the bottom end for simplicity, for dilute probes particles in a simple, low-viscosity liquid, $P(X, \tau)$ is a Gaussian in $X$, in which case in eq.\ \ref{eq:g1sanddisplacements} the terms in $q^{4}$, $q^{6}$, and higher all vanish. The field correlation function becomes
\begin{equation}
    g^{(1s)}(q,\tau) = N \exp( - q^{2} \langle (X(\tau))^{2}\rangle/2).
  \label{eq:g1sdilute}
\end{equation}
In such a system, Doob's Theorems\cite{doob1942a} guarantee that
\begin{equation}
     \langle (X(\tau))^{2}\rangle = 2 D \tau
  \label{eq:doob}
\end{equation}
so that the spectrum decays as a single exponential in time. Further details are found in Berne and Pecora\cite{berne1976a}, Chapter 5.

A word of caution: Berne and Pecora were writing for an audience with a solid knowledge of the physics and theoretical chemistry involved. Their Chapter 5 gives a treatment that only applies to dilute particles in simple solvents. They reasonably expected that their audience would recognize this. Berne and Pecora also treat results relevant to complex fluids in parts of their Chapters 10-12.

A claim that $P(X, \tau)$ is a Gaussian is often said to arise from the Central Limit Theorem, which gives the consequences if $\textbf{r}(t)$ is composed of a large number of small, independent steps. Instead, $\textbf{r}(t)$ is all or part of a single step. Under the conditions in which the Central Limit Theorem is valid, so that $P(X,\tau)$ is a Gaussian in $X$, eq.\ \ref{eq:doob} is equally sure to be valid.  Therefore, whenever the Central Limit Theorem is applicable, $g^{(1s)}(q,\tau)$ is sure to be a single pure exponential in $\tau$ as well as in $q^{2}$.

For the last four decades, there has been interest in studying optical probe diffusion  in complex fluids, fluids in which there are relaxations on the time and distance scales accessible to experimental study with QELSS.  However, if the fluid has relaxations on the time scale being studied, or even longer time scales, then $\textbf{r}(t)$ is emphatically \emph{not} composed of large numbers of independent steps. The Central Limit Theorem is totally irrelevant to measurements being made on time scales on which the fluid has relaxations.  In this interesting case, $P(X, t)$ may well not be a Gaussian, so (in the absence of an independent direct measurement showing that $P(X, t)$ is a Gaussian, for reasons other than being a Central Limit Theorem Gaussian) eq.\ \ref{eq:g1sdilute} can not be invoked.

Indeed, there is extensive experimental evidence showing that $P(X, t)$ is not Gaussian in complex and glassy fluids on time scales on which there are ongoing relaxations.\cite{kob2007a,phillies2015a,boon1980a,hansen1986a}:

First, particle tracking measurements can in some systems determine $P(X, t)$ directly, at least on longer time scales. Fifteen years ago, Apgar, et al.\cite{apgar2000a} and Tseng, et al.\cite{tseng2001a} measured $P(X, t)$ directly for probes in polymer solutions, unambiguously finding that $P(X, t)$ had a non-Gaussian form. Wang, et al.\cite{wang2009a,wang2012a}, and Guan, et al.\cite{guan2014a} measured more precisely the nature of the non-Gaussian deviations in $P(X, t)$. For small $X(t)$, $P(X, t)$ is close to a Gaussian, but at larger $\mid X(t)\mid$, $P(X,t)$ only decreases exponentially with increasing $\mid X(t) \mid$. Chaudhuri, et al.\cite{kob2007a} report similar 'fat tails' for glassy systems.  While some of the experimental systems were quite complex, Guan, et al.'s system\cite{guan2014a} was physically simple, namely it was a suspension of smaller colloidal hard spheres diffusing through a nondilute suspension of larger hard spheres.

Second, if $P(X, t)$ is a Central-Limit-Theorem Gaussian,  $\log(g^{(1s)}(q,t))$ must be linear in $q^{2}$ and in $t$.  For dilute probe particles in complex fluids such as polymer solutions, neither of these assumptions is true in general. For example, for polystyrene latex probes in hydroxypropylcellulose:water\cite{streletzky1999a,streletzky1999b,phillies2003a}, $\log(g^{(1s)}(q,t))$ is a sum of two or three stretched and unstretched exponentials. Furthermore, for some modes but not others, the mean relaxation rate $\langle \Gamma^{-1} \rangle^{-1}$ is not linear in $q^{2}$.

Third, supporting evidence that $P(X, t)$ is not a Central-Limit-Theorem Gaussian is provided by FCS measurements.
On one hand, transverse diffusion of macromolecules in cell membranes was studied by Wawrezinieck, et al.\cite{wawrez2005a} They found in their systems that the relaxation time $\tau_{D}$ was linear in $w^{2}$, but did not trend to zero as $w \rightarrow 0$, so the diffusion coefficient inferred from $\tau_{D}$ depended on $w$.  Their measurements show directly that $P(X, t)$ was not a Gaussian in $X(t)$. Furthermore, as shown by Schwille, et al.\cite{schwille1999a}, in complex fluids the mean-square displacement inferred from eq.\ \ref{eq:originalGt} does not always increase linearly in $t$.

As a practical experimental aside, if probe diffusion is nonGaussian, eqs.\ \ref{eq:g1sanddisplacements}, \ref{eq:FCSSpectrum2}, and \ref{eq:RICSGxt} are still valid.  By measuring $g^{(1)}(q,t)$ at multiple scattering vectors $q$, or by measuring $G(t)$ for a series of spot diameters $w$, $K_{2}(t)$,  $K_{4}(t)$, and higher-order terms may be accessible as functions of time.

\section{Generalized Langevin Equation}

\subsection{Introduction \label{section4A}}

This Section considers how the concentration dependence of diffusion coefficients can be obtained from a Generalized Langevin equation approach. The calculations here represent extensions of the correlation function descriptions of the previous section.  Because we insert mechanical models for particle motion and concrete forms for the direct and hydrodynamic interactions between the particles, we can make quantitative predictions for the dependences of $D_{m}$, $D_{s}$ and $D{p}$ on solute concentration and other solute properties.

In Section \ref{section4B} I show how various diffusion coefficients are related to the first cumulant of a relaxation spectrum. In order to interpret the integrals in Section \ref{section4B}, a model for particle motion that properly represents the forces between the particles must be supplied.  This representation is the generalized Langevin equation seen in Section \ref{section4C}. Section \ref{section4D} treats direct interactions between the diffusing particles, as treated with their interparticle potential energies.  Section \ref{section4E}  treats hydrodynamic interactions between particles, as represented by hydrodynamic interaction tensors. Claims that there is screening of the hydrodynamic interactions between macromolecules in solution are considered and refuted. Sections  \ref{section4F} and \ref{section4G} evaluate our expressions for $D_{m}$, including (Section \ref{section4G}) simple effects of direct and hydrodynamic interactions and (Section \ref{section4H}) dynamic friction terms. Section \ref{section4I} considers the long-range behavior of the Oseen tensor, in particular the difference between infinite and closed-system behaviors, leading to a microscopic replacement for the reference frame discussion. Section \ref{section4J} examines the wavevector dependence of $D_{m}$.  Section \ref{section4K} evaluates the self-diffusion coefficient $D_{s}$ and the probe diffusion coefficient $D_{p}$.

\subsection{\label{section4B} Diffusion Coefficients from Cumulants}

This section presents the general formulations that lead to extracting the mutual, self, and probe diffusion coefficients from the first cumulants of $g^{(1)}(q,t)$, $g^{(1s)}(q,t)$, and various numerical transforms, as obtained using QELSS, FPR, PGSE NMR, FCS, RICS, and related techniques. It should be emphasized: Cumulants are obtained from the short-time limits of various time derivatives. However, cumulants represent weighted averages over relaxations that decay on all time scales, short and long.  Suggestions that cumulant series cannot represent multimodal relaxations are incorrect. However, in some cases alternative expansions may be more useful.

Our starting point is eq. \ref{eq5} for the mutual diffusion coefficient, which may be rewritten as
\begin{equation}
     D_{m} q^{2} = - \lim_{\tau \rightarrow 0} \frac{\partial}{\partial \tau} \ln(g^{(1)}(q,\tau)).
     \label{eq:dmstart}
\end{equation}
For simplicity, in this section we take $\sigma_{i}=1$. From eqs. \ref{eq2}, \ref{eq6}, and \ref{eq5} the mutual diffusion
coefficient for identical particles is
\begin{equation}
D_{m}= - \frac{1}{q^{2}g^{(1)}(q,0)}\lim_{\tau \rightarrow 0}\frac{\partial}{\partial \tau}\left\langle\sum_{i,j=1}^{N}\exp(i\textbf{q}\cdot [ \textbf{r}_{i}(t) - \textbf{r}_{j}(t+\tau)])\right\rangle.
\label{eq10}
\end{equation}
$N$ is again the number of scattering particles in the system. Applying the identity $\textbf{r}_{j}(t+\tau)=\textbf{r}_{j}(t)+\int_{t}^{t+\tau}ds\  \textbf{v}_{j}(s)$, $\textbf{v}_{j}(s)$ being the velocity of the particle at time $s$, followed by a Taylor series expansion in $\textbf{q}\cdot\textbf{v}_{j}$, one has
\begin{equation}
D_{m}=-\frac{1}{q^{2}g^{(1)}(q,0)}\lim_{\tau \rightarrow 0} \frac{\partial}{\partial \tau} \left\langle \sum_{i,j=1}^{N} \exp[i\textbf{q}\cdot\textbf{r}_{ij}(t)]\sum_{n=0}^{\infty}\frac{[-i\textbf{q}\cdot\int_{T}^{t+\tau}ds\; \textbf{v}_{j}(s)]^{n}}{n!}\right\rangle
\label{eq11}
\end{equation}
where $\textbf{r}_{ij}=\textbf{r}_{j}(t)-\textbf{r}_{i}(t)$. Taking the derivative, the leading terms in \textbf{q} are
\begin{equation}
D_{m} = -\frac{1}{q^{2}g^{(1)}(q,0)}\lim_{\tau \rightarrow 0} \left\langle \sum_{i,j=1}^{N} \exp[i\textbf{q}\cdot\textbf{r}_{ij}(t)]\left(i\textbf{q}\cdot\textbf{v}_{j}(t+\tau) \right.\right.
\notag
\end{equation}
\begin{equation}
\left.\left.-\textbf{q}^{2} : \int_{t}^{t+\tau}ds\; \textbf{v}_{j}(s)\textbf{v}_{j}(t+\tau)+...\right)\right\rangle
\label{eq12}
\end{equation}
$N$ is again the order of the fit. The two velocities in the integral refer to the same particle. The above steps use only standard methods of freshman calculus: differentiation, integration, and expansion. No appeals to models of particle motion, time scales, or statistico-mechanical stationarity were invoked.

The process of linking $D_{m}$ to particle motions is readily duplicated for $D_{s}$ and for $D_{p}$, which are each the first cumulant of $g^{(1s)}(q,\tau)$, albeit in different systems. One may envision measuring $D_{s}$ by tracking an isolated particle in a uniform solution. [A QELSS measurement in homodyne mode requires that at least two particles be present; elsewise $S(q,t)$ is a constant.] For an experiment that tracks a single particle
\begin{equation}
g^{1s}(q,\tau)= \left\langle\sum_{i=1}^{N} \exp(i\textbf{q}\cdot[\textbf{r}_{i}(t)-\textbf{r}_{i}(t+\tau)]) \right\rangle.
\label{eq59}
\end{equation}
and
\begin{equation}
     D_{s} q^{2} = - \lim_{\tau \rightarrow 0} \frac{\partial}{\partial \tau} \ln\left(g^{(1s)}(q,\tau)\right).
     \label{eq:dsstart}
\end{equation}
The calculation of $D_{p}$ is notationally more complicated, because the probe and matrix particles may be entirely different in their natures, but eq.\ \ref{eq:dsstart} applies to both calculations. That calculation is postponed until Section \ref{section4K}.

\subsection{\label{section4C} Generalized Langevin Equation}

To make further progress, an adequate description of particle motion in solution is needed. Here particle motion will be characterized with a generalized Langevin equation. The original Langevin equation\cite{reif1965} described an isolated particle in an external potential
\begin{equation}
M\frac{d\textbf{v}(t)}{dt}=-f\textbf{v}(t) + \textbf{F}_B(t) - \nabla W(\textbf{r},t),
\label{eq13}
\end{equation}
where M and f are the particle mass and drag coefficient, and $W(\textbf{r},t)$ is the potential energy of the Brownian particle in the external field. $\nabla W$ may depend slowly on position, but only on length and time scales far longer than those over which $\textbf{F}_B$ varies.

Further interpretation of eq. \ref{eq13} requires a discussion of the significant time scales in the problem. If one applies a force to a particle, there is some short time $\sim \tau_{H}$ before the surrounding solvent molecules reach their steady-state behavior. Only for times $t > \tau_{H}$ is Stokes' Law behavior or anything similar expected. A second scale $\tau_{B} = m/f$ describes the time required for inertial relaxation of the Brownian particle. Rice and Gray\cite{rice1965} show $\tau_{B} \gg \tau_{H}$.

Over times $t \gg \tau_{B}$, which are the only times usually accessible to quasi-elastic light scattering, $M\frac{\partial\textbf{v}}{\partial t} \approx 0$. For $t\gg\tau_{B}$, the particle velocity $\textbf{v}$ in the Langevin equation may be divided as
\begin{equation}
\textbf{v} = \textbf{v}_B + \textbf{v}_D.
\label{eq14}
\end{equation}
The drift velocity is defined as
\begin{equation}
\textbf{v}_D = -f^{-1}\nabla W(\textbf{r});
\label{eq15}
\end{equation}
$\textbf{v}_D$ is the sedimentation velocity found in the Smoluchowski sedimentation-diffusion equation.

The Brownian velocity $\textbf{v}_B$ describes the solvent-driven motions of the particle over times $t>\tau_{H}$. We are considering real diffusing particles, not solutions to stochastic equations, so $\textbf{v}_B$ is a continuous variable with well-behaved derivatives and integrability.    $\textbf{v}_B$ arises from stress fluctuations in the solvent and at the solvent-particle interface, as described by the random force $\textbf{F}_B(t)$. $\textbf{v}_B$ is usually assumed to have a very short correlation time $\sim\tau_{B}$:
\begin{equation}
\langle\textbf{v}_B(t)\cdot\textbf{v}_B(0)\rangle = 0, \quad t \gg \tau_{B}.
\label{eq16}
\end{equation}
There are some technical complications related to $\textbf{v}(t)$ being unsteady, namely that Stokes' Law must be replaced by the Boussinesq equation. For a discussion of this point, see Chow and Hermans\cite{chow1973a}.

Unlike $\textbf{v}_B$, $\textbf{v}_D$ may have a slow secular time dependence, but is virtually constant over times $t \sim \tau_{B}$. Random motions are generally assumed to be decoupled from secular drift velocities $\textbf{v}_D$, so that the Kubo relation
\begin{equation}
\left\langle \int_{0}^tds\;\textbf{v}_B(s)\textbf{v}_B(0) \right\rangle=D\textbf{I}, \quad t\gg\tau_{B},
\label{eq17}
\end{equation}
is assumed to continue to apply if the particle is in a near-constant external potential.  Here $\textbf{I}$ is the identity tensor.

The Langevin equation was originally applied to diffusion in an external field, such as the fields encountered in a centrifuge or electrophoresis apparatus. In these systems, the Langevin approach and the Smoluchowski diffusion-sedimentation equation are equivalent, as seen in standard textbooks\cite{reif1965}. However, to apply the Langevin equation to compute the dynamics of mesoscopic particles in solution, one must make \emph{fundamental} reinterpretations of several terms of the equation:

First, the external potential $W(r)$ must be replaced by an $N$-particle interparticle potential $W_{N}(\textbf{r}^{N})$, where $r^{N} = (\textbf{r}_{1},\textbf{r}_2,. . . ,\textbf{r}_N)$. This reinterpretation, which appears in the literature as far back as the Kirkwood-Risemann \cite{kirkwood1948} papers on polymer dynamics, is not physically trivial. Unlike an external force $\nabla W(r)$, the intermacromolecular force $\nabla_{i}W_{N}(\textbf{r}^{N})$ on particle $i$ is strongly and rapidly time-dependent. Variations in the interparticle forces are correlated with Brownian displacements of individual particles, because Brownian displacements $\Delta\textbf{r}_B = \int ds\ \textbf{v}_B(s)$ in the positions of individual particles are large enough that $W(\textbf{r}^{N} + \Delta\textbf{r}_B^{N})\neq W_{N}(\textbf{r}^{N})$. Furthermore, Brownian-motion-induced changes in $W_{N}(\textbf{R}^{N})$ occur on the same time scale as Brownian motion.

Second, the position-independent drag coefficient $f$ must be replaced with an $N$-particle mobility tensor $\mu_{ij}\equiv\mu_{ij}(\textbf{r}^{N})$. $f$ may be concentration-dependent because the particle of interest interacts hydrodynamically with other particles in the system, but in eq. \ref{eq13} only an ensemble-average (``mean-field'') value of $f$ was used. In contrast to $f$, which depends only on macroscopic concentration variables, $\mu_{ij}$ depends on the current (and, in visco-elastic solvents not treated here, previous) positions $\{ \textbf{r}^{N} \}$ of the other particles in solution. In a system of interacting Brownian particles, Brownian motions and drift velocities are partially correlated, leading to complications with eq. \ref{eq17}, as discussed below.

Having made this fundamental reinterpretation of the Langevin equation, for a system of $N$ interacting Brownian particles, eq. \ref{eq14} becomes the Generalized Langevin Equation
\begin{equation}
\textbf{v}_{i}=\textbf{v}_{Bi}+\textbf{v}_{Di}.
\label{eq18}
\end{equation}
In this equation, each particle $i$ has its own velocity $\textbf{v}_{i}$ and Brownian and driven velocity components $\textbf{v}_{Bi}$ and $\textbf{v}_{Di}$. This equation looks exactly like eq.\ \ref{eq14}, but the above fundamental reinterpretations apply. We now consider what forces drive $\textbf{v}_{Bi}$ and $\textbf{v}_{Di}$, and how they are correlated.

\subsection{\label{section4D} Interparticle Potential Energies}

$\textbf{v}_{Di}$ is in part determined by the interparticle potential energy, and in part determined by the $N$-particle mobility tensor discussed in the next section. The total potential energy $W$ of the system is typically written as the sum of pair potentials $V$.  In real physical systems, three-body potentials that cannot be written as sums of pair potentials are undoubtedly important. For simple hard spheres of radius $a$ the pair potential energy is
\begin{eqnarray}
V(\textbf{r})=0,\;\;|r|>2a \label{eq30}\\
V(\textbf{r})=\infty,\;\;|r|<2a. \label{eq31}
\end{eqnarray}
Formally, $V(r)$ for a hard sphere lacks a derivative at $r=a$, so the force between two hard spheres is either zero or undefined, threatening mathematical complications. Physically, the force between a pair of particles is well-behaved everywhere; these complications should not arise in real systems. Final expressions for $D_{m}$ involve integrals over radial distribution functions $g(r)$, forces not appearing explicitly in the equations. $g(r)$ for a spherical macromolecule with no long-range interactions is not quite identical to $g(r)$ for hard spheres, because a real macromolecule is slightly compressible and delivers a well-defined force during a close encounter. The hard-sphere form for $g(r)$ should still be a good approximant to a real $g(r)$.

For charged hard spheres (often used\cite{biresaw1985,corti1981} in theoretical models to approximate the behavior of small micelles) in solutions that also contain added salt, a screened Coulomb potential
\begin{equation}
V(\textbf{r})=\frac{Q^{2}e^{-\kappa(r-2a)}}{4\pi\epsilon r(1+\kappa a)^{2}}
\label{eq32}
\end{equation}
is added to eq. \ref{eq30}. This equation gives the interaction between a single dielectric sphere and a point charge. Here $Q$ is the sphere charge, $a$ is the sphere radius, $\kappa$ is the Debye screening length, and $\epsilon$ is the ratio of the dielectric constants of the solvent and the sphere interior. Note that the potential energy between a dielectric sphere and a point ion is not the same as the potential energy  between two charged dielectric spheres; see Kirkwood and Schumaker\cite{kirkwood1952} and this author\cite{phillies1974}. For objects the size of micelles induced-dipole potential energies may be as important as the potential energy described by eq.\ \ref{eq32}.\cite{phillies1974} Some treatments of micelles also incorporate an intermicellar van der Waals potential, e.~g., refs.\ \onlinecite{biresaw1985,corti1981}.

\subsection{\label{section4E} Hydrodynamic Interactions; Hydrodynamic Screening}

In addition to the direct intermacromolecular interactions described by $W_{N}$, macromolecules in real solutions encounter two sorts of hydrodynamic interactions. First, when a particle is driven through solution by an outside force, the particle sets up a wake in the liquid around it. The wake drags along nearby particles, so applying a direct force to one particle causes nearby particles to move. Scattering of the wake by other particles in solution leads to higher-order many-particle hydrodynamic interactions. Second, because there are hydrodynamic interactions, the Brownian forces and velocities $\textbf{v}_{Bi}$ of nearby Brownian particles are strongly correlated, so that for $i\neq j$
\begin{equation}
\left\langle\int_{0}^t ds\;\textbf{v}_{Bi}(s)\textbf{v}_{Bj}(0)\right\rangle \equiv D_{ij} \neq 0,\;\;\;t\gg\tau_{B}.
\label{eq19}
\end{equation}
$D_{ij}$ is a two-particle diffusion tensor. In general, $D_{ij}\equiv D_{ij}(\textbf{r}^{N})$ depends on the relative positions of all particles in the system. Our eq.\ \ref{eq12} only uses the $i=j$ terms.

From the Kubo relation (eq. \ref{eq19}) one generally infers
\begin{equation}
\nonumber
\left\langle\sum_{i,j=1}^{N} \exp[i\textbf{q}\cdot\textbf{r}_{ij}(t)]\int_{t}^{t+\tau}ds\;\textbf{q}\cdot\textbf{v}_{Bj}(s)\textbf{q}\cdot\textbf{v}_{Bj}(t+\tau)\right\rangle =
\end{equation}
\begin{equation}
\left\langle\sum_{i,j=1}^{N} \exp[i\textbf{q}\cdot\textbf{r}_{ij}(t)]\textbf{q}\cdot D_{jj}\cdot{q}\right\rangle,\;\;\;\tau\gg\tau_{B}.
\label{eq20}
\end{equation}
Eq. \ref{eq20} embodies three implicit assumptions. First, $\exp[i\textbf{q}\cdot\textbf{r}_{ij}(t)]$ is taken to be essentially constant during the short interval within which $\int ds\;\textbf{v}_{Bj}(s)\textbf{v}_{Bj}(t+\tau)$ is nonzero. Second, the subtle correlations between $\textbf{r}_{j}$ and $\textbf{v}_{Bj}$ are assumed to decay between $t$ and $t+\tau$, so that $\textbf{r}_{j}(t)$ and $\textbf{v}_{Bj}(t+\tau)$ are not correlated. Third, in evaluating this equation, $\tau$ is sufficiently short that $D_{jj}$ can be evaluated using the initial particle positions.

As shown in ref. \cite{phillies1984}, while $\textbf{r}_{j}(t)$ and $\textbf{v}_{j}(t)$ are required by statistical mechanics to lack equal-time correlations, one may not assume that $\textbf{v}_{Bj}(t+s)$ and $\textbf{v}_{Dj}(t+\tau)$, the latter depending on the particle positions at time $t+\tau$,  are similarly uncorrelated. Failure to take account of correlations between Brownian and later driven velocities leads\cite{phillies1984} to erroneous expressions for $D_{m}$.

Finally,
\begin{equation}
\left\langle \sum_{i,j=1}^{N} \exp[i\textbf{q}\cdot\textbf{r}_{ij}(t)]i\textbf{q} \cdot \textbf{v}_{Bi}(t+\tau)\right\rangle = 0,\quad \tau\gg\tau_{B},
\label{eq21}
\end{equation}
because between $t$ and $t+\tau$ the Brownian velocity will have thermalized. At the time $t+\tau$, the Brownian velocity  $\textbf{v}_{Bj}(t+\tau)$ will have lost all correlation with $\textbf{r}_{ij}(t)$.

The direct velocity of particle $i$, as modified by hydrodynamic interactions, is
\begin{equation}
\textbf{v}_{Di} = \sum_{j=1}^{N} \mu_{ij}\cdot\textbf{F}_{j} = - \sum_{j=1}^{N} \mu_{ij}\cdot\nabla_{j} W_{N}(\textbf{r}^{N})
\label{eq22}
\end{equation}
In eq. \ref{eq22}, $\textbf{F}_{j}=-\nabla_{j}W_{N}(\textbf{r}^{N})$ is the direct force on particle $j$, while $\mu_{ij}$ is a hydrodynamic mobility tensor the allows a force on particle $j$ to create a motion of particle $i$. If $\tau$ is small, $\textbf{v}_{Di}$ is nearly constant over $(t,t+\tau)$, so $\int_{t}^{t+\tau}ds\;\textbf{v}_{Di}(s)\textbf{v}_{Di}(t+\tau)$ is linear in $\tau$. The drift velocity depends on the time-dependent relative positions of all particles, so $\textbf{v}_{Di}(t+\tau)$ depends on $\tau$.

For hard spheres in solution, there are explicit forms for the $\mu_{ij}$. From the work of Kynch\cite{kynch1959a}, Batchelor\cite{batchelor1976}, this author\cite{phillies1982}, and Mazur and van Saarloos\cite{mazur1982}, $\mu_{ij}$ can be expanded as
\begin{equation}
\mu_{ii}=\frac{1}{f_o}\left( \textbf{I}+\sum_{l, l \neq  i}\textbf{b}_{il}+\sum_{m, m \neq i {\rm or} l; m, neq i, l} \textbf{b}_{iml}+...\right)
\label{eq23}
\end{equation}
for the self terms and
\begin{equation}
\mu_{ij}=\frac{1}{f_o}\left( \textbf{T}_{ij} + \sum_{m\neq i,j}\textbf{T}_{imj}+...\right), \quad i\neq j
\label{eq24}
\end{equation}
for the distinct terms.

The leading terms of the b and T tensors are\cite{mazur1982}
\begin{equation}
\textbf{b}_{il} = -\frac{15}{4}\left(\frac{a}{r_{il}}\right)^4\hat{\textbf{r}}_{il}\hat{\textbf{r}}_{il},
\label{eq25}
\end{equation}

\begin{eqnarray}
\textbf{b}_{iml} = \frac{75a^7}{16r_{im}^{2}r_{il}^{2}r_{ml}^{2}} \{ [1-3(\hat{\textbf{r}}_{im}\cdot \hat{\textbf{r}}_{ml})^{2}][1-3(\hat{\textbf{r}}_{ml}\cdot \hat{\textbf{r}}_{li})^{2}] \notag\\
+6(\hat{\textbf{r}}_{im}\cdot \hat{\textbf{r}}_{ml})(\hat{\textbf{r}}_{ml}\cdot \hat{\textbf{r}}_{li})^{2}-6(\hat{\textbf{r}}_{im}\cdot \hat{\textbf{r}}_{ml})(\hat{\textbf{r}}_{ml}\cdot \hat{\textbf{r}}_{li})(\hat{\textbf{r}}_{li}\cdot \hat{\textbf{r}}_{im}) \} \hat{\textbf{r}}_{im} \hat{\textbf{r}}_{li}
\label{eq26}
\end{eqnarray}

\begin{equation}
\textbf{T}_{ij}=\frac{3}{4}\frac{a}{r_{ij}}[ \textbf{I}+\hat{\textbf{r}}_{ij}\hat{\textbf{r}}_{ij}]
\label{eq27}
\end{equation}

\begin{equation}
\textbf{T}_{iml}=-\frac{15}{8}\frac{a^4}{r_{im}^{2}r_{ml}^{2}}[\textbf{I}-3(\hat{\textbf{r}}_{im}\cdot \hat{\textbf{r}}_{ml})^{2}]\hat{\textbf{r}}_{im}\hat{\textbf{r}}_{ml},
\label{eq28}
\end{equation}
where $a$ is the sphere radius, $r_{ij}$ is the scalar distance between particles $i$ and $j$, $\hat{\textbf{r}}\hat{\textbf{r}}$ denotes an outer (dyadic) product, and only the lowest order term (in $\frac{a}{r}$) of each tensor is shown. See Mazur and van Saarloos\cite{mazur1982} for the higher-order terms.

$\textbf{b}_{ij}$ and $\textbf{T}_{ij}$ represent interactions between pairs of interacting spheres. $\textbf{T}_{ij}$ describes the displacement of particle $i$ due to a force applied to particle $j$, while $\textbf{b}_{ij}$ describes the retardation of a moving particle $i$ due to the scattering by particle $j$ of the wake set up by $i$. $\textbf{T}_{iml}$ and $\textbf{b}_{iml}$ describe interactions between trios of interacting spheres. $\textbf{T}_{iml}$ describes the displacement of particle $i$ by a hydrodynamic wake set up by particle $l$, the wake being scattered by an intermediate particle $m$ before reaching $i$. $\textbf{b}_{iml}$ describes the retardation of a moving particle $i$ due to the scattering, first by $m$ and then by $l$, of the wake set up by $i$. The effect of pair interactions on $D_{m}$ was treated by Batchelor\cite{batchelor1976} and a host of others; effects of three-body interactions were treated by this author\cite{phillies1982} and virtually simultaneously by Beenakker and Mazur\cite{beenakker1982}. An approach similar to that used to generate eqs.\ \ref{eq25}-\ref{eq28} allows calculation of the hydrodynamic interactions between 2, 3, or more random-coil polymers, as well as the concentration dependences of various polymer transport coefficients.\cite{philliespolymer}.

In terms of the above, the diffusion tensor is
\begin{equation}
D_{ij}=k_{B}T\mu_{ij}.
\label{eq29}
\end{equation}

It is sometimes proposed\cite{snook1983} that hydrodynamic interactions in many-body systems should be screened, i.~e., $\textbf{T}\sim r^{-1}\exp(-\kappa r)$. The term screening is meant to suggest an analogy with electrolyte solutions, in which the Coulomb potential has the form seen in eq.\ \ref{eq32} when a background electrolyte is present. Snook et al.\ \cite{snook1983} discuss how screened hydrodynamic interactions would modify $D_{m}$. The basis of the assertion that there is hydrodynamic screening is the correct observation that the hydrodynamic interaction and the Coulomb interaction are both $1/r$ interactions, the correct observation that the Coulomb interaction is screened in ionic solutions, and the incorrect conclusion that by analogy the hydrodynamic interaction must also therefore be screened. The analogy is immediately flawed in that the Coulomb interaction is a $1/r$ potential, while the Oseen interaction is a $1/r$ force, the longest-range force in nature. The falsity of the analogy is shown by the existence of gravity, as quantitatively described by Isaac Newton. Gravitating bodies have a bare $1/r$ interaction, but a many-particle gravitational system very certainly does not show screening of gravitational forces\cite{kandrup1984}.

Rigorous analyses of many-body hydrodynamics show conclusively that hydrodynamic interactions between Brownian particles are not screened. Beenakker and Mazur\cite{beenakker1983} resummed the complete many-body hydrodynamic interactions tensors, showing that intervening particles weaken the fundamental hydrodynamic interaction $\textbf{T}_{ij}$ but do not alter its range from $1/r$. The resummation included all ring diagrams, ring diagrams being the longest-range many-particle diagrams. While Beenakker and Mazur did not include all diagrams in their resummation, rigorous mathematical analogy with plasma theory shows that the more complex diagrams not included in their resummation are shorter-ranged than the ring diagrams, and cannot reduce the fundamental range of $\textbf{T}$ below $r^{-1}$.

For Brownian particles, deGennes\cite{degennes1976} is sometimes cited as demonstrating the importance of hydrodynamic screening.  Reference \cite{degennes1976} in fact only cites Freed and Edwards\cite{freed1974a}, rather than presenting a derivation, and must be read against statements earlier in ref.\ \onlinecite{degennes1976} that translation invariance assures that there is no hydrodynamic screening at long distances in macromolecule solutions. Ref.\ \onlinecite{freed1974a} has since been considerably refined by Freed and Perico\cite{freed1981a}, who conclude that hydrodynamic interactions are modified by polymers that are free to move through solution, but that hydrodynamic interactions are not screened.

Altenberger, et al.\cite{altenberger1985a,altenberger1988a} present an extended discussion of hydrodynamic screening in different systems. They show that hydrodynamic interactions in solution are not screened, at least on time scales sufficiently long that inertial effects may be neglected.   The absence of screening can be understood in terms of momentum conservation. In a sand bed, in which particles are held fixed, $\textbf{T}$ is known to be screened. Friction irreversibly transfers momentum out of the fluid flow into the sand matrix, the irreversible loss of momentum being directly responsible for hydrodynamic screening. In a solution, friction can still transfer momentum from the fluid into suspended macromolecules, thereby slowing the fluid and accelerating the macromolecules. However, in an equilibrium system, the macromolecules' velocity distribution is independent of time. Any momentum transferred from the fluid into solute macromolecules must on the average rapidly return to the fluid, so interactions between the solvent and dissolved macromolecules do not transfer momentum irreversibly from the fluid to the macromolecules. On the time scales that are accessible with QELSS or related techniques, hydrodynamic screening therefore does not exist in macromolecule solutions. Only if one omits the return of momentum from the particles to the fluid can one obtain screening for hydrodynamic interactions in solution. Altenberger, et al.\cite{altenberger1988a} present a detailed analysis of all claims, up to their time of publication, that there is hydrodynamic screening, showing why each of these claims is incorrect.

\subsection{\label{section4F} Application of the Model}

Eq.\ \ref{eq12} defines $D_{m}$ as the limit of a derivative as $\tau \rightarrow 0$. This section considers that derivative as applied to a system that follows the Generalized Langevin equation, which for a suspension of interacting macromolecules becomes
\begin{equation}
      \textbf{v}_{i} = \textbf{v}_{Bi} + \mu_{ii} \cdot (- \nabla_{i} W_{N}) + \sum_{j=1, j \neq i}^{N} \mu_{ij} (- \nabla_{j} W_{N})
      \label{eq:GLE1}
\end{equation}

It must be recognized that digital correlators only measure $g^{(1)}(q,\tau)$ for $\tau\gg\tau_{B}$. The experimentally accessible short-time limit of $g^{(1)}$ only reflects system behavior occurring well after the Brownian velocity components, other than the long-time tail, have been thermalized. The limit $\tau \rightarrow 0$  is thus actually a limit $\tau \rightarrow \epsilon$ for some $\epsilon \gg \tau_{H}, \tau_{B}$.  Combining eqs. \ref{eq12}, \ref{eq18}, \ref{eq19}, \ref{eq20}, \ref{eq21}, and \ref{eq22}, and eliminating terms that are linear or higher in $\tau$ one finds as an intermediate form
\begin{equation}
D_{m}=-\frac{1}{q^{2}g^{(1)}(q,0)} \; (\; \langle \sum_{i,j}\exp(i\textbf{q}\cdot\textbf{r}_{ij}(t))(-i\textbf{q}\cdot\textbf{v}_{Bi}
-i\textbf{q}\cdot\textbf{v}_{Di})  \notag
\end{equation}
\begin{equation}
     + \int_{t}^{t+\tau}ds\; \exp[i\textbf{q}\cdot\textbf{r}_{ij}(t)](\textbf{q}\cdot[\textbf{v}_{Bj}
(s)\textbf{v}_{Bj}(t+\tau) +\textbf{v}_{Dj}(s)\textbf{v}_{Bj}(t+\tau)  \notag
 \end{equation}
 \begin{equation}
     + \textbf{v}_{Bj}(s)\textbf{v}_{Dj}(t+\tau)  +\textbf{v}_{Dj}(s)\textbf{v}_{Dj}(t+\tau) ]\cdot\textbf{q})  \; \rangle \; ).
\label{eq33B}
\end{equation}

The first and last terms of the series of terms in $\textbf{v}_{Bj}$ and $\textbf{v}_{D}$  vanish, the first because between $t$ and $t+\tau$  $\textbf{v}_{Bj}(t+\tau)$ has thermalized and the last because it is linear in $\tau$ and vanishes as $\tau \rightarrow \epsilon$.  The fourth and fifth terms, the terms in $\textbf{v}_{Dj}\textbf{v}_{Bj}$ are the dynamic friction contribution to $D_{m}$, (the name was first used by Chandrasekhar\cite{chandrasekhar1942} in connection with a similar problem in stellar dynamics), whose individual terms are
\begin{equation}
\textbf{q}\cdot\Delta D_{ij}^m\cdot\textbf{q}= \left\langle\int_{t}^{t+\tau}ds\;\exp[i\textbf{q}\cdot\textbf{r}_{ij}]\textbf{q}\cdot[\textbf{v}_{Bj}
(s)\textbf{v}_{Dj}(t+\tau)+\textbf{v}_{Dj}(s)\textbf{v}_{Bj}(t+\tau)]\cdot\textbf{q}\right\rangle.
\label{eq35}
\end{equation}
and whose collective effect is
\begin{equation}
\textbf{q}\cdot \Delta D_{m}\cdot \textbf{q}= \sum_{i,j=1}^{N}\textbf{q}\cdot\Delta D_{ij}^m\cdot\textbf{q}
\label{eq36}
\end{equation}

$\Delta D_{m}$ describes the correlations between the Brownian displacements of each particle and the subsequent direct forces on it. $\Delta D_{m}$ is non-vanishing because, on any time scale sufficiently long that $S(q,\tau)$ is not a constant, the particles have moved from their initial positions. The expression for $D_{m}$ reduces to
\begin{equation}
D_{m}=-\frac{1}{q^{2}g^{(1)}(q,0)}\left(\left\langle\sum_{i,j}\exp(i\textbf{q}\cdot\textbf{r}_{ij}(t))(-i\textbf{q}\cdot\textbf{v}_{Di}-\textbf{q}\cdot D_{ii}\cdot \textbf{q})\right\rangle-\textbf{q}\cdot \Delta D_{m} \cdot \textbf{q}\right).
\label{eq33}
\end{equation}
An integration by parts displaces the points of application of the $\nabla_{i}$ operators implicit in $\textbf{v}_{Di}$; the $\nabla_{i}$ were shown explicitly in eq.\ \ref{eq:GLE1}. One finds
\begin{equation}
   D_{m}=-\frac{1}{q^{2}g^{(1)}(q,0)}\left( k_{B}T \left\langle\sum_{i,j,l}i\textbf{q}\nabla_{l}  : [\mu_{jl}\exp(i\textbf{q}\cdot\textbf{r}_{ij})]-\sum_{i,j}\textbf{q}\cdot\mu_{jj}\cdot\textbf{q}\exp[i\textbf{q}\cdot\textbf{r}_{ij}]\right\rangle
      -\textbf{q}\cdot\Delta D_{m}\cdot\textbf{q}\right).
\label{eq34}
\end{equation}

Eq.\ \ref{eq33} represents most clearly that $D_{m}$ arises from two sorts of motion, namely self-diffusion of individual solute molecules (terms in $D_{ii}$ and $\Delta D_{m}$) plus driven motion due to interparticle interactions (the term in $\textbf{v}_{Di}$). Drift velocities $\textbf{v}_{Di}$ are typically much smaller than Brownian velocities $\textbf{v}_{Bi}$ suggesting (incorrectly!) that the self-diffusion terms should dominate driven motion in eq. \ref{eq34}. However, drift velocities persist over long times, while Brownian velocities at a series of times add incoherently to the total particle motion. What is the persistence time of the drift velocity associated with a concentration fluctuation $a_{q}(t)$?  So long as $a_{q}(t)$ persists, the macromolecular concentration will be nonuniform.  A macromolecule in the concentration gradient corresponding to  $a_{q}(t)$ will on the average experience a net force that persists as long as the concentration gradient does. On the time scales observed by QELSS, the drift contribution to $D_{m}$ can be several times the self-diffusive contribution. As an experimental demonstration, note Doherty and Benedek's demonstration that removing the salt from a serum albumin solution can increase $D_{m}$ as measured by QELSS by severalfold over its high-salt value\cite{doherty1974a}.

Eq.\ \ref{eq34} simplifies. Ref.\ \onlinecite{phillies1984} demonstrates
\begin{equation}
   D_{m}= -\frac{1}{q^{2}g^{(1)}(q,0)}\left( \left\langle \sum_{i,j,l}\exp[i\textbf{q}\cdot\textbf{r}_{ij}]i\textbf{q}\nabla_{l}:[D_{jl}]+\sum_{i,j}(\textbf{q}\cdot D_{ij}\cdot\textbf{q})\exp[i\textbf{q}\cdot\textbf{r}_{ij}]\right\rangle
    + \textbf{q}\cdot\Delta D_{m} \cdot \textbf{q}\right)
\label{eq37}
\end{equation}
As emphasized by Felderhof\cite{felderhof1977}, $\nabla_{l}\cdot D_{jl}\neq 0$ if $D_{jl}$ is taken beyond $\mathcal{O} (\frac{a}{r})^5$. Eq.\ \ref{eq37} is totally consistent with the Smoluchowski equation if one interprets $D$ of the Smoluchowski equation, not as the Kubo-form coefficient of eq. \ref{eq19}, but as a dressed diffusion coefficient incorporating the dynamic friction correction\cite{phillies1984} $\textbf{q}\cdot\Delta D_{m} \cdot \textbf{q}$. Hess and Klein's\cite{hess1983} excellent review presents a different treatment of this same question.

The evaluation of eq.\ \ref{eq37} naturally separated into two parts.  In the next section, we evaluate the explicit ensemble average $\langle \cdots \rangle$.  The following section treats the dynamic friction term $\textbf{q}\cdot\Delta D_{m} \cdot \textbf{q}$.

\subsection{\label{section4G}Evaluation of $D_{m}$ for Hard Spheres}

For a solution of hard spheres, the most extensive evaluation of eq.\ \ref{eq37} is that of Carter and Phillies\cite{carter1985}; calculations of $D_{m}$ are also reported by Felderhof\cite{felderhof1978} and Beenakker and Mazur\cite{beenakker1982}.  Eq.\ \ref{eq37} can be written as a cluster expansion, i.\ e.\ as a sum of averages over distribution functions. The $n$-particle distribution function is
\begin{equation}
g^{(n)}(\textbf{r}_{1},\textbf{r}_2,...\textbf{r}_n)=V^{-N+n}\int d\{N-n\}\exp[-\beta(W_{N}-A)],
\label{eq38}
\end{equation}
where $V$ is the system volume, $A$ is the normalizing coefficient $V^{-N}\int d\{N\}\exp[-\beta(W^{N}-A)]=1$, and $d\{N-n\}$ denotes an integral over macroparticles $N-n+1,N-n+2,...,N$. The $g^{(n)}$, being concentration-dependent, have (except in electrolyte solutions) pseudovirial expansions
\begin{equation}
g^{(2)}(\textbf{r}_{12})=g^{(2,0)}(\textbf{r}_{12})+cg^{(2,1)}(\textbf{r}_{12})+c^{2}g^{(2,2)}(\textbf{r}_{12})+...
\label{eq39}
\end{equation}
\begin{equation}
g^{(3)}(\textbf{r}_{12},\textbf{r}_{13})=g^{(3,0)}(\textbf{r}_{12},\textbf{r}_{13})+cg^{(3,1)}(\textbf{r}_{12},\textbf{r}_{13})+...
\label{eq40}
\end{equation}
In these expansions the $g^{(n,i)}$ are concentration-independent.

The terms of eq.\ \ref{eq37} that contribute to $D_{m}$ through second order in $c$ may be rearranged as
\begin{equation}
D_{m}=\frac{D_{o}}{q^{2}g^{(1)}(q,0)}[Nq^{2}+I_A+I_B+...+I_G]
\label{eq41}
\end{equation}
in which
\begin{equation}
I_A=cN \int d\textbf{r}\; g^{(2)}(\textbf{r})\textbf{q}\cdot \textbf{b}_{12}(\textbf{r})\cdot\textbf{q}
\label{eq42}
\end{equation}
\begin{equation}
I_B=cN\int d\textbf{r}\;g^{(2)}(\textbf{r})\textbf{q}\cdot\textbf{T}_{12}(\textbf{r})\cdot\textbf{q}\exp(-i\textbf{q}\cdot\textbf{r})
\label{eq43}
\end{equation}
\begin{equation}
I_C=-cN\int d\textbf{r}\;g^{(2)}(\textbf{r})[\exp(-i\textbf{q}\cdot\textbf{r})i\textbf{q}\nabla :[\textbf{b}_{12}(\textbf{r})+\textbf{T}_{12}(\textbf{r})]+i\textbf{q}\nabla :\textbf{T}_{12}(\textbf{r})]
\label{eq44}
\end{equation}
\begin{equation}
I_D=c^{2}N\int d\textbf{r}d\textbf{s}\;g^{(3)}(\textbf{r},\textbf{s})\textbf{q}\cdot\textbf{b}_{123}(\textbf{r},\textbf{s})\cdot\textbf{q}
\label{eq45}
\end{equation}
\begin{equation}
I_E=c^{2}N\int d\textbf{r}d\textbf{s}\;g^{(3)}(\textbf{r},\textbf{s})\textbf{q}\cdot\textbf{T}_{123}(\textbf{r},\textbf{s})\cdot\textbf{q}\exp(-i\textbf{q}\cdot\textbf{s})
\label{eq46}
\end{equation}
\begin{equation}
I_F=-c^{2}N\int d\textbf{r}d\textbf{s}\;g^{(3)}(\textbf{r},\textbf{s})\exp(-i\textbf{q}\cdot\textbf{s})i\textbf{q}\nabla :[\textbf{b}_{12}(\textbf{r})+\textbf{T}_{12}(\textbf{r})]
\label{eq47}
\end{equation}
\begin{equation}
I_G=-c^{2}N\int d\textbf{r}d\textbf{s}\;g^{(3)}(\textbf{r},\textbf{s})[1+\exp(-i\textbf{q}\cdot\textbf{r})+\exp(-i\textbf{q}\cdot\textbf{s})][i\textbf{q}\nabla_{1}:\textbf{b}_{123}(\textbf{r},\textbf{s})+i\textbf{q}\nabla_3:\textbf{T}_{123}(\textbf{r},\textbf{s})]
\label{eq48}
\end{equation}
where $\textbf{r} \equiv \textbf{r}_{12}$ and $\textbf{s} \equiv \textbf{r}_{13}$. Replacing the $g^{(n)}$ with the $g^{(n,i)}$ of eqs.\ \ref{eq39} and \ref{eq40} transforms the $I_{i}$ into a pseudovirial expansion for $D_{m}$.

In the limit of low $q$, most of these integrals were evaluated by Carter\cite{carter1985} using predominantly analytic means or by Beenakker and Mazur\cite{beenakker1982} by Monte Carlo integration; Felderhof\cite{felderhof1978} obtained many of the $c^1$ corrections analytically. There are small differences between different calculations of $D_{m}$ because some authors\cite{beenakker1982,felderhof1978} omit the $\nabla\cdot\mu$ terms, while others treat reference frame corrections in different ways. Table I, taken from Carter et al.\ \cite{carter1985} gives the contributions of the $I_{i}$ to the first and second order concentration corrections to $D_{m}$. In Table I, the fundamental concentration unit is the solute volume fraction $\phi=4\pi a^3N/3$.

\centerline{\large\bf Table One}
Integrals for the concentration dependence of $D_{m}=D_{o}[1+K_{1}\phi+k_2\phi^{2}]/g^{(1)}(q,0)$.

\begin{tabular}{|r|r|r|}
\hline
Equation & $K_{1}$ & $k_2$\\
\hline
$I_{A}$ & -1.734 & -0.927\\
$I_{B}$& -5.707 & 13.574\\
$I_{C}$& -1.457 & 1.049\\
$I_{D}$& 0 & 1.80\\
$I_{E}$ & 0 & 6.69\\
$I_{F}$ & 0 & -4.1348\\
$I_{G}$ & 0 & 4.120\\
totals & -8.898 & 22.17\\
\hline
\end{tabular}

Details of the integrations appear in Carter et al.\ \cite{carter1985}. Several technical issues are noteworthy:

1) In eq.\ \ref{eq43}, the convergence of $g^{(2,0)}$ over the $(a/r)^1$ component of $T_{ij}$ is delicate. The Oseen tensor $\frac{3}{4}\frac{a}{r}(\textbf{I}+\hat{\textbf{r}}\hat{\textbf{r}})$ is only an approximation to the long-range part of the exact $T'_{il}$, inaccurate at extremely large distances because it does not enforce the physical requirement that the total volume flux of solvent and particles, across any surface that divides the container in twain, must vanish. Correction of the long-range behavior corresponds to the reference frame correction\cite{kirkwood1960}, which physically requires -- for a system with negligible volume of mixing -- that the volume flux of solute into a volume of space must cancel the simultaneous volume flux of solvent out of the same volume.The next subsection expands on this point. As shown in refs.\ \cite{phillies1982,phillies1981} and the next subsection, for a small hard solute the reference frame correction may be stated
\begin{equation}
\int d\textbf{r}\;([\hat{\textbf{q}}\cdot\textbf{T'}(\textbf{r})\cdot\textbf{q}]\exp(i\textbf{q}\cdot\textbf{r})+\phi_{l}\emph{H}(q))=0
\label{eq49}
\end{equation}
which lets one write
\begin{equation}
\hat{\textbf{q}}^{2}:\int d\textbf{r}\;g^{(2,0)}(r)\textbf{T'}(\textbf{r})=\hat{\textbf{q}}^{2}:\int d\textbf{r}\;\left[(g^{(2,0)}(r)-1)\frac{3a}{4r}(\textbf{I}+\hat{\textbf{r}}\hat{\textbf{r}})-\phi_{l}q^{2}\right]
\label{eq50}
\end{equation}
where $\phi_{l}$ is the fraction of the system's volume occupied by a single particle, $T'_{il}(\textbf{r}_{il})$ is the true long-range part of the hydrodynamic interaction tensor for a closed container, $\emph{H}(q)$ is the Fourier transform of the hydrodynamic shape of  the solute particle, and $\textbf{T'}(\textbf{r})$ has been approximated by $\textbf{T}(\textbf{r})$ over the short range within which $g(r) - 1 \neq 0$.  Eq.\ \ref{eq49} is basically physically equivalent to Oono and Baldwin's \cite{oono1986} method for removing the divergence from the perturbation series for $D_{m}$ of a solution of random-coil polymers.

2) A term of eq.\ \ref{eq43} is
\begin{equation}
I_{B2}=\frac{1}{2}c\int d\textbf{r}\;g^{(2,0)}(r)[q^{2}-3(\textbf{q}\cdot\hat{\textbf{r}})]\left(\frac{a}{r}\right)^3\exp(-i\textbf{q}\cdot\textbf{r})
\label{eq51}
\end{equation}
where the limits on $|r|$ come from $g^{(2,0)}$. If the $\exp(i\textbf{q}\cdot\textbf{r})$ were absent, the angular integral over $[q^{2}-3(\textbf{q}\cdot\hat{\textbf{r}})]$ would vanish, while the $\int^{\infty}r^{2}dr\;(a/r)^3$ would diverge; the integral without the $\exp(i\textbf{q}\cdot\textbf{r})$ is improper and has no meaningful value. On retaining the $\exp(i\textbf{q}\cdot\textbf{r})$ term, doing the integrals, and taking the small-$q$ limit at the end, one finds $I_{B2}=4\pi a^3cq^{2}/3$.

3) To evaluate $I_D...I_G$ analytically, Carter et al.\ \cite{carter1985} used a spherical harmonic expansion technique, originally due to Silverstone and Moats\cite{silverstone1977}, and introduced to statistical mechanics by this author\cite{phillies1981b,phillies1991}. The general objective of the expansion is to expand all functions in terms of spherical coordinates centered at some single point $\textbf{r}_{1}$. A function of $\textbf{r}_{23}$ is naturally expanded in terms of spherical harmonics centered at $\textbf{r}_2$ or $\textbf{r}_3$; expanding a function $f(\textbf{r}_{23})$ around $\textbf{r}_{1}$ appears unnatural. However, once all functions are expanded in terms of spherical harmonics centered at a single point $\textbf{r}_{1}$ together with polynomials in the scalar distances $r_{12},r_{13},...,r_{1n}$ from particle $1$, all angular integrals are integrals over products of spherical harmonics, which are basically trivial. By means of the spherical harmonic technique, a general $N$-particle cluster integral can be reduced from $3N$ to $N-1$ non-trivial integrations\cite{phillies1981b}. While the spherical harmonic expansion can yield an infinite series in the order $l$ of the harmonics, analytic calculations show that the infinite series typically converges exponentially rapidly (in $l$) to the correct answer\cite{phillies1991}. In equations \ref{eq45}-\ref{eq48}, matters are simpler. The spherical harmonic expansion here reduces $\int d\textbf{r}\ d\textbf{s}$ to a non-trivial two-dimensional integral.

For the dynamic structure factor of hard spheres, the result
\begin{equation}
\left(g^{(1)}(q,0)\right)^{-1}=1+8\phi+30\phi^{2}+...
\label{eq52}
\end{equation}
obtains.

Combining the above equations, for a solution of hard spheres at volume fraction $\phi$ one predicts for low $q$
\begin{equation}
D_{m}=D_{o}(1-0.898\phi-19.0\phi^{2}+...)+\hat{\textbf{q}}\cdot\Delta D_{m}\cdot\hat{\textbf{q}}.
\label{eq53}
\end{equation}
Some European authors\cite{felderhof1978,batchelor1976,beenakker1982} give an alternative form which is approximately
\begin{equation}
D_{m}=D_{o}(1+1.56\phi+0.91\phi^{2}+...).
\label{eq54}
\end{equation}
Eq.\ \ref{eq54} differs from eq.\ \ref{eq53} in its $\phi^1$ coefficient because the latter equation: neglects all terms in $\nabla\cdot\mu_{ij}$, assumes that the net solvent volume flux rather than the net total volume flux across closed surfaces must vanish, and ignores dynamic friction effects. A demonstration comparing our derivation of eq.\ \ref{eq54} with other published results as published in the open literature\cite{phillies1987} is given in Appendix A.

\subsection{\label{section4H}Dynamic Friction Effects}
Recognition of the importance of dynamic friction in diffusion can be traced back to Mazo\cite{mazo1965}, who analyzed results of Stigter et al.\ \cite{stigter1955} on $D_{s}$ of sodium lauryl sulfate micelles at various surfactant and background electrolyte concentrations. Schurr\cite{schurr1980-45,schurr1980-71} gives a related treatment of electrostatic effects in $D_{s}$. Beginning with the Einstein relation $D = k_{B}T/f$ and the Kirkwood\cite{kirkwood1946} fluctuation-dissipation equation
\begin{equation}
f=\frac{1}{3k_{B}T}\int_{0}^{\tau}\langle\textbf{F}(0)\cdot\textbf{F}(t)\rangle,
\label{eq66}
\end{equation}
Mazo showed that $f$ of a micelle has a component due to the micelle-micelle part of $\textbf{F}(t)$; his calculations\cite{mazo1965} on a simple model for a charged micelle found good agreement with experiment \cite{stigter1955}. these results were anticipated by the corresponding demonstration of Chandrasekhar and von Neumann\cite{chandrasekhar1942} of the existence of dynamic friction in the stellar dynamics of star clusters.

The Kirkwood equation suggests that interparticle forces should also augment the Stokes-Law drag coefficient $f_d$, which pertains to straight line unaccelerated motion. In the approximation $\textbf{T}_{ij}\approx 0$, this author\cite{phillies1977} found
\begin{equation}
\delta f_d(t)=\frac{k_{B}Tc_o}{(2\pi)^3}\int d\textbf{k}\;\frac{[h(k)]^{2}}{g^{(1)}(k)}(\textbf{k}\cdot\hat{\textbf{v}})^{2}\frac{\exp[-i(\textbf{k}\cdot\textbf{v}+\Gamma_{k})t]-1}{(\textbf{k}\cdot\textbf{v}+\Gamma_{k})t}
\label{eq67}
\end{equation}
where $\delta f_d(t)$ is the time-dependent increment to $f_d$; $\gamma_{k}=D_{m}k^{2}\approx k_{B}T[1-ch(k)](1-\phi)/f_o$; $\phi$, $c$, and $f_o$ are the solute volume fraction, concentration, and free-particle drag coefficient, respectively, and where $h(k) = \int d\textbf{r}\;[g^{(2)}(r)-1]\exp(i\textbf{k}\cdot\textbf{r})$. For hard spheres, eq.\ \ref{eq67} shows $f_d=f_o(1+\alpha_f\phi)$ for $\alpha_f=\frac{8}{3}$. For charged hard spheres with an auxiliary Debye interaction, $\alpha_f$ can become extremely large, as previously shown by Mazo. Similar considerations show the contributions of corresponding effects to the mutual diffusion coefficient, the self-diffusion coefficient, and the solution viscosity.\cite{phillies1981-2,phillies1983a,phillies1979a}.

The formalism developed above for $D_{m}$, $D_{s}$, and $D_{p}$ leads naturally to the terms $\Delta D_{m}$ of eq.\ \ref{eq35} \cite{phillies1981-2} and $\Delta D_{s}$ of eq.\ \ref{eq62} \cite{phillies1983a}. $\Delta D_{s}$ of eq.\ \ref{eq62} has essentially the same form as the Kirkwood equation \ref{eq66}, differing only in the derivation. In the limit of small $q$, $\exp(i\textbf{q}\cdot\textbf{r}_{ij})\approx 1$ so at small $q$ one has $\Delta D_{m} = \Delta D_{s}$. To lowest order in $q$, dynamic friction has the same effect on $D_{m}$ and $D_{s}$. Both $\Delta D$ terms are $\mathcal{O}(q^0)$ so neither $\Delta D_{m}$ nor $\Delta D_{s}$ vanishes at low $q$.

$\Delta D$ may be understood as arising from a caging effect, in which the motions of each particle are modified by interactions with its neighbors. No matter which way a particle moves (no matter what the orientation of $\Delta\textbf{x}_{Bi}=\int\textbf{v}_{Bi}(s)ds$), the particle's neighbors lag in responding to the particle's displacement. On the average, then, for any $\Delta \textbf{x}_{Bi}$ particle $i$ experiences a retarding force that drives the particle back towards its original position. However, by diffusion the cage recenters itself on the new location of the diffusing particle, the recentering being described expanding the particle's radial distribution function into its sinusoidal and cosinusoidal components, and allowing them to relax via diffusion toward their new equilibrium values.

Physically, dynamic friction acts by dispersing the random Brownian force $\textbf{F}_{Bi}$ on each particle over the nearby particles. In the presence of dynamic friction, the Brownian motion of each particle is driven by an average over the random forces on a particle and its neighbors. The probability distribution of the averaged force is narrower than the probability distribution of the individual $\textbf{F}_{Bi}$, so averaging the random force reduces $D_{m}$ and $D_{s}$.

The physical origin of dynamic friction is clarified by considering a particular interaction potential between two Brownian particles, namely a covalent bond. A covalent bond serves to average the random force on either particle in a dimer over both particles. A random force that would have displaced a free particle $1$ through $\textbf{R}_{1}$ will displace (up to corrections arising from hydrodynamic interactions) each particle of a covalently bonded pair through $\textbf{R}_{1}/ 2$. Superficially, while the covalent bond retards the motion of either particle, the bond does not appear to reduce the mass flux $m\delta x$ of diffusion; instead of moving a pair of particles each of mass $m$ through distances $\delta x_{1}$ and $\delta x_2$, the random force moves a particle of mass $2m$ through a distance $\frac{1}{2}(\delta x_{1} + \delta x_2)$. Less superficially, Brownian forces are random. If the free particles would have had random displacements $\textbf{R}_{1}$ and $\textbf{R}_2$, the displacement of either particle in a linked pair during the same time interval would be $\textbf{R}_{D} = (\textbf{R}_{1} + \textbf{R}_2)/ 2$. The random displacements $\textbf{R}_{1}$ and $\textbf{R}_2$ are independent random variables; they therefore add incoherently. The probability distribution for $\textbf{R}_{D}$ is thus narrower than the distributions for $\textbf{R}_{1}$ or $\textbf{R}_2$, so the random displacements of either particle in a dimer are less than the random displacements of a monomer. Forming a dimer reduces the mass flux arising from diffusion; in terms of the above treatment, if a monomer of mass $m$ moves a distance $\delta x$ the covalent bond causes a dimer of mass $2m$ to move a distance less than $\delta x/2$.

Several authors have suggested that dynamic friction should affect $D_{s}$, without necessarily affecting $D_{m}$. For example, Ackerson\cite{ackerson1976} applied the Mori formalism to solve the Smoluchowski equation,
\begin{equation}
\frac{\partial \rho(\textbf{r}^{N},t)}{\partial t}=\sum_{i,j=1}^{N}\frac{\partial}{\partial \textbf{r}_{i}}D_{ij}^{Sm}(\frac{\partial}{\partial\textbf{r}_{j}}-\beta\textbf{F}_{j})\rho(\textbf{r}^{N},t),
\label{eq68}
\end{equation}
obtaining a dynamic friction contribution to $D_{m}$ that vanishes if hydrodynamic interactions between particles are not included in the calculation. Here $\rho(\textbf{r}^{N},t)$ is the $N$-particle density function, $\textbf{F}_{j}$ is the direct force on particle $j$, and $D_{ij}^{Sm}$ is the Smoluchowski two-particle diffusion coefficient. In contrast, $\Delta D_{m}$ is non-zero for hard spheres with no hydrodynamic interactions. Implicit to Ackerson's analysis of the Smoluchowski equation is the assumption that the diffusion tensor $D_{ij}^{Sm}$ is a purely hydrodynamic object, all effects of the direct interactions being contained in the sedimentation term $D_{ij}\beta\textbf{F}_{j}$. However, comparison with the Langevin approach shows\cite{phillies1984} that $D_{ij}^{Sm}$ is properly interpreted as a dressed diffusion coefficient $D_{ij}^{Sm} \sim D_{ij}+\Delta D_{m}$, whose value depends in part on direct interactions. The Ackerson calculation is correct, except that it does not capture the part of dynamic friction already hidden with $D_{ij}^{Sm}$. The Smoluchowski and Langevin equation approaches therefore agree, except that the Langevin approach suggests how to calculate $\Delta D_{m}$ but the Smoluchowski approach must receive $D_{ij}^{Sm} \sim D_{ij}+\Delta D_{m}$ from an external source.

The cage size $l$, which is determined by the natural range of the interparticle interactions, is described by $g^{(2)}(r)$. For example, two charged spheres of radius $a$ in a solvent of Debye length $\kappa^{-1}$ have $l\approx a+\kappa^{-1}$, independent of the sphere concentration. The natural time scale on which $\delta f_d$ attains its long-time value is the time required for the particles to diffuse across a cage, which is $\tau_d=l^{2}f_o/k_{B}T$. At times $\ll\tau_d$, $\delta f_d\rightarrow 0$. The longer the range of interparticle interactions, the greater will be the time required for $\delta f_d$ to reach its limiting value. Since $l\sim c^0$, $\tau_d$ is independent of $c$. The typical distance between near neighbors (that is, the concentration $c$) determines the strength of the cage (so $\Delta D \sim c^1$), but does not directly affect the size of the cage (so $\tau_d\sim c^0$).

How does one interpret a time-dependent $\Delta D_{m}$ (and hence a time-dependent first cumulant $K_{1}$)? There are three natural time scales involved, namely $\tau_d$ defined above, the shortest time $\tau_{1}$ resolved by the correlator, and the time $\tau_q$ at which the quadratic correction $Ct^{2}$ to $S(q,t)=A-Bt+Ct^{2}$ becomes significant. $\tau_q$ is also the time scale on which particle displacements become comparable to $q^{-1}$, so that, e.~g., the difference between $\textbf{T}_{ij}(\textbf{r}_{ij}(\tau_q))$ and $\textbf{T}_{ij}(\textbf{r}_{ij}(0))$ is significant. For meaningful data, one must have $\tau_{1}\ll\tau_q$. If $\tau_d\gg\tau_q$ is the longest of these three times (as is found in the high-$q$ limit), the spectrum decays before particles encounter dynamic friction; $K_{1}$ is not modified by $\Delta D_{m}$. Conversely, if $\tau_d\ll\tau_{1}$, $\Delta D_{m}$ reaches its plateau value before any data is obtained; $K_{1}$ contains $\Delta D_{m}$ as a constant. If $\tau_{1}<\tau_d\leq\tau_q$, the time- dependence of $\Delta D_{m}$ will be visible in the spectrum, leading to a time dependence of the nominal $K_{1}$ defined above.

Hess and Klein\cite{hess1983} compute quantities equivalent to $\Delta D_{m}$ for a variety of systems, finding good agreement with the experiments of Gruener and Lehmann\cite{gruener1979} on QELSS of interacting polystyrene latex spheres. The model of Hess and Klein\cite{hess1983} predicts $S(q,t)\sim t^{-3/2}$ at long time. The experimental $S(q, t)\sim t^{-\alpha}$ for $\alpha\approx 1.5-1.2$, found by this author\cite{phillies1983z} for interacting polystyrene latex spheres in systems of low ionic strength, supports the Hess-Klein\cite{hess1983} model.

Fluctuation-dissipation calculations related to eq.\ \ref{eq66} predict an increment in solution viscosity from particle-particle interactions. In the limit in which eqs.\ \ref{eq35} and \ref{eq62} have been evaluated, $\Delta D_{m}=\Delta D_{s}$, but the viscosity increment is not the same as the increments to $D_{m}$ and $D_{s}$. Dynamic friction causes the Stokes-Einstein equation to fail, changes in $D$ and $\eta$ not being proportional to each other\cite{phillies1979a}.

\subsection{\label{section4I}Microscopic Treatment of Reference Frames}

Reference frame effects are undoubtedly familiar to small children. While using a bathtub, most children are all too aware that if they move rapidly from one end of the tub to the other, both the water and all the toys floating in the water move rapidly in the opposite direction. The objective of this Section is to restate this familiar observation of household physics in a slightly more elaborate mathematical form useful for the discussion of light scattering methods.
The restatement is originally found in Phillies and Wills\cite{phillies1981}, which is followed closely here.

The physical basis of reference frame corrections is the fundamental requirement that the volume flow of an incompressible solution across a closed boundary must vanish. Container walls\textemdash{}through which volume flow is impossible\textemdash{}may be part of the ``closed'' boundary. For a finite volume of an incompressible solution in a closed container, an equivalent statement of the fundamental requirement on volume flow is that volume flow across a plane that bisects the volume must vanish. Systems with non-zero volumes of mixing, in which the volume of the system is changed by concentration fluctuations, raise fundamental complications not included here.

The ultimate objective of this discussion is to verify that equation \ref{eq49}
\begin{equation}
\int_{V} d\textbf{r}_{i} \;\left[\hat{\textbf{q}}\cdot\textbf{T}'(\textbf{r}_{ij})\cdot\hat{\textbf{q}} \exp(i\textbf{q}\cdot\textbf{r}_{ij})+\phi_{j}H(q)\right]=0,
\label{eq125}
\end{equation}
is a correct statement of the reference frame correction. In this equation, $V$ is the container volume, $\hat{\textbf{q}}$ denotes the unit vector of $\textbf{q}$, $\phi_{j}$ is the volume fraction of the single particle $j$ which is driving the flow, $H(q)$ is the spatial Fourier transform of the hydrodynamic excluded volume of particle $j$, and $\textbf{T}'(\textbf{r}_{ij})$ is the exact hydrodynamic interaction tensor giving the flow of solution at $i$ arising from motions of particle $j$. For a solution containing $N$ identical solute particles, $\phi=N\phi_{j}$ is the volume fraction of solute in the solution.  The Oseen tensor  $\textbf{T}(\textbf{r}_{ij})$ is an approximation to $\textbf{T}'$ . The location of the particle driving the flow is $\textbf{r}_{j}$; $\textbf{r}_{i}$ labels all the locations in $V$.

The volume flux has two components. A moving particle transports its own volume across any surface it crosses. A moving particle also induces a volume flux in the surrounding medium. The volume flux of a particle across a surface $S$ is determined by the particle's velocity and cross-section within $S$. The volume flux of a particle $j$  across a surface $S$ may be written
\begin{equation}
      C_{j}(S)\; \hat{\textbf{s}}\cdot\textbf{v}_{j}.
\label{eq126}
\end{equation}
Here $S$ is a plane across the container, $\hat{\textbf{s}}$ is the normal to $S$, and $C_{j}(S)$ is the cross-sectional area of particle $j$ contained in the surface $S$. If particle $j$ is not intersected by $S$, $C_{j}(S)=0$.

A moving particle $j$ with velocity $\textbf{v}_{j}$ and location $\textbf{r}_{j}$ induces at the point $\textbf{r}_{i}$ in the surrounding fluid a fluid flow $\textbf{v}_{i}$:
\begin{equation}
\textbf{v}_{i}(\textbf{r}_{i})=\textbf{T}'(\textbf{r}_{ij})\cdot\textbf{v}_{j},
\label{eq127}
\end{equation}
This paper is based on linear hydrodynamics, in which fluid flows due to different particles simply add, so to first approximation the total flow at $\textbf{r}_{i}$ caused by all the particles in the system may be obtained by summing eq. \ref{eq127} on $j$. The flow across $S$ may be obtained by integrating $\textbf{v}_{i}\cdot\hat{\textbf{s}}$, the solvent flow perpendicular to $S$ over all points of $\textbf{r}_{i}$ in $S$.

Choose for a closed surface\textemdash{}across which the volume flux is required to vanish\textemdash{}some of the walls of the container, together with a plane $S$ that is perpendicular to the scattering vector $\textbf{q}$. Across this surface, the volume flow is
\begin{equation}
  C_{j}(S)\hat{\textbf{s}}\cdot\textbf{v}_{j}+ \int_{S}dS \left[\hat{\textbf{s}}\cdot\textbf{T}'(\textbf{r}_{ij})\cdot\textbf{v}_{j}\right] = 0,
\label{eq128}
\end{equation}
where the $\int dS$ includes all points in the plane $S$.  There are no flows across any of the walls of the container.

For $\hat{\textbf{q}}||\hat{\textbf{s}}$, the unit vectors $\hat{\textbf{q}}$ and $\hat{\textbf{s}}$ are interchangeable. Within the plane $S$, $\exp(i\textbf{q}\cdot\textbf{r}_{ij})$ is a constant.  As a result
\begin{equation}
\int_{S} d\textbf{r}_{i} \; \hat{\textbf{s}}\cdot\textbf{T}'(\textbf{r}_{ij})\cdot\textbf{v}_{j} \exp(i\textbf{q}\cdot(\textbf{r}_{j}-\textbf{r}_{i}))= - \int_{S} d\textbf{r}_{i} \; \tilde{C}(\textbf{r}_{i})  \exp(i\textbf{q}\cdot(\textbf{r}_{j}-\textbf{r}_{i}))  \hat{\textbf{s}}\cdot\textbf{v}_{j},
\label{eq129}
\end{equation}
Here $\tilde{C}(\textbf{r}_{i})$ is $1$ or $0$ depending on whether or not particle $j$ has part of its volume at the point $\textbf{r}_{i}$ in the surface $C$.  Since particle $j$ is in general of finite extent, the statement that particle $j$ has a cross-section that is intersected by $S$ does not imply that the center of mass $\textbf{r}_{j}$ of particle $j$ must lie in $S$.

On the lhs of eq. \ref{eq129} (neglecting points near the walls), the only tensors available to form $\textbf{T}'$ are $I$ and $\textbf{rr}$, so by symmetry only the $\hat{\textbf{q}}$ component of $\textbf{v}_{j}$ can contribute to the integral. Therefore, within the integral on the left hand side one may replace $\textbf{T}'\cdot\textbf{v}_{j}$ with $\textbf{T}'\cdot\hat{\textbf{q}}\hat{\textbf{q}}\cdot\textbf{v}_{j}$.

Also, the fluid volume may be decomposed into an (infinite) series of planes parallel to $S$. A sum over the contents of all these planes includes the entire volume of the container. Integrating eq. \ref{eq129} over all these planes, the rhs of eq. \ref{eq129} gives the spatial Fourier transform $\phi_{j} H(q)$ of the particle volume. Choosing the origin to be at $\textbf{r}_{i}$, $C_{j}$ is implicitly a function of $\textbf{r}_{ij}$ rather than $\textbf{r}_{i}$, so it is useful to define
\begin{equation}
\int_{V} d\textbf{r}_{i}\;\exp(i\textbf{q}\cdot\textbf{r}_{ij}) \tilde{C}(\textbf{r}_{i}) )=V_{j}H(q),
\label{eq130}
\end{equation}
where $H(0)\equiv 1$. Defining $\phi_{j}=V_{j}/V$, and noting that the rhs of eq. \ref{eq130} is independent of all position coordinates, one may write $V_{j}=\int d\textbf{r}_{i}\;V_{j}/V$; here $\textbf{r}_{i}$ is a dummy variable of integration over a constant. Thus, noting $\int_{all planes}dS\;\int_{S}d\textbf{r}_{i}\equiv\int_{V}d\textbf{r}_{i}$ and $\hat{\textbf{s}}\equiv\hat{\textbf{q}}$, one has
\begin{equation}
\int_{V}dr_{i}\;\hat{\textbf{q}}\cdot\textbf{T}'(\textbf{r}_{ij})\cdot\hat{\textbf{q}}\exp(i\textbf{q}\cdot\textbf{r}_{ij)}(\hat{\textbf{q}}\cdot\textbf{v}_{j})
=-V_{j}H(q)(\hat{\textbf{q}}\cdot\textbf{v}_{j}),
\label{eq131}
\end{equation}

Eq. \ref{eq131} is true for arbitrary $\hat{\textbf{q}}\cdot\textbf{v}_{j}$, so
\begin{equation}
\int_{V} d\textbf{r}_{i}\;\left[\hat{\textbf{q}}\cdot\textbf{T}'(\textbf{r}_{ij})\cdot\hat{q}e^{i\textbf{q}\cdot\textbf{r}_{ij}}+\phi_{j}H(q)\right]=0,
\label{eq132}
\end{equation}
completing the desired demonstration. Finally, $\textbf{T} \approx \textbf{T}'$ except at large distances, where $\textbf{T}' \approx \textbf{T} \approx 0$, so eq.\ \ref{eq132} can be subtracted from eq.\ \ref{eq43} to obtain eq.\ \ref{eq50}.

\subsection{\label{section4J}Wavevector Dependence of $D_{m}$}

Equation \ref{eq53} is to be understood as the long-wavelength $(q\rightarrow 0)$ limit of the more general forms of eqs.\ \ref{eq41}-\ref{eq48}. In the earlier equations $I_A$ and $I_D$ have only a simple $q^{2}$ dependence, but all other integrals contributing to $D_{m}$ contain exponential factors $\exp(-i\textbf{q}\cdot\textbf{r})$. The exponentials can give $D_{m}$ a dependence on $|q|$. If $q$ is small enough, $\exp(-i\textbf{q}\cdot\textbf{r})$ is approximately 1 or $1-i\textbf{q}\cdot\textbf{r}$; the $I_{i}$ are then all $\sim q^{2}$, so that $D_{m} \sim q^0$. However, if $q$ is not small (for hard spheres, if $qa\ge0.1$ or so), the exponentials modulate the kernels of the integrals; the $I_{i}$ are no longer simply proportional to $q^{2}$. Equivalently, at large $q$ the mutual diffusion coefficient becomes $q$-dependent.

The wavevector dependence of $D_{m}$ is significant when the scattering length $q^{-1}$ becomes comparable with the effective range $a_e$ of interparticle forces (this length is, roughly speaking, the distance over which $g^{(2)}(r)$ differs appreciably from unity). The third length in the problem, namely the mean distance between nearest-neighbor particles (sometimes erroneously described as the mean interparticle distance), affects the strength of interparticle interactions, but does not directly change the wavevector dependence. Changes in concentration do indirectly change the wavevector dependence by changing $g(r)$. Erroneously?  A typical distance between two particles is not the mean nearest-neighbor distance, it is half the distance across the container. In many systems, $qa_e \ll 1$, so $D_{m}$ is perceptibly independent of $q$.

Pusey et al.\cite{pusey1972} reported the first observation of a $D_{m}$ that is $q$-dependent due to interparticle forces, namely $D_{m}$ of charged R17 virus in nearly pure water. At nearly the same time, Altenberger and Deutch\cite{altenberger1973} demonstrated that $D_{m}$ from the QELSS spectrum of a macromolecule solution contains a concentration- and wavevector-dependent correction term arising from interparticle interactions. These results have been substantially extended. For highly-charged spheres in salt-free water, particles can be sufficiently far apart that hydrodynamic interactions are nearly negligible even though electrostatic forces remain large. Under these conditions the $q$-dependence of $D_{m}$ is predicted to arise primarily from the $g^{(1)}(q,0)^{-1}$ term of eq.\ \ref{eq41}, so that $D_{m}g^{(1)}(q,0)$ should be independent of $q$. This final prediction is confirmed by work of Gruener and Lehman\cite{gruener1979}.

It is sometimes said that a $q$-dependence of the mutual diffusion coefficient implies that the underlying process is not diffusive, by which is meant that Fick's Law $\textbf{J}(\textbf{r},t)=-D_{m}\nabla c(\textbf{r},t)$ must be replaced by
\begin{equation}
\textbf{J}(\textbf{r},t)=-\int d\textbf{R}\;D_{m}(\textbf{r}-\textbf{R})\nabla c(\textbf{R},t).
\label{eq55}
\end{equation}
The non-local diffusion tensor $D_{m}(\textbf{r}-\textbf{R})$ replaces the local diffusion coefficient $D_{m}$. The non-local diffusion tensor has a maximum range $\delta R$ over which it is effective. If one is only concerned with diffusion through distances much larger than $\delta R$, $D_{m}(\textbf{r}-\textbf{R})$ and $D_{m}$ are indistinguishable. Equivalently, if $q^{-1}\gg \delta R$, $D_{m}$ from QELSS will be perceptibly independent of $q$.

Eq.\ \ref{eq41} becomes especially interesting in the limit $\textbf{q}\rightarrow\infty$. In this limit, factors $\exp(-i\textbf{q}\cdot\textbf{r})$ and $\exp(-i\textbf{q}\cdot\textbf{s})$ oscillate rapidly with respect to the remainder of the integrands of eqs.\ \ref{eq42}-\ref{eq48}, so integrals over these exponentials vanish. For the same reason, at large $q$ the distinct terms of $g^{(1)}(q, 0)$ tend to zero. The remainder of eq.\ \ref{eq48} vanishes by symmetry, so
\begin{equation}
\lim_{\textbf{q}\rightarrow\infty}D_{m}=D_{o}\left[1+ c \int d\textbf{r}\;(\hat{\textbf{q}}\cdot\textbf{b}_{12}\cdot\hat{\textbf{q}})g^{(2)}(\textbf{r})+ c^{2} \int d\textbf{r}d\textbf{s}\;(\hat{\textbf{q}}\cdot\textbf{b}_{123}\cdot\hat{\textbf{q}})g^{(3)} (\textbf{r},\textbf{s})\right]+\hat{\textbf{q}}\cdot\Delta D_{m}\cdot\hat{\textbf{q}}.
\label{eq56}
\end{equation}

Our analysis of $\Delta D_{m}$ reveals that $\Delta D_{m}$ arises from a ``caging'' or ``averaging'' effect that becomes significant when the diffusing particles have diffused through distances comparable with the distance over which $g^{(2)}(r)-1$ is non-zero. In the $q \rightarrow \infty$ limit, particle motion is only observed at very short times, because at longer times $S(q,\tau)$ has decayed into the noise in the spectrum. At very short times, the particle has only moved through very short distances. If particles only move over the short distances over which particle motion is observed in the $q\rightarrow\infty$ limit, $\Delta D_{m}$ is not effective at retarding particle motion;  $\lim_{q\rightarrow\infty}\Delta D_{m}\rightarrow 0$.

Eq.\ \ref{eq55} may be written
\begin{equation}
\lim_{\textbf{q}\rightarrow\infty}D_{m}=D_{o}(1+k_{1\infty}\phi+k_{2\infty}\phi^{2}).
\label{eq57}
\end{equation}
$\textbf{b}_{123}$ is a three-particle term, which first contributes to $D_{m}$ at the $\phi^{2}$ level. The $(a/r)^7$ approximations to the $\textbf{b}$ give $k_{1\infty}=-1.734$ and $k_{2\infty}=0.873$. Batchelor's\cite{batchelor1976} treatment of these tensors gets $k_{1\infty}=-1.83$.

\subsection{\label{section4K}Self Diffusion Coefficient $D_{s}$ and Probe Diffusion Coefficient $D_{p}$}

$D_{s}$ and $D_{p}$ both measure the diffusion of a single particle through a uniform solution.  They differ in that the self-diffusion coefficient refers to the diffusion of a single particle through a solution of other particles of the same species, while the probe diffusion coefficient refers to the diffusion of a given particle through a solution of particles of some other species. $D_{s}$ and $D_{p}$ are variously measured using FPR, PGSE NMR, FCS, RICS, QELSS, and macroscopic tracer diffusion techniques, with some techniques only measuring $D_{s}$ and others only measuring $D_{p}$. For these experiments, and taking $\sigma_{i}^{2} = 1$ and thus $g_{1s}(q,0)=N$  for mathematical clarity,
\begin{equation}
D_{s}\equiv - \lim_{\tau\rightarrow 0+}\frac{\partial}{\partial \tau}\ln(g^{(1s)}(\tau))= \left\langle -\frac{1}{Nq^{2}}\sum_{i=1}^{N}(-i\textbf{q}\cdot\textbf{v}_{i}(\tau)-
\textbf{q}^{2}:\int_{t}^{t+\tau}ds\;\textbf{v}_{i}(s)\textbf{v}_{i}(t+\tau)+...)\right\rangle.
\label{eq60}
\end{equation}
Replacing $\textbf{v}_{i}(t)$ with its Brownian and direct components, and applying eqs.\ \ref{eq18}-\ref{eq29},
\begin{equation}
D_{s}=D_{o}\hat{\textbf{q}}\cdot\left(\textbf{I}+\frac{1}{N}\sum_{i,l=1,l\neq i}^{N}\langle\textbf{b}_{il}\rangle+\frac{1}{N}\sum_{i,l,m=1;l\neq m\neq i}^{N} \langle\textbf{b}_{ilm}\rangle\right)\cdot\hat{\textbf{q}}+\hat{\textbf{q}}\cdot\Delta D_{s}\cdot\hat{\textbf{q}}
\label{eq61}
\end{equation}
where
\begin{equation}
\Delta D_{s}=N^{-1}\left\langle\sum_{i=1}^{N} \int_{0}^t ds\;\left(\textbf{v}_{Bi}(s)\textbf{v}_{Di}(t)+\textbf{v}_{Di}(s)\textbf{v}_{Bi}(t)\right)\right\rangle.
\label{eq62}
\end{equation}
As $q\rightarrow\infty$, $\Delta D_{s}$ vanishes, because particles have not yet moved a distance comparable to the range of $g^{(2)}(r)$. Eqs.\ \ref{eq61} and \ref{eq56} then reveal that in the high-$q$ limit the concentration dependences of $D_{s}$ and $D_{m}$ are the same. Furthermore, to $\mathcal{O}(q^{2})$, $\Delta D_{m}$ includes only the self  $(i=j)$ terms of $\Delta D_{ij}^m$, which are the same as the individual terms of $\Delta D_{s}$ (eq.\ \ref{eq62}), so to $\mathcal{O}(q^{2})$ the dynamic friction contributions to $D_{s}$ and $D_{m}$ are equal.

Optical probe experiments are based on ternary solvent: matrix: probe systems. In applying QELSS to such a system, in many cases one works in the limit that the probe species, while dilute, completely dominates scattering, while the matrix species scatters negligible amounts of light, even if it is concentrated. In other cases, measurement of the QELSS spectrum of the probe:matrix solution and of the probe-free matrix solution, followed by subtraction of the latter from the former at the field correlation function level, removes the matrix scattering from the scattering by the mixture, permitting successful isolation of the probe spectrum\cite{streletzky1995a}.

The above formalism can evaluate $D_{p}$ if some minor adaptations are made. Namely, in eq.\ \ref{eq2}, for a probe experiment the scattering lengths $\sigma_{i}$ assumes two values: $\sigma_{i}=1$ for probe molecules, and $\sigma_{i}\approx 0$ for matrix molecules. Sums over particles may include either matrix or probe molecules. The $N_{M}$ matrix molecules are labelled $\{1,2,...,N_{M}\}$, while the $N_p$ probe molecules are labelled $N_{M} + 1,N_{M} + 2,...,N_{M} + N_p$, the total number of molecules being $N_{T} = N_{M}+N_p$. The matrix and probe concentrations are $c_{M}=N_{M} / V$ and $c_p=N_p/V$, respectively; for dilute-probe results only terms of lowest order in $N_p$ or $c_p$ are retained. Since the matrix and probe species are not necessarily the same, the hydrodynamic interaction tensors $\textbf{b}$ and $\textbf{T}$ and the radial distribution functions $g^{(n,i)}$ may have different values for probe-probe, probe-matrix, and matrix-matrix molecular pairs.

Under these conditions, eq.\ \ref{eq37} gives us the probe diffusion coefficient
\begin{equation}
D_{p}=\frac{1}{q^{2}g^{(1)}(q,0)}\left(\langle-\sum_{i,j=1}^{N_p}\sum_{l=1}^{N_{T}}\sigma_{i}\sigma_{j}\exp(i\textbf{q}\cdot\textbf{r}_{ij})i\textbf{q}
\nabla_{l}:[D_{jl}]+\sum_{i,j=1}^{N_p}\sigma_{i}\sigma_{j}\textbf{q}\cdot D_{ij}\cdot\textbf{q}\exp[i\textbf{q}\cdot\textbf{r}_{ij}]\rangle\right. \notag
\end{equation}
\begin{equation}
\left.+\sum_{i,j=1}^{N_p}\sigma_{i}\sigma_{j}\textbf{q}\cdot\Delta D_{ij}^c\cdot\textbf{q}\right).
\label{eq63}
\end{equation}
Terms with $i,j>N_p$ vanish because $\sigma_{i},\sigma_{j}\approx 0$ for matrix molecules. While matrix molecules are optically inert, they are still hydrodynamically active, so neither $\textbf{b}_{il}$ nor $\textbf{T}_{il}$ vanishes if $l$ refers to a matrix molecule.

The significance of eq.\ \ref{eq63} becomes clearer if $\sum_{i,j}$ is resolved into self and distinct components, and limited to terms referring to no more than two particles, viz:
\begin{equation}
D_{p}=\frac{-D_{o}}{q^{2}g^{(1)}(q,0)}\left\{\sum_{i=1}^{N_p}\sigma_{i}^{2}\left(
-\langle\textbf{q}\cdot\textbf{I}\cdot\textbf{q}-\sum_{l=1}^{N_{T}}(\textbf{q}
\cdot\textbf{b}_{il}\cdot\textbf{q})+i\textbf{q}\nabla_{i}:[\textbf{I}+
\sum_{m=1}^{N_{T}}\textbf{b}_{mi}]\right.\right.\notag
\end{equation}
\begin{equation}
\left.\left.+i\textbf{q}\sum_{l=1}^{N_{T}}\nabla_{l}:[\textbf{T}_{il}]\rangle-\textbf{q}
\cdot\frac{\Delta D_{s}}{D_{o}}\cdot\textbf{q}\right)\right.+\notag
\end{equation}
\begin{equation}
\left.\sum_{i\neq j=1}^{N_p}\sigma_{i}\sigma_{j}\left[\langle\exp(i\textbf{q}\cdot\textbf{r}_{ij})
(-\textbf{q}\cdot\textbf{T}_{ji}\cdot\textbf{q}+i\textbf{q}\nabla_{i}:
[\textbf{T}_{ji}]+i\textbf{q}\nabla_{j}:[\textbf{I}])\rangle-\textbf{q}\cdot\frac{\Delta D_{ij}^c}{D_{o}}\cdot\textbf{q}\right]\right\}.
\label{eq64}
\end{equation}
In this equation, $D_{o}$ is the single-particle diffusion coefficient. Terms in $\sigma_{i}^{2}i\textbf{q}\nabla$ vanish by symmetry. Terms in $\sigma_{i}\sigma_{j}\exp(i\textbf{q}\cdot\textbf{r}_{ij})$ are only non-zero if two probe particles are close enough to interact (close enough that their $g^{(2,0)}\neq 1$). If the probes are dilute, this event rarely happens, so the $\sigma_{i}\sigma_{j}$ terms are negligible with respect to the $\sigma_{i}^{2}$ terms. With these reductions, $D_{p}$ becomes
\begin{equation}
D_{p}=\frac{-D_{o}}{q^{2}g^{(1)}(q,0)}\left[\sum_{i=1}^{N_p}\sigma_{i}^{2}\left[-\textbf{q}\cdot\textbf{I}\cdot\textbf{q}-\sum_{l=1}^{N_{T}}\textbf{q}\cdot\langle\textbf{b}_{il}\rangle\cdot\textbf{q}-\textbf{q}\cdot\frac{\Delta D_{s}}{D_{o}}\cdot\textbf{q}\right]\right].
\label{eq65}
\end{equation}
Through $\mathcal{O}(c^1)$, the concentration effects on $D_{p}$, $D_{s}$, and $D_{m}(q\rightarrow\infty)$ are now seen to be identical.

\section{Other Approaches}

\subsection{Coupling of Concentration and Energy-Density Fluctuations}
A pure simple liquid such as water has an intrinsic spectrum arising from scattering of light by propagating pressure fluctuations and non-propagating energy-density fluctuations\cite{berne1976a}. Scattering from sound waves creates Brillouin peaks, whose centers are displaced from the incident light frequency. Scattering from energy fluctuations (sometimes described as ``entropy'' or ``temperature'' fluctuations\cite{mandelbrot1989}) leads to a central ``Rayleigh'' line with width proportional to the thermal conductivity $\Xi$. If a solute is added, the spectrum of the pure fluid gains a further line arising from concentration fluctuations; this mass-diffusive line is the spectrum treated above. It is almost always true that scattering by the pure liquid is much weaker than scattering by concentration fluctuations, so that concentration scattering dominates the spectrum.

In general, concentration and energy fluctuations are coupled, as by the Soret effect (a temperature gradient driving a mass flux) and the Dufour effect (a concentration gradient driving heat flow). Correlations in pressure, energy-density, and concentration fluctuations are treated by Mountain and Deutch, and by Phillies and Kivelson\cite{mountain1969,phillies1979z}. Cross-coupling of energy and concentration diffusion modifies the relaxation rate of heat- and mass- diffusive modes, and creates an additional spectral line (of integrated intensity zero) due to static cross-correlations between energy and concentration fluctuations. If $\Xi q^{2}\gg D_{m}q^{2}$, and if solvent scattering is weak (conditions readily satisfied by macromolecule solutions), the spectrum decouples, so that the relaxation of concentration fluctuations is determined entirely by $D_{m}$.

\subsection{Smoluchowski and Mori-Zwanzig Formalisms}

The purpose of this short section is to put in one place discussions on the Smoluchowski diffusion-sedimentation equation and outcomes from the Mori-Zwanzig formalism.  The Smoluchowski equation
\begin{equation}
\frac{\partial \rho(\textbf{r}^{N},t)}{\partial t}=\sum_{i,j=1}^{N}\frac{\partial}{\partial \textbf{r}_{i}}D_{ij}^{Sm}(\frac{\partial}{\partial\textbf{r}_{j}}-\beta\textbf{F}_{j})\rho(\textbf{r}^{N},t),
\label{eq68a}
\end{equation}
connects the time dependence of the $N$-particle probability distribution function $\rho(\textbf{r}^{N},t)$ to a set of external forces $\textbf{F}_{j}$ on the particles and to a diffusion coefficient $D_{ij}^{Sm}$.  The diffusion coefficient is usually taken to be the hydrodynamic diffusion coefficient $D_{ij} = k_{B} T \mu_{ij}$,  The diffusion-sedimentation equation was obtained for systems in which the forces are imposed externally, so that there are no correlations between the forces and the Brownian displacements. In this case the dynamic friction term vanishes, an outcome that is known experimentally\cite{mazo1965,stigter1955} to be incorrect for diffusing systems.  In order apply the Smoluchowski equation to a diffusing system. for starters one needs to reinterpret $D_{ij}^{Sm}$ to included dynamic friction terms. Such an approach is not considered further here.

In the Mori-Zwanzig formalism, transport coefficients are taken to arise from force-force correlation functions (memory kernels) that describe the temporal evolution of the projected force. In conventional Mori-Zwanzig calculations, one sets up the transport equations, but at the key point the memory kernels for the Mori-Zwanzig projected forces are simply identified as the appropriate transport coefficients.  The Mori-Zwanzig formalism yields a transport coefficient, but the meaning of that coefficient is defined by the Mori-Zwanzig equation.  The transport coefficient's value is obtained by fitting experimental measurements to solutions to the Mori-Zwanzig solutions.  The symbols used for Mor-Zwanzig transport coefficients are the same as the symbols used in classical measurements, but that similarity may be shallow.

One can simply say that the Mori memory kernel corresponding to mutual diffusion is ${\cal D}_{ij}$ for some ${\cal D}_{ij}$, but having done so one cannot also claim that ${\cal D}_{ij}$ is $k_{B} T \mu_{ij}$ with $\mu_{ij}$ being the hydrodynamic $\mu$. That would be overdefining ${\cal D}_{ij}$.  Any expression other than $k_{B} T \mu_{ij}$ that has the correct dimensions and $q$-dependence is also consistent with the unevaluated Mori Kernel. In particular, the rhs of eq.\ \ref{eq33B}, including the dynamic friction terms of eq.\ \ref{eq36} and the terms in $\nabla \mu_{ij}$, is the correct evaluation for the lead terms in $q$ of the Mori kernel.

Alternatives to simple identification of particular memory kernels as particular transport coefficients do exist.  A Mori kernel has been extracted from computer simulations on a simple system\cite{phillies1995s}.  The memory kernel for the projected forces, including the use of the projected time evolution operator, has been evaluated analytically by Nasto in a special case\cite{nasto2007a}. Direct evaluations of the Mori memory kernel to determine hydrodynamic transport coefficients corresponding to $D_{ij}^{Sm}$ have not been made.

Finally, we note the notion of using the Mori formalism to generate solutions of the Smoluchowski equation. This approach is not justifiable in terms of derivations of the Mori formalism.  The Smoluchowski equation has a diffusion term, driving concentration fluctuations to decay toward zero, but unlike the Langevin equation has no random force term . As a results, solutions of the Smoluchowski equation are not stationary in time; they show, e.g., $a_{q}(t)$ decaying to zero. In contrast, derivations of the Mori equation explicitly or implicitly (for example, by assuming that certain Laplace transforms exist) assume that the Mori equation solutions are stationary in time. In consequence, one can formally insert the Smoluchowski equation into the Mori formalism, but the meaning of the outcome of such a process is unclear.

\section{Discussion}

\subsection{\label{section5a}Implications for QELSS Measurements}

The diffusion coefficients measured by QELSS, FPR, FCS, RICS, PGSE NMR, inelastic neutron scattering, and other techniques all depend on the solute concentration. The treatment above shows how these dependences may be calculated. The Stokes-Einstein equation
\begin{equation}
    D=\frac{k_{B}T}{6\pi\eta a}.
\label{eq58}
\end{equation}
is sometimes used to interpret $D_{m}$ from QELSS measurements. For dilute neutral spheres, radius $a$, diffusing through a simple solvent having a modest viscosity $\eta$ not too much larger than the viscosity of water, this equation can be appropriate. The equation is assuredly invalid for macromolecules diffusing at elevated concentrations. Equivalently, $D_{m}$, $\eta$, $T$, and eq.\ \ref{eq58} cannot be combined to calculate the hydrodynamic radius of non-dilute diffusing spheres. $D_{m}$ in non-dilute solutions is nearly certainly not independent of solute concentration\cite{phillies1976}, though over limited ranges of $c$ the dependence may be small.

We considered in detail the mutual and self diffusion coefficients of neutral hard spheres.  The lead term in the dependence of $D_{m}$ on solute volume fraction $\phi$ (eq.\ \ref{eq53}) is small.  There are large hydrodynamic and thermodynamic contributions to  $dD_{m}/dc$, but for neutral hard spheres at concentrations that are not too large these contributions almost cancel. Changing the intermacromolecular interactions, for example by charging the spheres,  will change the hydrodynamic and thermodynamic contributions to $dD_{m}/dc$. There is no reason to expect the changes in the different contributions to cancel. For systems that are not neutral hard spheres $D_{m}$ may depend appreciable on concentration. Indeed, the dependence found for charged macromolecules can be much larger than the dependence for neutral spheres\cite{doherty1974a,biresaw1985}.

Particle sizes in concentrated solution have sometimes been estimated from QELSS data.  The particle shapes and interactions must be known. The inferred radii depend on the detailed theoretical model in use.\cite{biresaw1985,corti1981}. Measurements of probe diffusion by polystyrene latex spheres of different sizes diffusing through micelles, followed by interpretation applying the above hydrodynamic models, have allowed determination of the (substantial) water content of Triton X-100 and other micelles.\cite{streletzky1995a}

van Megen and Underwood\cite{van1990a} report a light scattering method  that can determine the distinct part of the field correlation function $g^{(1d)}(q,t)$.  Their approach is based on continuous variation in the solvent index of refraction, coupled with the use of two colloidal species having the same size and surface properties, but different cores and hence indices of refraction.  The same physical approach is found in inelastic neutron scattering, using deuterated and hydrogenated macromolecular species and a series of mixed deuterated and hydrogenated solvents.

\subsection{Comparison of $D_{m}$ and $D_{s}$ with Experiment}

This section treats experimental data suitable for testing the aforementioned theoretical models. There is also an extensive literature\cite{bartlett1990} on diffusion, crystallization, and glass formation in concentrated sphere suspensions. However, the concentration regime in which spheres vitrify, and the concentration regime in which the above calculations are likely to be valid without extension to higher order in $\phi$, are not overlapping.

The best study of $D_{m}$ at low $q$ in a hard-sphere system appears to be that of Mos et al. \cite{mos1985}, who used homodyne coincidence spectroscopy (HCS) to measure $D_{m}$. Homodyne coincidence spectroscopy \cite{phillies1981-3} is a two-detector QELSS experiment, in which a sample is simultaneously illuminated with two incident laser beams, the scattered light is collected by two detectors placed on opposite sides of the scattering volume, and the intensity-intensity time cross-correlation function is measured. As was first shown by this author\cite{phillies1981-4}, the homodyne coincidence spectrometer differs from a conventional one-beam one-detector QELSS instrument in that an HCS spectrometer is substantially immune to multiple scattering artifacts. While multiply-scattered light reaches each HCS detector, only the single-scattered light has cross-correlations from one detector to the other, so the only time-dependent component of an HCS spectrum is that due to single scattering.

Mos et al.\cite{mos1985} studied colloidal silica spheres, prepared by the Stoeber process\cite{stoeber1968}, stearylated\cite{iler1979}, and suspended with sonication in xylene and toluene at concentrations of 0 - 180 $g/L$. Over this density range, $D_{m}$ exhibits a linear decline with increasing $c$, i.e.,
\begin{equation}
D_{m}=D_{o}(1+\alpha\phi),
\label{eq69}
\end{equation}
Using the estimated density of $1.75$g/cm$^3$, Mos et al.'s data for xylene solutions shows $\alpha=-0.86$; in toluene solutions, $\alpha=-1.2$ was found. Here $D_{o}$ is the zero-concentration limit of $D_{m}$ and $\phi$ is the sphere concentration in volume fraction units. The total change in $D_{m}$ over the observed concentration range is ca.\ 10\% of $D_{o}$. Inferring from the data on $D_{m}$ and $r_{H}$ a 3-5\% uncertainty in Mos's individual measurements of $D_{m}$, the uncertainty in $\alpha$ must be roughly $\pm0.1$. The theoretical $\alpha$ of eq.\ \ref{eq53} is -0.9, in excellent agreement with these experiments.

The largest uncertainty in the experiment is the identification of stearylated silica particles, $r_{H}\approx$ 370\AA, with a C$_{18}$ chain coating, as hard spheres. If the particles had weak attractive or repulsive interactions, in addition to their hard-sphere interaction, the concentration dependence of $D_{m}$ would be expected to be different. An attractive interaction usually makes $\alpha$ more negative, while a repulsive interaction usually makes $\alpha$ more positive. Thus, if one wished to believe the Batchelor-Felderhof solution\cite{batchelor1976,felderhof1977} of the diffusion problem (which leads to $\alpha = +1.56$ for hard spheres), one could claim that stearylated silica particles attract each other weakly, thereby reducing their $\alpha$ from the Batchelor-Felderhof value to the value found experimentally. Mos, et al., in fact make this interpretation for their spheres. Caution is needed to avoid circular arguments. Mos, et al.,\cite{mos1985} began with the assumption that the Batchelor-Felderhof calculation is correct, and therefore sensibly inferred from their observed negative $\alpha$ that their spheres must attract each other weakly.  However, they do not in their paper adduce other evidence that the spheres had an attractive potential, in addition to the hard sphere potential. If eq.\ \ref{eq53} were correct, the $\alpha$ found for stearylated silica spheres would have the value appropriate for hard spheres, consistent with the physical expectation that these uncharged spherical objects that form stable solutions in non-polar solvents should be close to true hard spheres.

Kops-Werkhoven et al. \cite{kops-werkhoven1981} used QELSS on stearylated silica spheres in cyclohexane to determine $D_{m}$ at low $q$, finding $\alpha = +1.56$, in agreement with the Batchelor-Felderhof theory\cite{batchelor1976,felderhof1977}. The spheres used by Kops-Werkhoven et al.\cite{kops-werkhoven1981} and by Mos et al.\cite{mos1985} are chemically the same, so in first approximation both groups either did or did not study hard spheres. The sphere-sphere interactions could differ slightly, because the two studies used different organic solvents. Kops-Werkhoven et al. also employed a contrast-matching technique to estimate $\alpha_{s}$ in
\begin{equation}
D_{s}=D_{o}(1+\alpha_{s}\phi+...),
\label{eq70}
\end{equation}
reporting $\alpha_{s} = -2.7\pm0.5$, which is not in good agreement with the theoretical (hydrodynamic) estimate $\alpha_{s} \approx -1.83$. The agreement of the experimental $\alpha_{s}$ might improve if dynamic friction at some level were included in the theoretical estimate for $\alpha_{s}$.

For the stearylated silica spheres studied by both groups, Mos et al.\cite{mos1985} demonstrated in a non-polar solvent having an only slightly less favorable index-of-refraction match than the solvent used by Kops-Werkhoven et al.\cite{kops-werkhoven1981} ($\delta n\approx 0.05$ in toluene, vs. $\delta n \approx 0.015$ in cyclohexane, based on cycloheptane being an index-matching fluid for these spheres) that artifactual positive values of $\alpha_{m}$ can arise from multiple scattering. Mos et al.\cite{mos1985} further showed that multiple scattering is important in sphere:toluene mixtures. Mos et al.'s measurements on silica:nonpolar solvent mixtures, which employed HCS (a technique that is immune to multiple scattering artifacts), are therefore preferable to ref.\ \cite{kops-werkhoven1981}'s QELSS measurements. The difference in $\alpha_{m}$ between the two sets of measurements is small but significant, implying weak multiple scattering.

Qiu, et al.,\cite{qiu1990} report $D_{m}$ of polystyrene latices in water at elevated concentrations (up to $\phi\approx0.45$), using diffusing wave spectroscopy (DWS) to study highly turbid solutions. DWS is a multiple-scattering technique that measures $D_{m}$ at very large effective $q$. However, the underlying theory for DWS assumes that the displacement probability distribution $P(x,\tau)$ for the moving particles is a Gaussian, which need not be the case\cite{phillies2015a}, so some caution is needed in interpreting these results. The lattices in the study had $r_{H}$ of 0.206 $\mu$ and 0.456$\mu$; DWS determined $D_{m}$ for particle motions $\leq 50$\AA. The latex particles are here diffusing distances $\ll r_{H}$, so in this experiment $t\ll\tau_{d}$ and the dynamic friction term $\Delta D_{s}\approx 0$. Qiu et al.\cite{qiu1990} report $D_{m}=D_{o}(1-(1.86\pm0.07)\phi)$, i.e. $k_{1s}\approx-1.86\pm0.07$, in good agreement with Batchelor's \cite{batchelor1976} theoretical estimate $k_{1\infty}\approx-1.83$ and with our estimate $k_{1\infty}\approx-1.734$. In contrast, in Kops-Werkhofen's determinations\cite{kops-werkhoven1981} of $\alpha_{s}$, particles diffused distances $\approx r_{H}$, so in refs.\ \cite{kops-werkhoven1981} $\Delta D_{s}$ could have been substantial, perhaps explaining why Kops-Werkhofen, et al., found a more negative value for $\alpha_{s}$ than did Qiu, et al.

The microscopic and continuum models have been extended by Borsali and collaborators\cite{benmouna1987} to systems containing a solvent and two physically distinct random-coil polymer species. Benmouna, Borsali, and collaborators\cite{benmouna1987} compute spectra of ternary systems in which one or both components is concentrated and in which one or both components scatter light. Their polymer models are somewhat remote from the emphases of this review, but the agreement between their predictions and their experiments is excellent, supporting the belief that the above results are fundamentally correct.

\appendix

\section{Other Methods for Calculating $D_{m}$}
The purpose of this Appendix is to compare possible methods for computing $D_{m}$. Relevant methods include those proposed by Batchelor\cite{batchelor1976}, Felderhof\cite{felderhof1978}, and this author\cite{carter1985,phillies1984}. These calculations are all correct, but do not all correspond to the same experiment or physical quantity. While I have previously shown\cite{phillies1987} why refs.\ \onlinecite{batchelor1976} and \onlinecite{felderhof1978} do not find the concentration dependence of $D_{m}$ as measured by light scattering, this demonstration has not been widely understood.  For example there has sometimes been a misapprehension that refs.\ \onlinecite{batchelor1976}, \onlinecite{felderhof1978}, \onlinecite{carter1985} and \onlinecite{phillies1984} only disagree about dynamic friction effects. In fact these works disagree as to the correct form for the concentration dependence of $D_{m}$, even if dynamic friction is neglected. It will here be shown that each reference gives an algebraically correct calculation of some diffusion coefficient, but that only refs.\ \onlinecite{phillies1984} and \onlinecite{carter1985}  calculate the mutual diffusion coefficient $D_{m}$ that is measured by QELSS.

Experimentally, light scattering spectroscopy is directly sensitive to collective coordinates determined by particle positions. Specifically, the instantaneous scattered field is proportional to the instantaneous value of the $k^{th}$ spatial Fourier component $\epsilon_{k}(t)$ of the local index of refraction, where $\epsilon_{k}(t)$ is in turn proportional to the $k^{th}$ spatial Fourier component $a_{k}(t)$ of the concentration of scattering particles. Here $k$ is the scattering vector selected by the source and detector positions. In light scattering spectroscopy, $D_{m}$ is obtained from the temporal evolution of $a_{k}(t)$, namely
\begin{equation}
D_{m}=\frac{\lim_{t\rightarrow 0}\left(\frac{d}{dt}\right)\langle a_{-k}(0)a_{k}(t)\rangle}{-k^{2}\langle a_{k}(0)a_{k}(0)\rangle}.
\label{eq133}
\end{equation}
In a real experiment, eq.\ \ref{eq133} is typically applied via a cumulant expansion of $S(k, t)$.

In contrast to eq.\ \ref{eq133}, Batchelor\cite{batchelor1976} obtains $D_{m}$ by evaluating the flux of particles due to an applied steady (thermodynamic) force, the flux being related to the diffusion coefficient by
\begin{equation}
\textbf{J}=-D_{m}\nabla c.
\label{eq134}
\end{equation}
In ref.\ \cite{batchelor1976} $D_{m}$ was obtained by generalizing the Einstein expression
\begin{equation}
D=k_{B}T(\textbf{b})
\label{eq135}
\end{equation}
for $D_{m}$, $\textbf{b}$ being the mobility tensor. In Einstein's original analysis, $\textbf{b}$ was a constant. In Batchelor's generalization of Einstein's analysis, $\textbf{b}$ is given by a microscopic expression which depends on the relative positions of the particles in the system. $D_{m}$ is a macroscopic quantity which does not depend explicitly on particle positions. To obtain a macroscopic $D_{m}$ from a microscopic $\textbf{b}$, Batchelor took an ensemble average of $\textbf{b}$ over possible particle configurations. Batchelor's generalization of the Einstein form for $D_{m}$ was therefore
\begin{equation}
D_{m}=k_{B}T\langle\textbf{b}\rangle.
\label{eq136}
\end{equation}

While the calculation of $D_{m}$ as $k_{B}T\langle \textbf{b} \rangle$ is mathematically correct, QELSS does not determine $D_{m}$ by measuring a flux and a concentration gradient, and taking a ratio. Instead, $D_{m}$ from a QELSS measurement is obtained from the first cumulant in an expansion of $g^{(1)}(q,t)$, i.\ e., from the time rate of change of a concentration. While eq.\ \ref{eq136} might happen to give the value of $D_{m}$ that is obtained by light scattering, eqs.\ \ref{eq133} and \ref{eq134} are not the same. Any disagreement between results obtained from these equations is properly resolved in favor of the equation that correctly models the quantity that is measured experimentally by QELSS, this being eq.\ \ref{eq133}.

References \cite{carter1985} and \cite{felderhof1978}  both made assumptions equivalent to assuming that particle motions are correctly described by the Smoluchowski sedimentation equation
\begin{equation}
\frac{dc(r,t)}{dt}=\textbf{S}c(r,t),
\label{eq137}
\end{equation}
where the Smoluchowski operator is
\begin{equation}
\textbf{S}=\nabla\cdot(\textbf{D}\cdot\nabla+\textbf{D}\cdot\textbf{F}/k_{B}T).
\label{eq138}
\end{equation}
Here $\textbf{D}$ is the diffusion tensor (as distinct from the inferred diffusion coefficient $D_{m}$) and $\textbf{F}$ is the applied force on a particle. The dynamic friction effects discussed above, which are not at issue here, may be incorporated in $\textbf{D}$, so that eq.\ \ref{eq137} is formally valid no matter whether or not dynamic friction effects are present. Equation \ref{eq138} manifestly includes terms in $\nabla\cdot[\textbf{D}]$, which are non-vanishing if $\textbf{D}$ depends on position, or if $\textbf{D}$ depends on concentration and $c$ depends on position.

Just as eq.\ \ref{eq135} was applied to the problem by interpreting $\textbf{b}$ as a microscopic mobility tensor, so also eqs.\ \ref{eq137} and \ref{eq138} were applied in refs.\ \cite{carter1985} and \cite{felderhof1978}  by giving $\textbf{D}$, $\textbf{F}$, and $c(r,t)$ a microscopic interpretation. In particular, $c(r,t)$ was replaced with $a_{k}(t)$ and its microscopic representation in terms of particle positions, namely
\begin{equation}
a_{k}(t)=\sum_{j=1}^{N}\exp(i\textbf{k}\cdot\textbf{r}_{j}),
\label{eq139}
\end{equation}
the sum being over all $N$ particles in the system. $D_{m}$ is then obtained from some ensemble average over the short-time limit of $S$.

How do the various applications of eq.\ \ref{eq138} differ from each other? Felderhof's calculation \cite{felderhof1977} of $D_{m}$ sets the Smoluchowski equation in the form
\begin{equation}
\frac{\partial a(\textbf{r}_{1},t)}{\partial t}=D_{o}bf\nabla_{1}\cdot\left[\nabla_{1}a(\textbf{r}_{1},t)+c\beta\int\nabla_{1}Vg^{(2)}a(\textbf{r}_2,t)d\textbf{r}_2\right]\notag
\end{equation}
\begin{equation}
+c\nabla_{1}\cdot\int\textbf{b}g^{(2)}d\textbf{r}_2\cdot\nabla_{1}a(\textbf{r}_{1},t)+c_o\nabla_{1}\cdot\int\textbf{T}\cdot g^{(2)}\nabla_2a(\textbf{r}_2,t)d\textbf{r}_2+\mathcal{O}(c^{2}).
\label{eq140}
\end{equation}
where $a(\textbf{r}_{1},t)$ is the local density at $\textbf{r}_{1}$, $V$ is the interparticle potential, $g^{(2)}=g^{(2)}(\textbf{r}_2-\textbf{r}_{1})$ is the equilibrium pair distribution function, and $\nabla_{i}$ is the gradient with respect to particle $i$, the particles being the particle of interest $1$ and a neighboring particle $2$. As interacting particles must be relatively close together, the concentration gradients at their locations should be roughly equal, i.\ e.
\begin{equation}
\nabla_{1}a(\textbf{r}_{1},t)=\nabla_2a(\textbf{r}_2,t)
\label{eq141}
\end{equation}
for a pair of interacting particles. (This is a small-k approximation.) Rearrangement of eq.\ \ref{eq140} gives
\begin{equation}
\frac{da(\textbf{r}_{1},t)}{dt}=D_{o}\nabla_{1}\cdot\left[\nabla_{1}a(\textbf{r}_{1},t)+c\beta\int d\textbf{r}_2\nabla_{1}V(\textbf{r}_{12})g^{(2)}(\textbf{r}_{12})a(\textbf{r}_2,t)\right]\notag
\end{equation}
\begin{equation}
+c\int(\textbf{b}+\textbf{T}):\nabla_{1}^{2}a(\textbf{r}_{1},t)g^{(2)}(\textbf{r}_{12})d\textbf{r}_2
\label{eq142}
\end{equation}
as evaluated in ref.\ \cite{felderhof1978}, plus terms such as
\begin{equation}
c\int\nabla_{1}\cdot[\textbf{T}]g^{(2)}(\textbf{r}_{12})\cdot\nabla_2a(\textbf{r}_2,t)d\textbf{r}_2
\label{eq143}
\end{equation}
which vanish because $\int\nabla\cdot\textbf{T}d\textbf{r}$ is odd in $\textbf{r}$. Reference \cite{felderhof1978} thus finds that terms in $\nabla\cdot(\textbf{T}+\textbf{b})$ do not contribute to $D_{m}$.

References \cite{felderhof1978}, \cite{carter1985}, and \cite{phillies1984} (that is, equations \ref{eq142} and \ref{eq34}) disagree because they obtain $D_{m}$ by taking ensemble averages over different functions. In eq.\ \ref{eq34}, $D$ was obtained from an average having the general form
\begin{equation}
\langle a_{-k}(0)\textbf{S}'a_{k}(t)\rangle
\label{eq144}
\end{equation}
in which $\textbf{S}'$ is a time evolution operator related to $\textbf{S}$, while the $a_{k}$ are statistical weights, appearing because the contribution of a particular particle configuration to the measured $D_{m}$ is weighted by the light scattering intensity contributed by that particle configuration.

In contrast to eq.\ \ref{eq144}, eqs.\ \ref{eq142} and \ref{eq143} obtain $D_{m}$ from an ensemble average over the algebraic kernel of $\textbf{S}$, with no factors of $a_{k}$ included within the average. The kernel of the microscopic $\textbf{S}$ does give the concentration changes to be expected from a given microscopic particle configuration, so $\langle\textbf{S}\rangle$ does give an average rate of change for $a_{k}(t)$; also, $\langle\textbf{S}\rangle$ has no divergence ($\nabla\cdot\textbf{T}$) terms. However, QELSS obtains the \emph{light-scattering-intensity-weighted} average (the $z$-average) temporal evolution of $a_{k}(t)$, not the unweighted average. States which scatter no light make no contribution to the observed temporal evolution of $a_{k}(t)$. The $z$-weighting in eq.\ \ref{eq144} arises from factors $a_{-k}(0)$ and $a_{k}(t)$. To obtain a properly $z$-weighted average, factors of $a_{k}$ must be included in the ensemble average. Including these factors replaces, e.~g., eq.\ \ref{eq143} by
\begin{equation}
c\int d\textbf{r}_2\nabla_{1}\cdot(e^{i\textbf{k}\cdot\textbf{r}_{12}})\nabla\cdot[\textbf{T}],
\label{eq145}
\end{equation}
a term previously seen in eq.\ \ref{eq34}. Unlike the term eq.\ \ref{eq143}, the term $\hat{\textbf{k}}\nabla\cdot\textbf{T}\exp(i\textbf{k}\cdot\textbf{r})$ of eq.\ \ref{eq145} includes parts that are not odd in $\textbf{r}$ and do not vanish on performing the integral. QELSS measures $\langle a(0)da(t)/dt\rangle$, not $J$, so the appropriate microscopic average for $D_{m}$ is that of eq.\ \ref{eq34}, not the form of eqs.\ \ref{eq135} or \ref{eq142}, but \ref{eq135} or \ref{eq142} do represent diffusion coefficients which some experiment measures.

\section{A Partial Bibliography--Theory of Particle Diffusion}

The following is an incomplete bibliography of pre-1990 theoretical papers on the diffusion of mesoscopic particles (colloids, proteins, micelles) at elevated concentrations. I have added a few more recent papers. I make no claim of completeness, though I have tried to be sure that most major lines of work are represented. Additions and corrections are welcome.

\begin{enumerate}
\item Ackerson, Bruce J.\ ``Correlations for Interacting Brownian Particles. II'', \emph{J.\ Chem.\ Phys.} 69, 684-690 (1978).
\item Adelman, S.\ A.\ ``Hydrodynamic Screening and Viscous Drag at Finite Concentration'', \emph{J.\ Chem.\ Phys.} 68, 49-59 (1978).
\item Adler, R.\ S.\ and Freed, K.\ F.\ ``On Dynamic Scaling Theories of Polymer Solutions at Nonzero Concentrations'', \emph{J.\ Chem.\ Phys.} 72, 4186-4193 (1980).
\item Allison, S.\ A., Chang, E.\ L., and Schurr, J.\ M.\ ``The Effects of Direct and Hydrodynamic Forces on Macromolecular Diffusion'', \emph{Chem.\ Phys.} 38, 29-41 (1979).
\item Altenberger, A.\ R. ``Generalized Diffusion Processes and Light Scattering from a Moderately Concentrated Solution of Spherical Macroparticles'', \emph{Chem.\ Phys.} 15, 269-277 (1976).
\item Altenberger, A.\ R.\ and Deutch, J.\ M.\ ``Light Scattering from Dilute Macromolecular Solutions'', \emph{J.\ Chem.\ Phys.} 59, 894-898 (1973).
\item Altenberger, A.\ R.\ and Tirrell, M.\ ``Friction Coefficients in Self-Diffusion, Velocity Sedimentation, and Mutual Diffusion'', \emph{J.\ Polym.\ Sci.\ Polym.\ Phys.\ Ed.} 22, 909-910 (1984).
\item Altenberger, A.\ R.\ and Tirrell, M.\ ``Comment on 'Remarks on the Mutual Diffusion of Brownian Particles' '', \emph{J.\ Chem.\ Phys.} 84, 6527-6528 (1986).
\item Altenberger, A.\ R., Tirrell, M.\ and Dahler, J.\ S.\ ``Hydrodynamic Screening and Particle Dynamics in Porous Media, Semidilute Polymer Solutions and Polymer Gels'', \emph{J.\ Chem.\ Phys.} 84, 5122-5130 (1986).
\item Altenberger, A.\ R.\ and Dahler, J.\ S.\ ``On the Kinetic Theory and Rheology of a Solution of Rigid-Rodlike Macromolecules'', \emph{Macromolecules} 18, 1700-1710 (1985).
\item Altenberger, A.\ R.\ and Dahler, J.\ S.\ ``Rheology of Dilute Solutions of Rod-Like Macromolecules'', \emph{Int.\ J.\ of Thermophysics} 7, 585-597 (1986).
\item Altenberger, A.\ R., Dahler, J.\ S., and Tirrell, M.\ ``On the Theory of Dynamic Screening in Macroparticle Solutions'', \emph{Macromolecules} 21, 464-469 (1988).
\item Altenberger, A.\ R.\ ``On the Rayleigh Light Scattering from Dilute Solutions of Charged Spherical Macroparticles'', \emph{Optica Acta} 27, 345-352 (1980).
\item Altenberger, A.\ R., Dahler, J.\ S., and Tirrell, M.\ ``A Statistical Mechanical Theory of Transport Processes in Charged Particle Solutions and Electrophoretic Fluctuation Dynamics'', \emph{J.\ Chem.\ Phys.} 86, 4541-4547 (1987).
\item Altenberger, A.\ R.\ ``On the Theory of Generalized Diffusion Processes'', \emph{Acta Phys.\ Polonica} A46, 661-666 (1974).
\item Alley, W.\ E.\ and Alder, B.\ J.\ ``Modification of Fick's Law'', \emph{Phys.\ Rev.\ Lett.} 43, 653-656 (1979).
\item Arauz-Lara, J.\ L.\ and Medina-Noyola, M.\ ``Theory of Self-Diffusion of Highly Charged Spherical Brownian Particles'', \emph{J.\ Phys.\ A: Math.\ Gen.} 19, L117-L121 (1986).
\item Batchelor, G.\ K.\ ``Brownian Diffusion of Particles with Hydrodynamic Interaction'', \emph{J.\ Fluid Mech.} 74, 1-29 (1976).
\item Batchelor, G.\ K.\ ``Diffusion in a Dilute Polydisperse System of Interacting Spheres'', \emph{J.\ Fluid Mech.} 131, 155-175 (1983).
\item Beenakker, C.\ W.\ J.\ and Mazur, P.\ ``Self-Diffusion of Spheres in a Concentrated Suspension'', \emph{Physica} 120A, 388-410 (1983).

\item Beenakker, C.\ W.\ J.\ and Mazur, P.\ ``Diffusion of Spheres in a Concentrated Suspension II'', \emph{Physica} 126A, 349-370 (1984).
\item Beenakker, C.\ W.\ J.\ and Mazur, P.\ ``Is Sedimentation Container-Shape Dependent?'', \emph{Phys.\ Fluids} 28, 3203-3206 (1985).
\item Beenakker, C.\ W.\ J.\ ``Ewald Sum of the Rotne-Prager Tensor'', \emph{J.\ Chem.\ Phys.}
\item Beenakker, C.\ W.\ J.\ and Mazur, P.\ ``Diffusion of Spheres in Suspension: Three-Body Hydrodynamic Interaction Effects'', \emph{Phys.\ Lett.} 91A, 290-291 (1982).
\item Boissonade, J.\ ``The Screening Effect in Suspensions of Freely Moving Spheres'', \emph{J.\ Physique-Lett.} 43, L371-L375 (1982).
\item Carton, J.-P., Dubois-Violette, E., and Prost, J.\ ``Brownian Diffusion of a Small Particle in a Suspension I. Excluded Volume Effect'', \emph{Phy.\ Lett.} 86A, 407-408 (1981).
\item Carton, J.-P., Dubois-Violette, E., and Prost, J.\ ``Brownian Diffusion of a Small Particle in a Suspension, II. Hydrodynamic Effect in a Random Fixed Bed'', \emph{Physica} 119A, 307-316 (1983).
\item Chaturvedi, S.\ and Shibata, F.\ ``Time-Convolutionless Projection Operator Formalism for Elimination of Fast Variables. Application to Brownian Motion'', \emph{Z.\ Physik B} 35, 297-308 (1979).
\item Chow, T.\ S., and Hermans, J.\ J.\ ``Random Force Correlation Function for a Charged Particle in an Electrolyte Solution'' \emph{J.\ Coll.\ Interface Sci.} 45, 566-572 (1973).
\item Cichocki, B., and Hess, W.\ ``On The Memory Function for the Dynamic Structure Factor of
Interacting Brownian Particles'', \emph{Physica} 141 A, 475-488 (1987).
\item Combis, P., Fronteau, J., and Tellez-Arenas, A.\ ``Introduction to a Brownian Quasiparticle Model'', \emph{J.\ Stat.\ Phys.} 21, 439-446 (1979).
\item Cohen, E.\ G.\ D., De Schepper, I.\ M., and Campa, A.\ ``Analogy Between Light Scattering of Colloidal Suspensions and Neutron Scattering of Simple Fluids'', \emph{Physica} 147A, 142-151 (1987).
\item Cukier, R.\ I.\ ``Diffusion of Brownian Spheres in Semidilute Polymer Solutions'', \emph{Macromolecules} 17, 252-255 (1984).
\item Deulin, V.\ I.\ ``The Effect of Molecular Interaction on the Intrinsic Viscosity'', \emph{Makromol.\ Chem.} 180, 263-265 (1979).
\item Dieterich, W.\ and Peschel, I.\ ``Memory Function Approach to the Dynamics of Interacting Brownian Particles'', \emph{Physica} 95A, 208-224 (1979).
\item Deutch, J.\ M., and Oppenheim, I.\ ``Molecular Theory of Brownian Motion for Several Particles'', \emph{J.\ Chem.\ Phys.} 54, 3547-3555 (1971).
\item Felderhof, B.\ U., and Jones, R.\ B.\ ``Faxen Theorems for Spherically Symmetric Polymers in Solution'', \emph{Physica} 93A, 457-464 (1978).
\item Felderhof, B.\ U.\ ``Force Density Induced on a Sphere in Linear Hydrodynamics'', \emph{Physica} 84A, 557-568 (1976).
\item Felderhof, B.\ U.\ ``Diffusion of Interacting Brownian Particles'', \emph{J.\ Phys.\ A: Math.\ Gen.} 11, 929-937 (1978).
\item Felderhof, B.\ U.\ ``The Contribution of Brownian Motion to the Viscosity of Suspensions of Spherical Particles'', \emph{Physica} 147A, 533-543 (1988).
\item Gaylor, K., Snook, I., and van Megen, W.\ ``Comparison of Brownian Dynamics with Photon Correlation Spectroscopy of Strongly Interacting Colloidal Particles'', \emph{J.\ Chem.\ Phys.} 75, 1682-1689 (1981).
\item Gaylor, K., Snook, I., and van Megen, W.\ ``Brownian Dynamics of Many-body Systems'', \emph{J.\ Chem.\ Soc.\ Faraday Trans.\ II} 76, 1067-1078 (1980).
\item Gaylor, K., Snook, I., van Megen, W., and Watts, R.\ O.\ ``Dynamics of Colloidal Systems: Time-dependent Structure Factors'' \emph{J.\ Phys.\ A: Math.\ Gen.} 13, 2513-2520 (1980).

\item Golden, K.\ I., and De-xin, L.\ ``Dynamical Three-Point Correlations and Quadratic Response Functions in Binary Ionic Mixture Plasmas'', \emph{J.\ Stat.\ Phys.} 29, 281-307 (1982).
\item Harris, S.\ ``Self-Diffusion Coefficient for Dilute Macromolecular Solutions'', \emph{J.\ Phys.\ A: Math.\ Gen.} 10, 1905-1909 (1977).
\item Harris, S.\ ``Ion Diffusion and Momentum Correlation Functions in Dilute Electrolytes'', \emph{Molecular Phys.} 26, 953-958 (1973).
\item Hatziavramidis, D., and Muthukumar, M.\ ``Concentration Dependent Translational Self-Friction Coefficient of Rod-like Macromolecules in Dilute Suspensions'', \emph{J.\ Chem.\ Phys.} 83, 2522-2531 (1985).
\item Hanna, S., Hess, W., and Klein, R.\ ``Self-Diffusion of Spherical Brownian Particles with Hard-Core Interaction'', \emph{Physica} 111A, 181-199 (1982).
\item Hess, W., and Klein, R.\ ``Self-Diffusion Coefficient of Charged Brownian Particles'', \emph{J.\ Phys.\ A: Math.\ Gen.} 15, L669-L673 (1982).
\item Hess, W., and Klein, R.\ ``Theory of Light Scattering from a System of Interacting Brownian Particles'', \emph{Physica} 85A, 509-527 (1976).
\item Klein, R., and Hess, W.\ ``Dynamical Properties of Charged Spherical Brownian Particles'', \emph{Lecture Notes in Physics, Proc.\ International Conf.\ Ionic Liquids, Molten Salts, Polyelectrolytes}, Berlin, June 22-25, 1982.
\item Klein, R., and Hess, W.\ ``Mass Diffusion and Self-diffusion Properties in Systems of Strongly Charged Spherical Particles'', \emph{Faraday Discuss.\ Chem.\ Soc.} 76, 137-150 (1983).
\item Hess, W., and Klein, R. ``Long-time Versus Short-time Behavior of a System of Interacting Brownian Particles'', \emph{J.\ Phys.\ A: Math.\ Gen.} 13, L5-L10 (1980).
\item Hess, W., and Klein, R. ``Dynamical Properties of Colloidal Systems I. Derivation of Stochastic Transport Equations'', \emph{Physica} 94A, 71-90 (1978).
\item Hess, W. and Klein, R. ``Dynamical Properties of Colloidal Systems II. Correlation and Response Functions'', \emph{Physica} 99A, 463-493 (1979).
\item Hess, W. ``Wavevector and Frequency Dependent Longitudinal Viscosity of Systems of Interacting Brownian Particles'', \emph{Physica A} 107 190-200 (1981).
\item Hess, W., and Klein, R., ``Dynamical Properties of Colloidal Systems, III. Collective and Self-Diffusion of Interacting Charged Particles'', \emph{Physica A} 105, 552-576 (1981).
\item Hubbard, J.\ B., and McQuarrie, D.\ A.\ ``Mobility Fluctuations and Electrophoretic Light Scattering from Macromolecular Solutions'', \emph{J.\ Stat.\ Phys.} 52, 1247-1261 (1988).
\item Jones, R.\ B.\ and Burfield, G.\ S. ``An Analytical Model of Tracer Diffusivity in Colloid Suspensions'', \emph{Physica} 133A, 152-172 (1985).
\item Jones, R.\ B.\ ``Diffusion of Tagged Interacting Spherically Symmetric Polymers'', \emph{Physica} 97A, 113-126 (1979).
\item Jones, R.\ B., and Burfield, G.\ S.\ ``Memory Effects in the Diffusion of an Interacting Polydisperse Suspension'', \emph{Physica} 111A, 562-576 (1982).
\item Klein, R., and Hess, W. ``Dynamics of Suspensions of Charged Particles'', \emph{J.\ de Physique, Colloques} 46, pp.C3-211-C3-222 (1985).
\item Kirkwood, J.\ G., Baldwin, R.\ L., Dunlop, P.\ J., Gosting, L.\ J., and Kegeles, G.\ ``Flow Equations and Frames of Reference for Isothermal Diffusion in Liquids'', \emph{J.\ Chem.\ Phys.} 33, 1505-1513 (1960).
\item Kynch, G.\ J. ``The Slow Motion of Two or More Spheres Through a Viscous Fluid'', \emph{J. Fluid Mech.} 5, 193-208 (1959).
\item Ladd, A.\ J.\ C. ``Hydrodynamic Interactions and the Viscosity of Suspensions of Freely Moving Spheres'', \emph{J.\ Chem.\ Phys.} 90, 1149-1157 (1989).

\item Ladd, A.\ J.\ C. ``Hydrodynamic Interactions in a Suspension of Spherical Particles'', \emph{J.\ Chem.\ Phys.} 88, 5051-5063 (1988).
\item Lee, W.\ I.\ and Schurr, J.\ M.\ ``Effect of Long-Range Hydrodynamic and Direct Intermacromolecular Forces on Translational Diffusion'', \emph{Chem.\ Phys.\ Lett.} 38, 71-74 (1976).
\item London, R.\ E.\ ``Force-Velocity Cross Correlations and the Langevin Equation'', \emph{J.\ Chem.\ Phys.} 66, 471-479 (1977).
\item Marqusee, J.\ A.\ and Deutch, J.\ M.\ ``Concentration Dependence of the Self-Diffusion Coefficient'', \emph{J.\ Chem.\ Phys.} 73, 5396-5397 (1980).
\item Masters, A.\ J.\ ``Time-scale Separations and the Validity of Smoluchowski, Fokker-Planck and Langevin Equations as Applied to Concentrated Particle Suspensions'', \emph{Molecular Phys.} 59, 303-317 (1986).
\item Mazo, R.\ M.\ ``On the Theory of Brownian Motion. I. Interaction Between Brownian Particles'', \emph{J.\ Stat.\ Phys.} 1, 89-98, (1969).
\item Mazo, R.\ M.\ ``On the Theory of the Concentration Dependence of the Self-Diffusion Coefficient of Micelles'', \emph{J.\ Chem.\ Phys.} 43, 2873-2877 (1965).
\item Mon, C.\ Y.\ and Chang, E.\ L.\ ``Concentration Dependence of Diffusion of Interacting Spherical Brownian Particles'' (Preprint only) (1978).
\item Ohtsuki, T.\ ``Dynamical Properties of Strongly Interacting Brownian Particles-II. Self-Diffusion'', \emph{Physica} 110A, 606-616 (1982).
\item Phillies, G.~D.~J. ``Effect of Intermacromolecular Interactions on Diffusion. III. Electrophoresis in Three-Component Solutions'', \emph{J.\ Chem.\ Phys.} 59, 2613-2617 (1973).
\item Phillies, G.~D.~J. ``Effect of Intermacromolecular Interactions on Diffusion. I. Two-Component Solutions'', \emph{J.\ Chem.\ Phys.} 60, 976-982 (1974).
\item Phillies, G.~D.~J. ``Effect of Intermacromolecular Interactions on Diffusion. II. Three-Component Solutions'', \emph{J.\ Chem.\ Phys.} 60, 983-989 (1974).
\item Phillies, G.~D.~J. ``Excess Chemical Potential of Dilute Solutions of Spherical Polyelectrolytes'', \emph{J.\ Chem.\ Phys.} 60, 2721-2731 (1974).
\item Phillies, G.~D.~J. ``Continuum Hydrodynamic Interactions and Diffusion'', \emph{J.\ Chem.\ Phys.} 62, 3925-3962 (1975).
\item Phillies, G.~D.~J. ``Fluorescence Correlation Spectroscopy and Non-Ideal Solutions'', \emph{Biopolymers} 14, 499-508 (1975).
\item Phillies, G.~D.~J. ``Contribution of Slow Charge Fluctuations to Light Scattering from a Monodisperse Solution of Macromolecules'', \emph{Macromolecules} 9, 447-450 (1976).
\item Phillies, G.~D.~J. ``Diffusion of Spherical Macromolecules at Finite Concentration'', \emph{J.\ Chem.\ Phys.} 65, 4334-4335 (1976).
\item Phillies, G.~D.~J. ``Heterodyne Coincidence Spectroscopy: Direct Probe of Three-Body Correlations in Fluids'', \emph{Molecular Physics} 32, 1695-1702 (1976).
\item Phillies, G.~D.~J. ``On the Contribution of Non-Hydrodynamic Interactions to the Concentration Dependence of the Drag Coefficient of Rigid Macromolecules'', \emph{J.\ Chem.\ Phys.} 67, 4690-4695 (1977).
\item Phillies, G.~D.~J. ``On the Contribution of Direct Intermacromolecular Interactions to the Viscosity of a Suspension of Hard Spheres'', \emph{J.\ Chem.\ Phys.} 71, 1492-1494 (1979).
\item Phillies, G.~D.~J., and Kivelson, D.\ ``Theory of Light Scattering from Density Fluctuations in a Two-Component Reacting Fluid'', \emph{Molecular Physics} 38, 1393-1410 (1979).
\item Phillies, G.~D.~J. ``Contribution of Nonhydrodynamic Interactions to the Concentration Dependence of the Friction Factor of the Mutual Diffusion Coefficient'', \emph{J.\ Chem.\ Phys.} 74, 2436-2440 (1981).
\item Phillies, G.\ D.\ J., and Wills, P.\ R.\ ``Light Scattering Spectrum of a Suspension of Interacting Brownian Macromolecules'', \emph{J.\ Chem.\ Phys.} 75, 508-514 (1981).
\item Phillies, G.~D.~J. ``Suppression of Multiple Scattering Effects in Quasi-Elastic Light Scattering by Homodyne Cross-Correlation Techniques'', \emph{J.\
    Chem.\ Phys.}  74, 260-262 (1981).
\item Phillies, G.~D.~J. ``Experimental Demonstration of Multiple-Scattering Suppression in Quasi-Elastic-Light-Scattering by Homodyne Coincidence Techniques'', \emph{Phys.\ Rev.\ A} 24, 1939-1943 (1981).
\item Phillies, G.~D.~J. ``Reactive Contribution to the Apparent Translational Diffusion Coefficient of a Micelle'', \emph{J.\ Phys.\ Chem.} 85, 3540-3541 (1981).
\item Phillies, G.~D.~J. ``Interpretation of Micelle Diffusion Coefficients'', \emph{J.\ Coll.\ Interface Sci.} 86, 226-233 (1982).
\item Phillies, G.~D.~J. ``The Second Order Concentration Corrections to the Mutual Diffusion Coefficient of Brownian Macroparticles'', \emph{J.\ Chem.\ Phys.} 77, 2623-2631 (1982).
\item Phillies, G.~D.~J. ``Non-Hydrodynamic Contribution to the Concentration Dependence of the Self Diffusion of Interacting Brownian Macroparticles'', \emph{Chemical Physics} 74, 197-203 (1983).
\item Phillies, G.~D.~J. ``Hidden Correlations and the Behavior of the Dynamic Structure Factor at Short Times'', \emph{J.\ Chem.\ Phys.} 80, 6234-6239 (1984).
\item Phillies, G.~D.~J. ``Why Does the Generalized Stokes-Einstein Equation Work?'', \emph{Macromolecules} 17, 2050-2055 (1984).
\item Phillies, G.~D.~J. ``Translational Drag Coefficients of Assemblies of Spheres with Higher-Order Hydrodynamic Interactions'', \emph{J.\ Chem.\ Phys.} 81, 4046-4052 ( 1984).
\item Carter, J.~M. and Phillies, G.~D.~J. ``Second-Order Concentration Correction to the Mutual Diffusion Coefficient of a Suspension of Hard Brownian Spheres'', \emph{J.\ Phys.\ Chem.} 89, 5118-5124 (1985).
\item Phillies, G.~D.~J.  ``Comment on `Remarks on the Mutual Diffusion Coefficient of Brownian Particles''', \emph{J.\ Chem.\ Phys.} 84, 5972-5973 (1986).
\item Ullmann, G.~S.\ and Phillies, G.~D.~J. ``Diffusion of Particles in a Fluctuating Fluid with Creeping Flows'', \emph{J.\ Phys.\ Chem.} 90, 5473-5477 (1986).
\item Phillies, G.~D.~J. ``Numerical Interpretation of the Concentration Dependence of Micelle Diffusion Coefficients'', \emph{J.\ Coll.\ Interface Sci.} 119, 518-523 (1987).
\item Phillies, G.~D.~J. and Kirkitelos, P.~C. ``Higher-Order Hydrodynamic Interactions in the Calculation of Polymer Transport Properties'', \emph{Journal of Polymer Science B: Polymer Physics} 31, 1785-1797 (1993).
\item Phillies, G.~D.~J. ``Dynamics of Brownian Probes in the Presence of Mobile or Static Obstacles'', \emph{Journal of Physical Chemistry} 99 4265-4272 (1995).
\item Phillies, G.~D.~J. ``Low-Shear Viscosity of Non-Dilute Polymer Solutions from a Generalized Kirkwood-Riseman Model'', \emph{Journal of Chemical Physics} 116, 5857-5866 (2002).
\item Merriam, S.~C.  and Phillies, G.~D.~J., ``Fourth-Order Hydrodynamic Contribution to the Polymer Self-Diffusion Coefficient'', \emph{Journal of Polymer Science B} 42, 1663-1670 (2004).
\item Phillies, G.~D.~J. ``Interpretation of Light Scattering Spectra in Terms of Particle Displacements'', \emph{Journal of Chemical Physics} 122 224905 1-8 (2005).
\item Phillies, G.~D.~J. ``Interpretation of Pulsed-Field-Gradient NMR in Terms of Particle Displacements'', \emph{arxiv:cond-mat.soft}/ (2011).
\item Phillies, G.~D.~J. ``Position-Displacement Correlations in QELSS Spectra of Non-Dilute Colloids'', \emph{Journal of Chemical Physics} 137, 124901 1-4 (2012)
\item Phillies, G.~D.~J. ``Interpretation of Quasielastic Scattering Spectra of Probe Species in Complex Fluids'', \emph{Journal of Chemical Physics}, 139, 034902 (2013).
\item Phillies, G.~D.~J. ``The Gaussian Diffusion Approximation for Complex Fluids is Generally Invalid'',  \emph{arXiv}:1406.4894 (2013).
\item Phillies, G.~D.~J. ``The Gaussian Diffusion Approximation is Generally Invalid in Complex Fluids'', \emph{Soft Matter} 11, 580-586 (2015).
\item Pomeau, Y.\ and Resibois, P.\ ``Time Dependent Correlation Functions and Mode-Mode Coupling Theories'', \emph{Phys.\ Reports} 19, 63-139 (1975).
\item Pusey, P.~N., Fijnaut, H.~M., and Vrij, A.\ ``Mode Amplitudes in Dynamic Light Scattering by Concentrated Liquid Suspensions of Polydisperse Hard Spheres'', \emph{J.\ Chem.\ Phys.} 77, 4270-4281 (1982).
\item Pusey, P.\ N. and Tough, R.\ J.\ A., ``Langevin Approach to the Dynamics of Interacting Brownian Particles'', \emph{J.\ Phys.\ A: Math.\ Gen.} 15, 1291-1308 (1982).
\item Rowe, A.~J. ``The Concentration Dependence of Transport Processes: A General Description Applicable to the Sedimentation, Translation Diffusion, and Viscosity Coefficients of Macromolecular Solutes'', \emph{Biopolymers} 16, 2595-2611 (1977).
\item Schurr, J.~M. ``The Thermodynamic Driving Force in Mutual Diffusion of Hard Spheres'', \emph{Chem.\ Phys.} 65, 217-223 (1982).
\item Schurr, J.~M. ``Dynamic Light Scattering and Mutual Diffusion in Non-Ideal Systems, One- and Multi-component Spherical Solutes'', \emph{J. Chem.\ Phys.} 111, 55-86 (1986).
\item Schurr, J.~M. ``The Fluctuating-Force Formulation of Friction Drag Coefficients'', \emph{Chem.\ Phys.} 71, 101-104 (1982).
\item Schurr, J.~M. ``A Theory of Electrolyte Friction on Translating Polyelectrolytes'', \emph{Chem.\ Phys.} 45, 119-132 (1980).
\item Schurr, J.~M. ``Theory of Dynamic Light Scattering by Polymers and Gels'', \emph{Chem.\ Phys.} 30, 243-247 (1978).
\item Snook, I.\ and van Megen, W.\ ``Calculation of the Wave-Vector Dependent Diffusion Constant in Concentrated Electrostatically Stabilized Dispersions'', \emph{J.\ Coll.\ Interface Sci.} 100, 194-202 (1984).
\item Titulaer, U.~M. ``Corrections to the Smoluchowski Equation in the Presence of Hydrodynamic Interactions'', \emph{Physica} 100A, 251-265 (1980).
\item Tokuyama, M., and Oppenheim, I. ``On the Theory of Concentrated Hard-Sphere Suspensions'', \emph{Physica A} 216, 85-119 (1995).
\item Tokuyama, M. ``Nonequilibrium Effects on Slow Dynamics in Concentrated Colloidal Suspensions'', \emph{Phys.\ Rev.\ E} 54, R1062-R1065 (1996).
\item Tokuyama, M., Enomoto, Y., and Oppenheim, I. ``Slow Dynamics of Nonequilibrium Density Fluctuations in Hard-Sphere Suspensions'', \emph{Phys.\ Rev.\ E} 55 R29-R32 (1997).
\item Tokuyama, M. ``Slow Dynamics of Nonequilibrium Density Fluctuations in Hard-Sphere Suspensions'', \emph{Progr.\ Theor.\ Physics Supplement (Japan)} 126, 43-47 (1997).
\item Tokuyama, M. ``Theory of Slow Dynamics in Highly Charged Colloidal Suspensions'', \emph{Phys.\ Rev.\ E} 58, R2729-R2732 (1998).
\item Tough, R.\ J.\ A.\ ``Self Diffusion in a Suspension of Interacting Brownian Particles'', \emph{Molecular Phys.} 46, 465-474 (1982).

\item Tough, R.\ J.\ A., Pusey, P.\ N., Lekkerkerker, H.\ N.\ W., and van den Broeck, C.\ ``Stochastic Descriptions of the Dynamics of Interacting Brownian Particles'', \emph{Molecular Phys.} 59, 595-619 (1986).
\item Tsang, T.\ ``Dynamics of Interacting Brownian Particles'', \emph{Chem.\ Phys.\ Lett.} 112, 220-223 (1984).
\item Van Saarloos, W.\ and Mazur, P.\ ``Many-Sphere Hydrodynamic Interactions II. Mobilities at Finite Frequencies'', \emph{Physica} 120A, 77-102 (1983).
\item Vink, H.\ ``Comments on the Onsager Reciprocal Relations in the Frictional Formalism of Non-equilibrium Thermodynamics'', \emph{Acta Chemica Scandinavica} A38, 335-336 (1984).
\item Vink, H.\ ``Viscous Effects in the Frictional Formalism of Non-equilibrium Thermodynamics'', \emph{J.\ Chem.\ Soc., Faraday Trans.} 1, 79, 2355-2358 (1983).
\item Weissman, M.\ B.\ and Ware, B.\ R.\ ``Applications of Fluctuation Transport Theory'', \emph{J.\ Chem.\ Phys.} 68, 5069-5076 (1978).
\item Wolynes, P.\ G.\ and Deutch, J.\ M.\ ``Dynamical Orientation Correlations in Solution'', \emph{J.\ Chem.\ Phys.} 67, 733-741 (1977).

\end{enumerate}

\end{document}